\documentclass[fleqn,usenatbib]{mnras}

\usepackage[T1]{fontenc}
\usepackage{graphicx}
\usepackage[cmex10]{amsmath}
\usepackage{array}
\usepackage{fixltx2e}
\usepackage{url}
\usepackage{amsfonts}
\usepackage{lmodern}
\usepackage{algpseudocode}
\usepackage{algorithmicx}
\usepackage{algorithm}
\usepackage{float}
\usepackage{xfrac}
\usepackage{multirow}
\usepackage{balance}
\usepackage{media9}
\usepackage{afterpage}

\newcommand{\bc}{}
\newcommand{\ec}{}

\DeclareMathOperator{\prox}{prox}

\DeclareMathOperator{\soft}{\op{S}}
\DeclareMathOperator{\identity}{\op{I}}

\DeclareMathOperator{\proj}{\op{P}}
\DeclareMathOperator*{\argmin}{argmin}

\newcommand{\ds}{\displaystyle}
\newcommand{\hquad}{\;\:}

\newcommand{\bs}{\boldsymbol}
\newcommand{\bm}[1]{\boldsymbol{\mathsf{#1}}}
\newcommand{\bb}{\mathbb}
\newcommand{\mc}{\mathcal}
\newcommand{\diff}{\mathrm{d}\hspace{-0.1ex}}
\newcommand{\acp}[1]{(\uppercase{#1})}
\newcommand{\ac}[1]{\uppercase{#1}}
\newcommand{\sw}[1]{\textsc{#1}}
\newcommand{\op}[1]{\boldsymbol{\mc{#1}}}

\algblock[Block]{Block}{EndBlock}
\algblockdefx[Block]{Block}{EndBlock}%
[1]{{#1}}%
[1]{{#1}}

\newcommand{\Given}[1]{\State{\bf given} {#1}}
\newcommand{\Output}[1]{\State{\bf output} {#1}}

\newcommand{\RepeatFor}[1]{\Repeat {\bf~for} {#1}}

\newcommand{\ParForD}[2]{\Block{{#1} {\bf distribute} {#2} {\bf and do in parallel}}}
\newcommand{\EndParForD}[1]{\EndBlock{\bf end and gather} {#1}}

\newcommand{\ParFor}[1]{\Block{{#1} \bf do~in~parallel}}
\newcommand{\EndParFor}{\EndBlock{\bf end}}

\newcommand{\Set}[1]{\Block{{#1} \bf set}}
\newcommand{\EndSet}{\EndBlock{\bf end}}

\title[Preconditioned primal-dual algorithm]{An accelerated splitting algorithm for radio-interferometric imaging: when natural and uniform weighting meet}

\author[A. Onose et al.]{
Alexandru Onose\thanks{E-mail: a.onose@hw.ac.uk},
Arwa Dabbech
and Yves Wiaux
\\
Institute of Sensors, Signals and Systems, Heriot-Watt University, Edinburgh EH14 4AS, UK\\
}

\date{Accepted XXX. Received YYY; in original form ZZZ}

\pubyear{2017}

\begin{document}
\label{firstpage}
\pagerange{\pageref{firstpage}--\pageref{lastpage}}
\maketitle

\begin{abstract}
Next generation radio-interferometers, like the Square Kilometre Array, will acquire \bc large \ec amounts of data with the goal of improving the size and sensitivity of the reconstructed images by orders of magnitude.
The efficient processing of large-scale data sets is of great importance.
We propose an acceleration strategy for a recently proposed primal-dual distributed algorithm. 
A preconditioning approach can incorporate into the algorithmic structure both the sampling density of the measured visibilities and the noise statistics.
Using the sampling density information greatly accelerates the convergence speed, especially for highly non-uniform sampling patterns, while relying on the correct noise statistics optimises the sensitivity of the reconstruction.
In connection to \sw{clean}, our approach can be seen as including in the same algorithmic structure both natural and uniform weighting, thereby simultaneously optimising both the resolution and the sensitivity.
The method relies on a new non-Euclidean proximity operator for the data fidelity term, that generalises the projection onto the $\ell_2$ ball where the noise lives for naturally weighted data, to the projection onto a generalised ellipsoid incorporating sampling density information through uniform weighting.
Importantly, this non-Euclidean modification is only an acceleration strategy to solve the convex imaging problem with data fidelity dictated only by noise statistics.
We \bc show \ec through simulations with realistic sampling patterns the acceleration obtained using the preconditioning.
We also investigate the algorithm performance for the reconstruction of the 3C129 radio galaxy from real visibilities and compare with multi-scale \sw{clean}, showing better sensitivity and resolution.
Our \sw{matlab} code is available online on GitHub.
\end{abstract}


\ec
\section{Introduction}

Radio-interferometry \acp{ri} is a technique that permits the observation of radio emissions with great sensitivity and angular resolution. 
It provides valuable data for many research directions in astronomy, cosmology or astrophysics~\citep{Thompson2007}.
The next-generation radio telescopes, like the planned Square Kilometre Array~\citep[\ac{ska};][]{Dewdney2009}, are expected to push the sensitivity further to achieve a dynamic range of six or seven orders of magnitude and to reconstruct large, giga-pixel size, images.
To achieve such a feat, the amount of data to be acquired will be huge and the signal processing techniques from \ac{ri} need to be revisited and reinvented.
Fast specialised algorithmic solvers are being developed~\citep{Onose2016, Deguignet2016, Ferrari2014,Yatawatta2015, Yatawatta2016, Carrillo2014} and vigorous research is being directed towards tackling the challenges of both \ac{ri} imaging and \ac{ri} calibration~\citep{Rau2009, Wijnholds2014}.

The \ac{ska}, whose construction is scheduled to start in 2018, will be comprised of a huge number of antennas, approximately $131~000$ low frequency elements and 197 dishes for medium frequency~\citep{Dewdney2009, Broekema2015}.
With an expected number of $65~000$ frequency bands of operation, the data rates estimates will be in the terabits per second range~\citep{Broekema2015} and will present a challenge for both the communication infrastructure and signal processing.
The current standard algorithmic solvers, belonging to the \sw{clean} family  \citep{Hogbom1974, Schwab1984, Bhatnagar2004, Cornwell2008a}, do not scale well to such tremendous data sizes.

Recently, convex optimisation techniques coupled with compressive sensing models~\citep{Wiaux2009a, Li2011, Carrillo2012, Garsden2015} have been shown to potentially outperform the standard state-of-the-art \sw{clean} imaging algorithms.
Such methods typically approach the imaging problem by minimising a convex objective function defined as a sum of multiple terms: several data terms dependent on the measured data (the visibilities), and a number of regularisation priors usually promoting sparsity or smoothness in an appropriate domain and positivity.
This is a global approach, all algorithms searching for the unique solution that minimises the convex objective function.

Besides the reconstruction quality, the processing speed is of great interest with fast and parallelizable algorithms having been recently proposed \citep{Carrillo2014, Ferrari2014, Yatawatta2015, Onose2016}.
Such approaches come in contrast with the standard \sw{clean} methods which employ local procedures and rely on greedy updates and other signal pre-processing steps, like the \ac{ri} weighting used to mitigate the effects produced by an unbalanced density profile of the sampling strategy.
For algorithms that work directly in image space, like \sw{clean}, the type of \ac{ri} weighting is very important and affects the overall image reconstruction results \citep{Briggs1995, Boone2013, Yatawatta2014}.
Natural weighting provides controlled noise statistics with the aim of maximising the sensitivity. 
Uniform weighting reduces the side-lobes of the point spread function by scaling the visibilities with the inverse sampling density and provides better resolution at the cost of lowered sensitivity.
Since any weighting other than natural essentially biases the data, \sw{clean} is not able to maximise both resolution and sensitivity.
To mitigate this, intermediate robust weighting \citep{Briggs1995} or adaptive weighting schemes \citep{Yatawatta2014} have also been proposed and serve as a tradeoff between resolution and sensitivity.

Convex optimisation methods \citep{Carrillo2012, Carrillo2014} that impose constraints directly in visibility space work with naturally weighting data.
Such approaches can optimise both the resolution and sensitivity, which is impossible to achieve with \sw{clean} and its evolutions.
An unbalanced density profile of the sampling strategy does not influence the final solution of the convex optimisation problem. It can have however a potentially significant detrimental effect on the convergence speed of the algorithmic structures. 

We study herein an acceleration strategy of the primal-dual \acp{pd} algorithmic structure recently proposed by \cite{Onose2016}.  It can incorporate sampling density information into the algorithmic structure to achieve faster convergence speed for non uniform visibility distributions in $u$--$v$ space.
We propose the use of a preconditioning strategy that improves the convergence speed significantly, making the \ac{pd} approach even more appealing for the large-scale signal processing associated with the future radio telescopes.
We rely on the same convex optimisation problem from \cite{Onose2016} but introduce a non-euclidian, skewed, proximity step that uses a preconditioning matrix reminiscent of the uniform weighting used by \sw{clean} and the other \ac{ri} imaging methods that work in image space.
Intuitively, to link with the behaviour of \sw{clean}, such an approach maintains the sensitivity of the natural weighting but achieves the resolution of the uniformly weighted data.

We show through simulations the acceleration achieved using the preconditioning strategy for simulated random, \ac{ska} and \ac{vla} coverages.
A study of the computational burden of the non-euclidian proximity step is also included.
We also showcase the reconstruction capabilities of the algorithm using real interferometric data of the 3C129 radio galaxy and compare with \sw{clean}.
The observations were performed for two 50 MHz channels using the \ac{vla} in configuration B and C.

The remainder of this article is organised as follows.
Section \ref{sec-ri-intro} introduces the \ac{ri} problem and briefly reviews the current existing standard solvers.
Section \ref{sec-opt-ri} presents the main convex optimisation problem we associate with the image reconstruction and introduces the tools used by the preconditioned \ac{pd} solver.
Sections \ref{sec-algo} details the proposed preconditioned \ac{pd} algorithm and the acceleration strategy.
Extensive simulations and results are presented in Section \ref{sec-sims}.
Section \ref{sec-conc} presents our final remarks and future work directions.

\section{Radio-interferometric imaging}
\label{sec-ri-intro}
In \ac{ri}, the measured data, the visibilities, are produced by an array of geographically separated antennas that are paired to measure radio emissions from a given area of the sky.
Under the simplifying assumptions of non-polarised monochromatic \ac{ri} imaging, the measurement equation for a measured visibility point $y(\bs{u})$ can be stated as
\begin{equation}
	y(\bs{u}) = \int D(\bs{l},\bs{u}) x(\bs{l}) e^{-2 i \pi \bs{u} \cdot \bs{l}} \diff^2 \bs{l},
	\label{measurement-eq}
\end{equation}
with the direction dependent effects (\ac{dde}s) that affect the measurements, modelled through $D (\bs{l}, \bs{u})$.
Here, we denote by $\bs{u} = (u,v)$, the projected baseline components in the orthogonal plane relative to the line of sight.
The observed sky brightness is described in the same coordinate system, with coordinates $(l, m)$.
We denote $\bs{l} = (l,m)$.
The well-known $w$ component effect, associated with the baseline components in the line of sight, is a known \ac{dde}. Unknown \ac{dde}s related to primary beam and ionospheric effects are assumed to have been properly calibrated so that we consider here a pure imaging problem.

The reconstruction algorithms work with a discretised version of the inverse problem (\ref{measurement-eq}).
This resolves to the linear measurement equation
\begin{equation}
	\bs{y} = \bm{\Phi} \bs{x} + \bs{n},
	\label{inverse-problem}
\end{equation}
where $\bs{x} \in \bb{R}^N$ is the unknown intensity image of interest of which $M$ visibility measurements $\bs{y} \in \bb{C}^M$ are taken by the radio telescope array. 
The measurements are corrupted by additive noise $\bs{n}$, each component $n_e$ assumed to have a known variance $\sigma = \sigma_e, \forall e$.
The measurement operator $\bm{\Phi}=\bm{\Theta}\bm{G}\bm{F}\bm{Z}$ is a linear map from the image space to the visibility domain.
It is composed of the matrix $\bm{G}\in \bb{C}^{M \times nN}$ containing compact support interpolation kernels \citep{Fessler2003} and modeling the DDEs, an $n$-oversampled Fourier operator $\bm{F} \in \bb{C}^{nN\times nN}$ and an oversampling and scaling operator $\bm{Z} \in \bb{R}^{nN \times N}$ that pre-compensates for the interpolation \citep{Fessler2003}.
If the original visibilities are affected by noise with different variances, $\sigma_{e_1} \neq \sigma_{e_2}$ for some $e_1$ and $e_2$, a diagonal matrix $\bm{\Theta}$ with diagonal elements $\theta_{e, e} = \frac{1}{\sigma_e}$ is used to whiten the noise.
This is equivalent to the natural weighting performed in \ac{ri}.

\subsection{The CLEAN method}
The inverse problem defined by (\ref{inverse-problem}) has been thoroughly studied and various deconvolution methods have been proposed.
 The standard imaging algorithms, belonging to the \sw{clean} family, perform a greedy non-linear deconvolution based on local iterative beam removal \citep{Hogbom1974, Schwarz1978, Schwab1984, Thompson2007}.
They rely on a sparsity prior on the solution implicitly introduced through the greedy, pixel by pixel, image reconstruction procedure. 
This resembles the matching pursuit (MP) algorithm \citep{Mallat1993}. It can also be seen as a regularised gradient descent method that minimises the residual norm $\| \bs{y} - \bm{\Phi} \bs{x} \|_2^2$ via a gradient descent subject to an implicit sparsity constraint on $\bs{x}$ \citep{Rau2009},
\begin{equation}
	\bs{x}^{(t)} = \bs{x}^{(t-1)} + \op{T} \Big( \bm{\Phi}^\dagger \big( \bs{y} - \bm{\Phi} \bs{x}^{(t-1)} \big) \Big).
\end{equation}
%We denote here by $\bm{\Phi}^\dagger$ the adjoint of the linear operator $\bm{\Phi}$.
\bc The notation $^\dagger$ denotes the adjoint of a linear operator. \ec
Multiple versions and improvements have been suggested, multi-scale \sw{clean} \citep{Cornwell2008a}, adaptive scale \sw{clean} \citep{Bhatnagar2004}.
In parallel with \sw{clean}, the maximum entropy method solvers \citep{Ables1974, Gull1978, Cornwell1985} have been proposed but in practice \sw{clean} was favoured.

\subsection{Convex optimisation algorithms}
Recently, convex optimisation methods are beginning to gain traction in \ac{ri} and offer improved reconstruction quality and speed over the classical \sw{clean} approaches \citep{Wiaux2009a, Wiaux2009b, Wenger2010, Li2011, Carrillo2012, Carrillo2014, Ferrari2014, Yatawatta2015, Garsden2015, Dabbech2015, Onose2016}.
They approach the imaging problem under the framework of compressed sensing \acp{cs}.
Such methods add a regularisation of the ill-posed reconstruction problem in the form of a prior that assumes a low dimensional signal model \citep{Donoho2006, Candes2006}.
Seen through the \ac{cs} framework, the signal of interest $\bs{x}$ is considered to have a sparse representation, $\bs{x}=\bm{\Psi} \bs{\alpha}$ with $\bs{\alpha}\in \bb{C}^{D}$ containing only a few nonzero elements \citep{Fornasier2011}. The dictionary $\bm{\Psi} \in \bb{C}^{N \times D}$ is usually a collection of wavelet bases or, more generally, an over-complete frame.

An analysis-based approach \citep{Elad2007} to recover the signal of interest $\bs{x}$ by solving the ill-posed inverse problem (\ref{inverse-problem})  can be formally stated as \citep{Carrillo2012, Carrillo2013, Carrillo2014, Onose2016}
\begin{equation}
	\min_{\bs{x}} \| \bm{\Psi}^\dagger \bs{x} \|_0 \hquad \rm{subject~to} \hquad \|\bs{y} - \bm{\Phi} \bs{x}\|_2 \leq \epsilon \hquad \rm{and} \hquad \bs{x} \in \bb{R}_+^{N}.
	\label{analysis-l1-problem}
\end{equation}
The sparsity averaging reweighed analysis \acp{sara} sparsity prior \citep{Carrillo2012}, used as the sparsity dictionary $\bm{\Psi}$, has been shown to be a good sparsity basis.
Since the solution $\bs{x}$ is an intensity image, a reality and positivity prior is also assumed.
Data fidelity is enforced by constraining the residual to belong to an $\ell_2$ ball defined given an estimate $\epsilon$ of the noise affecting the measurements.
Synthesis-based approaches have also been proposed \citep{Wiaux2009a, Wiaux2009b, McEwen2011}.

The $\ell_0$ norm is non-convex and thus the problem defined in (\ref{analysis-l1-problem}) is intractable.
By replacing the $\ell_0$ norm with its closest convex relaxation, the $\ell_1$ norm, and by reformulating the constraints from (\ref{analysis-l1-problem}) with the use of the indicator function\footnote{
The indicator function $\iota_{\mc{C}}$ of a convex set $\mc{C}$ is defined as
\begin{equation}
	(\forall \bs{z}) \quad \iota_{\mc{C}} (\bs{z}) \overset{\Delta}{=} \left\{ \begin{aligned}
					0 & \quad \bs{z} \in \mc{C} \\
					+\infty & \quad \bs{z} \notin \mc{C}.
				   \end{aligned} \right. \nonumber
\end{equation}} $\iota_{\mc{C}}$
we can state a basic minimisation problem as
\begin{equation}
	\min_{\bs{x}} f(\bs{x}) + l(\bm{W}^\dagger \bm{\Psi}^\dagger \bs{x}) + h(\bm{\Phi}\bs{x}).
	\label{basic-min-problem}
\end{equation}
The function $f = \iota_{\bb{R}^N_+}$ introduces the reality and positivity requirements for the recovered solution, the function $l = \| \cdot \|_1$ represents the sparsity inducing prior and $h(\bs{z}) = \iota_{\mc{B}}  (\bs{z}), \mc{B} = \{ \bs{z} \in \bb{C}^M: \| \bs{z} - \bs{y} \|_2 \leq \epsilon \}$ is the data fidelity term constraining the residual to be situated in the $\ell_2$ ball \bc $\mc{B}$ \ec defined by the noise level $\epsilon$.
%The notation $^\dagger$ denotes the adjoint of a linear operator.
A re-weighted $\ell_1$ approach \citep{Candes2008} is generally used to approximate the $\ell_0$ norm by imposing the weights $\bm{W}$ on the operator $\bm{\Psi}$ and solving sequentially several $\ell_1$ problems with different $\bm{W}$.
This basic minimisation problem \citep{Carrillo2012, Onose2016} has been approached using several state-of-the-art algorithmic solvers: the simultaneous direction method of multipliers \citep{Carrillo2014}, the alternating direction method of multipliers and a \ac{pd} algorithm with forward-backward iterations \citep{Onose2016}.

The forward-backward iterative structure is one of the main pillars used in the algorithmic structure presented herein.
We can view it as being conceptually extremely close to the major-minor cycle structure of \sw{clean}.
Consider one of the most basic approaches, the unconstrained version of the minimisation problem (\ref{analysis-l1-problem}), namely %
$
\min_{\bs{x}} \| \bm{W}^\dagger \bm{\Psi}^\dagger \bs{x} \|_1 + \rho \|\bs{y} - \bm{\Phi} \bs{x}\|^2_2
$,
with $\rho$ a free parameter.
This can be solved using forward-backward iterations by performing a gradient step together with a proximal step \citep{Moreau1965},%
\begin{equation}
	\prox_g (\bs{z}) \overset{\Delta}{=} \argmin_{\bar{\bs{z}}} g(\bar{\bs{z}}) + \frac{1}{2} \| \bs{z} - \bar{\bs{z}}\|_2^2.
	\label{proximity-operator-basic}
\end{equation}
The forward gradient step consists in doing a step in the opposite direction to the gradient of the $\ell_2$ norm of the residual.
This is essentially equivalent to a major cycle of \sw{clean}. 
In this particular case, the proximal step is a simple soft-thresholding operation in the given basis $\bm{W}^\dagger \bm{\Psi}^\dagger$ \citep{Combettes2007b}.
It consists in decreasing the absolute values of all the coefficients of $\bm{W}^\dagger \bm{\Psi}^\dagger\bs{x}$ that are above a certain threshold by the threshold value, and setting to zero those below the threshold.
Such an approach is very similar to the minor cycle of \sw{clean}, with the soft-threshold value being an analogous to the loop gain factor. 
\sw{clean} iteratively builds up the signal by picking up parts of the most important coefficients until the residuals become negligible.
The soft-thresholding acts globally by removing small and insignificant coefficients, on all signal locations simultaneously.
As such, \sw{clean} can be intuitively understood as a very specific version of the forward-backward algorithm.

\section{Forward-backward  \ac{pd} algorithm}
\label{sec-opt-ri}

We continue by reviewing the minimisation problem and the randomised \ac{pd} algorithm \citep{Condat2013, Vu2013, Pesquet2015} recently proposed for \ac{ri} by \cite{Onose2016}, on which this work relies.
It solves a primal, block wise, minimisation problem similar to (\ref{basic-min-problem}),
\begin{equation}%
	\min_{\bs{x}} f(\bs{x}) + \gamma \sum_{i=1}^b l_i(\bm{W}_i^\dagger\bm{\Psi}^\dagger_i\bs{x}) + \sum_{j=1}^d h_j(\bm{\Phi}_j\bs{x}),
	\label{split-min-problem}
\end{equation}
together with its dual formulation \citep{Bauschke2011},
\begin{equation}%
	\begin{aligned}%
		\min_{\substack{\bs{u}_i \\ \bs{v}_j}} f^* \Bigg(\!\!-\sum_{i=1}^b \bm{\Psi}_i \bm{W}_i \bs{u}_i - & \sum_{j=1}^d \bm{\Phi}^\dagger_j \bs{v}_j \Bigg) + \\
		& + \frac{1}{\gamma} \sum_{i=1}^b l_i^*(\bs{u}_i) + \sum_{j=1}^d h_j^*(\bs{v}_j).
	\end{aligned}
	\label{split-min-dual-problem}
\end{equation}
Here, since the $\ell_1$ norm is additively separable, we have split the over-complete sparsity basis into $b$ parts, $\bm{\Psi} = \begin{bmatrix} \bm{\Psi}_1 & \ldots & \bm{\Psi}_b \end{bmatrix}$.
The weighting matrix $\bm{W}$ is also split to produce a weight matrix $\bm{W}_i$ for each $\bm{\Psi}_i$.
The scalar $\gamma$ is a free configuration parameter and only affects the convergence speed \citep{Onose2016}.
The functions from (\ref{split-min-problem}) are defined block wise but similarly to (\ref{basic-min-problem}).
Thus, the functions $l_i = \| \cdot \|_1$ represent the sparsity inducing prior and $h_j(\bs{z}) = \iota_{\mc{B}_j}  (\bs{z}), \mc{B}_j = \{ \bs{z} \in \bb{C}^{M_j}: \| \bs{z} - \bs{y}_j \|_2 \leq \epsilon_j \}$ are the data fidelity terms constraining the residual to be situated in $\ell_2$ balls defined by the noise level $\epsilon_j$, for each part of the visibility data $\bs{y}_j$.
The notation $^*$ denotes the Legendre-Fenchel conjugate function.\footnote{
The Legendre-Fenchel conjugate function $g^*$ of a function $g$ is%
\begin{equation}%
	(\forall \bs{v}) \qquad g^*(\bs{v}) \overset{\Delta}{=}  \sup_{\bs{z}} \bs{z}^\dagger \bs{v} - g(\bs{z}). \nonumber
\end{equation}}

\subsection{Distributed problem formulation}

We work in a setup where the visibility data are split into $d$ blocks, such that
\begin{equation}
    		\bs{y} = \begin{bmatrix}
                		\bs{y}_1 \\
                		\vdots \\
                		\bs{y}_d
                	\end{bmatrix}, \qquad\qquad
                	\bm{\Phi} = \begin{bmatrix}
                		\bm{\Phi}_1 \\
                		\vdots \\
                		\bm{\Phi}_d
                	\end{bmatrix}
                	= \begin{bmatrix}
                		\bm{\Theta}_1\bm{G}_1 \bm{M}_1\\
                		\vdots \\
                		\bm{\Theta}_d\bm{G}_d \bm{M}_d
                	\end{bmatrix}  \bm{F}\bm{Z},
        	\label{data-split}
\end{equation}
to allow for distributed and parallelised processing \citep{Carrillo2014, Onose2016}.
We also rely on the fact that $\bm{G}$ is composed of compact support kernels and introduce the matrices $\bm{M}_j \in \bb{R}^{nN_j \times nN}$ to select only the parts of the discrete Fourier plane involved in computations for block $j$. Each block operator $\bm{G}_j \in \bb{C}^{M_j \times nN_j}$ requires partial Fourier information, namely only $nN_j$ coefficients \citep{Onose2016}. The diagonal matrix $\bm{\Theta}$ is also split accordingly.

The inverse problem (\ref{inverse-problem}) was therefore be rewritten for each data block as%
\begin{equation}%
\bs{y}_j = \bm{\Phi}_j \bs{x} + \bs{n}_j,
\end{equation}
with $\bs{n}_j$ being the part of the noise associated with the measurements $\bs{y}_j$ and with $\bm{\Phi}_j$ the associated linear operator.

\subsection{The re-weighted $\ell_1$ approach}
A re-weighted $\ell_1$ \citep{Candes2008} serves to approximate the $\ell_0$ norm by solving successive $\ell_1$ penalised problems.
The weights $\bm{W}^{(k)}_i$, at step $k$, are computed based on the solution $\bs{x}^{(k-1)}$ from the previously solved problem from step $k-1$ such that
\begin{equation}
\op{D}_e\left(\bm{W}^{(k)}_i\right) = \frac{\omega^{(k)}}{\omega^{(k)} + \left( \left| \bm{\Psi}^\dagger_i \bs{x}^{(k-1)} \right| \right)_{e}},
\label{reweighting}
\end{equation}
with the operator $\op{D}_e$ denoting diagonal element $e$.
The parameter $\omega^{(k)}$ is decreased from a preset value at each re-weight step.
This ensures that, after several such steps, if the values of the $e$th coefficient $\left( \left| \bm{\Psi}^\dagger_i \bs{x}^{(k)} \right| \right)_{e}$ are large, the penalty applied is decreased towards $0$.
The small coefficients, smaller than $\omega^{(k)}$, are still being largely penalised.
Thus, this iterative procedure removes the bias introduced by the $\ell_1$ relaxation of the sparsity constraint.
This procedure is summarised as Algorithm \ref{reweighted-pd}.
Note that each call to Algorithm~\ref{alg-primal-dual}, which will be detailed in the following sections, should use the past primal and dual solutions, from step $k-1$, as initialisation in order to warm start the convergence.

\begin{algorithm}[t]
\caption{Re-weighting scheme.}
\label{reweighted-pd}

\begin{algorithmic}[1]
\small

\Given{$\omega^{(0)}, \bs{x}^{(0)}, \tilde{\bs{x}}^{(0)}, \bs{u}_i^{(0)}, \bs{v}_j^{(0)}, \tilde{\bs{u}}^{(0)}_i, \tilde{\bs{v}}^{(0)}_j, \bm{W}^{(0)}_i$}

\RepeatFor{$k=1,\ldots$}
\State {$
\left[\bs{x}^{(k)}, \tilde{\bs{x}}^{(t)}, \bs{u}_i^{(k)}, \bs{v}_j^{(k)}, \tilde{\bs{u}}^{(k)}_i, \tilde{\bs{v}}^{(k)}_j \right ] = \mathrm{Algorithm~\ref{alg-primal-dual}} ~\big( \cdots \big) 
$}
\State {\bf set} $\omega^{(k)}$ {\bf smaller than} $\omega^{(k-1)}$
\State {$\forall j$ {\bf set} $\bm{W}^{(k)}_i$ {\bf according to} (\ref{reweighting})}
\Until {{\bf convergence}}
\Output{$\bs{x}^{(k)}$}
\end{algorithmic}
\end{algorithm}

\subsection{Proximity operators}
As previously mentioned, the \ac{pd} algorithm \citep{Pesquet2015} relies on forward-backward iterations \citep{Komodakis2015} to deal with the non smooth terms present in both the primal minimisation problem (\ref{split-min-problem}) and its dual formulation (\ref{split-min-dual-problem}).
The forward step corresponds to a gradient-like step and the backward step is an implicit sub-gradient-like step performed through the use of the proximity operator \citep{Moreau1965}.

Using the definition (\ref{proximity-operator-basic}), the proximity operator associated with the function $f$ in (\ref{split-min-problem}) has a closed form solution and becomes the projection
\begin{equation}
	\Big( \proj_{\mc{C}}(\bs{z}) \Big)_e  \overset{\Delta}{=} \left\{ 
	\begin{aligned}
		\Re(z_e) & \qquad \Re(z_e) > 0 \\
		0 \quad & \qquad \Re(z_e) \leq 0
	\end{aligned}\right. \quad \forall e
	\label{proj-plus}
\end{equation}
onto the positive real orthant.
Similarly, the proximity operator for the sparsity prior functions $l_i$ is the component wise soft-thresholding operator
\begin{equation}
	\Big( \soft_{\alpha}(\bs{z}) \Big)_e \overset{\Delta}{=} \left\{ 
	\begin{aligned}
		\ds \frac{z_e\big\{ | z_e | - \alpha \big\}_{+}}{| z_e |} & \qquad | z_e | > 0\\
		\ds 0 \qquad \quad & \qquad | z_e | = 0\\
	\end{aligned}\right. \quad \forall e,
	\label{prox-L1}
\end{equation}
for a given threshold $\alpha$.
For the data fidelity terms $h_j$, the proximity operator has a closed form as the projection onto an $\ell_2$ ball $\mc{B}_j$,%
\begin{equation}
	\proj_{\mc{B}_j}(\bs{z})  \overset{\Delta}{=} \left\{ 
	\begin{aligned}
		\epsilon_j \frac{\bs{z} - \bs{y}_j}{\|\bs{z} - \bs{y}_j\|_2} + \bs{y_j} & \qquad \|\bs{z} - \bs{y}_j\|_2 > \epsilon_j\\
		\bs{z} \qquad \qquad & \qquad \|\bs{z} - \bs{y}_j\|_2 \leq \epsilon_j.
	\end{aligned}\right.
	\label{proj-L2}
\end{equation}%
More details can be found in \cite{Onose2016}, which proposed the \ac{pd} algorithm for solving (\ref{split-min-problem}) and (\ref{split-min-dual-problem}) in the absence of any preconditioning strategy.

\section{Accelerated forward-backward \ac{pd} algorithm}
\label{sec-algo}

The structure of the proposed algorithm, presented in Algorithm \ref{alg-primal-dual}, is based on \cite{Pesquet2015}. It is similar to that of the \ac{pd} algorithm proposed for \ac{ri} by \cite{Onose2016}.
As before, we solve concurrently both the primal minimisation problem (\ref{split-min-problem}) and its dual formulation (\ref{split-min-dual-problem}).
Forward-backward iterations, consisting of a gradient descent step coupled with a proximal update, are used to update both the primal and the dual variables.
The key difference that accelerates the convergence speed is the use a new non-Euclidean proximity operator for the data fidelity to replaces the projection onto the $\ell_2$ ball, used in \cite{Onose2016}, with a projection onto a generalised ellipsoid that incorporates both the noise statistics and sampling density information.
\bc By incorporating the sampling density information, the algorithm can make a larger step towards the final solution at each iteration. \ec
This acceleration strategy changes only the forward-backward step associated with the data fidelity terms, the rest of the updates remain the same as in \cite{Onose2016}.
In analogy with \sw{clean}, \bc the algorithm \ec can be understood as being composed of complex \sw{clean}-like forward-backward steps performed in parallel in multiple data, prior and image spaces \cite{Onose2016}.

\subsection{Non-euclidean proximity operator}

A generalisation of the proximity operator allows us to use additional prior information about the data when performing the computations associated with the data fidelity terms $h_j$, in order to accelerate the convergence speed.
It offers a broad flexibility in the way the data fidelity is enforced throughout the iterations.

Thus, we rely on the generalised proximity operator relative to a metric induced by a strongly positive, self-adjoint\footnote{A linear operator $\bm{U}$ is said to be strongly positive and self-adjoint if $\langle \bs{x} | \bm{U} \bs{x} \rangle \geq \alpha \| \bs{x}\|^2_2, \forall \bs{x}, \forall \alpha > 0$ and $\bm{U}^\dagger = \bm{U}$, respectively.}  linear operator $\bm{U}$ \citep{Hiriart1996},
\begin{equation}
	\prox^{\bm{U}}_g (\bs{z}) \overset{\Delta}{=} \argmin_{\bar{\bs{z}}} g(\bar{\bs{z}}) + \frac{1}{2} (\bs{z} - \bar{\bs{z}})^\dagger \bm{U} (\bs{z} - \bar{\bs{z}}).
	\label{generalised-proximity-operator}
\end{equation}
The standard definition from (\ref{proximity-operator-basic}) is found when $\bm{U} = \bm{I}$.
A generalisation of the Moreau decomposition provides the link between the proximity operators of a function $g$ and that of its conjugate $g^{*}$ \citep{Combettes2014, Pesquet2015} for any operator $\bm{U}$,%
\begin{equation}
	\prox^{\bm{U}^{-1}}_{\alpha g^*} (\bs{z})= \Big( \identity - \alpha \bm{U} \prox^{\bm{U}}_{\alpha^{-1} g} \Big) \big(\alpha^{-1} \bm{U}^{-1} \bs{z}\big),
	\label{generalised-proximity-operator-decomposition}
\end{equation}%
and allows for a facile way of computing the proximity operators for the conjugate functions.

We choose the preconditioning matrices $\bm{U}_j$ to be diagonal, with positive, non-zero diagonal elements and thus positive definite and invertible.
It results directly from (\ref{generalised-proximity-operator}) that  
\begin{equation}
	\begin{aligned}
	& \prox^{\bm{U}_j}_{h_j} (\bs{z}) = \argmin_{\bar{\bs{z}}} h_j(\bar{\bs{z}}) + \frac{1}{2} (\bs{z} - \bar{\bs{z}})^\dagger \bm{U}_j (\bs{z} - \bar{\bs{z}}) \\
	& \hquad					    = \argmin_{\bar{\bs{z}}} h_j(\bar{\bs{z}}) + \frac{1}{2} \Big(\bm{U}_j^{\frac{1}{2}}\bs{z} - \bm{U}_j^{\frac{1}{2}}\bar{\bs{z}}\Big)^\dagger  \Big(\bm{U}_j^{\frac{1}{2}}\bs{z} - \bm{U}_j^{\frac{1}{2}}\bar{\bs{z}}\Big).
	\end{aligned}
	\label{elip-proj-1}
\end{equation}
By making the variable change $\bs{s}=\bm{U}_j^{\frac{1}{2}}\bs{z}$ and  $\bar{\bs{s}}=\bm{U}_j^{\frac{1}{2}}\bar{\bs{z}}$ we can rewrite (\ref{elip-proj-1}) as
\begin{equation}
	\begin{aligned}
	& \prox^{\bm{U}_j}_{h_j}  \Big(\bm{U}_j^{-\frac{1}{2}}\bs{s}\Big) = \bm{U}_j^{-\frac{1}{2}} \bigg( \argmin_{\bar{\bs{s}}} h_j\Big(\bm{U}_j^{-\frac{1}{2}}\bar{\bs{s}}\Big) + \\
	& \qquad\qquad\qquad +  \frac{1}{2} (\bs{s} - \bar{\bs{s}})^\dagger  (\bs{s} - \bar{\bs{s}}) \bc \bigg) \ec = \bm{U}_j^{-\frac{1}{2}} \proj_{\mc{E}_j} (\bar{\bs{s}}).
	\end{aligned}
	\label{elip-proj-2}
\end{equation}
Here we have denoted by $\proj_{\mc{E}_j}$ the projection onto a generalised ellipsoid $\mc{E}_j= \{ \bar{\bs{s}} \in \bb{C}^{M_j}: \| \bm{U}_j^{-\frac{1}{2}} \bar{\bs{s}} - \bs{y}_j \|_2 \leq \epsilon_j \}$ associated with the preconditioned matrix $\bm{U}_j$ and the data fidelity function $h_j$.
This formulation serves as a generalisation of the way data fidelity is enforced \citep{Carrillo2014, Onose2016}.
\bc Note that the minimisation problem (\ref{split-min-problem}) and its dual formulation (\ref{split-min-dual-problem}) do not change when the generalised proximity operator (\ref{generalised-proximity-operator}) is used. This only affects the way convergence is achieved. Thus, if $\bm{U}_j = \bm{I}$, the constraints that the residual should belong to the $\ell_2$ balls $\mc{B}_j$ is enforced such that the Euclidian distance from the starting point $\bm{\Phi}_j\bs{x}$ and the ball $\mc{B}_j$ is minimised. This results in the simple projection onto the $\ell_2$ ball $\mc{B}_j$ from (\ref{proj-L2}). If instead a different metric $\bm{U}_j \neq \bm{I}$ is used, the projection becomes skewed and the Euclidian distance to the ball $\mc{B}_j$ is not minimised anymore. However, the new projection point still satisfies $\| \bm{\Phi}_j\bs{x} - \bs{y}_j \|_2 \leq \epsilon_j $. This can be expressed as the projection onto the ellipsoid $\mc{E}_j$ with the resulting projection point moved to the $\ell_2$ ball by the application of $\bm{U}_j^{-\frac{1}{2}}$ in (\ref{elip-proj-2}). \ec

\begin{algorithm}[t]
\caption{Forward-backward algorithm for solving (\ref{elip-proj-1}).}
\label{alg-fb-proj}

\begin{algorithmic}[1]
\small

\Given{$\bar{\bs{z}}^{(0)}, \mu$}

\RepeatFor{$t=1,\ldots$}

\State $\ds \bar{\bs{z}}^{(t)} =  \proj_{\mc{B}_j} \Big(\bar{\bs{z}}^{(t-1)} - \mu \bm{U}_j \big(\bar{\bs{z}}^{(t-1)} - \bs{z}\big)\Big)$

\Until {\bf convergence}
\end{algorithmic}
\end{algorithm}

For a generic metric $\bm{U}_j \neq \bm{I}$, an iterative procedure is required to compute the proximity operator $\prox^{\bm{U}_j}_{h_j}$. 
We propose a forward-backward approach that works directly with the definition of the proximity step (\ref{elip-proj-1}).
It performs a gradient step, with step $\mu$, in the direction of the smooth term $\frac{1}{2} (\bs{z} - \bar{\bs{z}})^\dagger \bm{U}_j (\bs{z} - \bar{\bs{z}})$ followed by the application of the proximity operator for the function $h_j$, which is the projection (\ref{proj-L2}).
This is formally presented as Algorithm \ref{alg-fb-proj}.
The step size $\mu$ must satisfy $\mu \leq \sfrac{1}{\|\bm{U}_j\|^2_{\mathrm{S}}}$. 
Since the preconditioning matrix $\bm{U}_j$ is diagonal, we have $\|\bm{U}_j\|_{\mathrm{S}} = \max_e\big(\op{D}_e(\bm{U}_{j})\big)$ with the operator $\op{D}_e$ selecting the $e$th diagonal element of $\bm{U}_j$.

Faster converging proximal gradient algorithms for solving (\ref{generalised-proximity-operator}) may be employed \citep{Tseng2008}. 
However, for simplicity we limit the presentation herein to the forward-backward approach presented as Algorithm \ref{alg-fb-proj}.
Alternatively, we can compute the projection $\proj_{\mc{E}_j}$ onto the ellipsoid $\mc{E}_j$ and then estimate $\prox^{\bm{U}_j}_{h_j} (\bs{z})$ as in (\ref{elip-proj-2}).
A very fast iterative approach was developed by \cite{Dai2006} for any choice of metric $\bm{U}_j$.
It requires an initial point on the feasible region, which, due to $\bm{U}_j$ being positive definite and invertible, can be easily computed using Algorithm \ref{alg-fb-proj}. Note that this is not the case for a general operator $\bm{U}_j$, for which the derivations form (\ref{elip-proj-1}) and (\ref{elip-proj-2}) are not guaranteed to hold.

\begin{algorithm}[t]
\caption{Preconditioned forward-backward \ac{pd}.}
\label{alg-primal-dual}

\begin{algorithmic}[1]
\small

\Given{$\bs{x}^{(0)}, \tilde{\bs{x}}^{(0)}, \bs{u}_i^{(0)}, \bs{v}_j^{(0)}, \tilde{\bs{u}}^{(0)}_i, \tilde{\bs{v}}^{(0)}_j, \bm{W}_i, \bm{U}_j, \epsilon_j, \kappa, \tau, \eta, \zeta, \lambda$}

\RepeatFor{$t=1,\ldots$}
\State {\bf generate sets} $\mc{P} \subset \{1,\ldots, b\}$ {\bf and} $\mc{D} \subset \{1,\ldots, d\}$

\State $\ds \tilde{\bs{a}}^{(t)} = \bm{F}\bm{Z} \tilde{\bs{x}}^{(t-1)}$
\Set{$\forall j \in \mc{D}$}
	\State $\ds \bs{a}_j^{(t)} = \bm{M}_j \tilde{\bs{a}}^{(t)}$
\EndSet
\Block{\bf run simultaneously}
\ParForD {$\forall j \in \mc{D}$}{$\bs{a}_j^{(t)}$}
	\State $\ds \bar{\bs{v}}_j^{(t)} = \Bigg(\!\identity - \bm{U}_j \bm{U}_j^{-\frac{1}{2}} \proj_{\mc{E}_j}\!\!\! \Bigg) \! \Big(\bm{U}_j^{-1} \bs{v}_j^{(t-1)} \!+ \bm{\Theta}_j\bm{G}_j \bs{a}^{(t)}_j \!\Big)$
	\State $\ds \bs{v}_j^{(t)} = \bs{v}_j^{(t-1)} + \lambda \left( \bar{\bs{v}}_j^{(t)} - \bs{v}_j^{(t-1)} \right)$
	\State $\ds \tilde{\bs{v}}^{(t)}_j = \bm{G}_j^\dagger \bm{\Theta}^\dagger_j \bs{v}^{(t)}_j$
\EndParForD{$\tilde{\bs{v}}^{(t)}_j$}
\Set{$\forall j \in \{1, \ldots d\} \setminus \mc{D}$}
	\State $\ds \bs{v}^{(t)}_j = \bs{v}^{(t-1)}_j$
	\State $\ds \tilde{\bs{v}}^{(t)}_j = \tilde{\bs{v}}^{(t-1)}_j$
\EndSet
\ParFor{$\forall i \in \mc{P}$}
	\State $\ds \bar{\bs{u}}_i^{(t)} = \Bigg( \identity - \soft_{\kappa \|\bm{\Psi}\bm{W}\|^2_{\rm{S}}} \!\! \Bigg) \Big( \bs{u}_i^{(t-1)} + \bm{W}_i^\dagger \bm{\Psi}_i^\dagger  \tilde{\bs{x}}^{(t-1)} \Big)$
	\State $\ds \bs{u}_i^{(t)} = \bs{u}_i^{(t-1)} + \lambda \left( \bar{\bs{u}}^{(t)} - \bs{u}_i^{(t-1)} \right)$
	\State $\ds \tilde{\bs{u}}^{(t)}_i = \bm{\Psi}_i \bm{W}_i \bs{u}^{(t)}_i$
\EndParFor
\Set{$\forall i \in \{1, \ldots b\} \setminus \mc{P}$}
	\State $\ds \bs{u}^{(t)}_i = \bs{u}^{(t-1)}_i$
	\State $\ds \tilde{\bs{u}}^{(t)}_i = \tilde{\bs{u}}^{(t-1)}_i$
\EndSet
\EndBlock{\bf end}
\State $\ds \bar{\bs{x}}^{(t)} \! = \proj_{\mc{C}} \! \Bigg(\! \bs{x}^{(t-1)}  - \tau  \Big(\!  \eta \bm{Z}^\dagger \bm{F}^\dagger \! \sum_{j=1}^d \!   \bm{M}_j^\dagger \tilde{\bs{v}}_j^{(t)} \! +\! \zeta \sum_{i=1}^b \!  \tilde{\bs{u}}_i^{(t)} \Big)\!\!\Bigg)$
\State $\ds \bs{x}^{(t)} \!= \bs{x}^{(t-1)} + \lambda \left( \bar{\bs{x}}^{(t)} - \bs{x}^{(t-1)} \right)$
\State $\ds \tilde{\bs{x}}^{(t)} = 2\bar{\bs{x}}^{(t)} - \bs{x}^{(t-1)}$
\Until {{\bf convergence}}
\Output{$\bs{x}^{(t)}, \tilde{\bs{x}}^{(t)}, \bs{u}_i^{(t)}, \bs{v}_j^{(t)}, \tilde{\bs{u}}^{(t)}_i, \tilde{\bs{v}}^{(t)}_j$}
\end{algorithmic}
\end{algorithm}

\subsection{The preconditioned algorithmic structure}
All the updates associated with the dual variables $\bs{v}^{(t)}_j$ and $\bs{u}^{(t)}_i$ from (\ref{split-min-dual-problem}) are performed in Algorithm \ref{alg-primal-dual} in parallel in steps $9$--$13$ and $18$--$22$, respectively.
Randomisation is supported given a probabilistic construction of the active sets $\mc{P}$ and $\mc{D}$.
Thus, only a part of the dual variables are updated per iteration, the rest remaining unchanged as in steps $14$--$17$ and $23$--$26$.
The forward-backward updates rely on the Moreau decomposition (\ref{generalised-proximity-operator-decomposition}) to compute the proximity operator associated with the conjugate functions $l_i^*$ and $h_j^*$ relying on the proximity operator of the functions $l_i$ and $h_j$.
The resulting updates become the soft-thresholding (\ref{prox-L1}) for the prior dual variables $\bs{u}^{(t)}_i$ from step $19$ and the skewed projection (\ref{elip-proj-2}) onto the ellipsoid $\mc{E}_j$ for the data fidelity dual variables $\bs{v}^{(t)}_j$ from step $10$.
For the soft-thresholding, we perform a re-parametrisation similar to the one performed in \cite{Onose2016}.
Since $\gamma$ is a free parameter, we replace the resulting algorithmic soft-threshold size $\frac{\gamma}{\zeta}$ with $\kappa \|\bm{\Psi}\bm{W}\|^2_{\rm{S}}$ to produce a operator independent configuration parameter $\kappa$.
The parameter $\kappa$ is only linked to the scale of the unknown image to be recovered.
The application of the operators $\bm{G}_j^\dagger\bm{\Theta}^\dagger_j$ and $\bm{\Psi}_i\bm{W}_i$ is also performed in parallel, in steps $12$ and $21$.
The contribution of all the dual variables is then used to update the primal variable, the image of interest $\bs{x}^{(t)}$ in steps $28$--$29$.
This is a forward-backward step which, through the use of the Moreau decomposition, resumes to the projection (\ref{proj-plus}) onto the positive orthant presented in step $28$.

\subsection{The epiphany: when natural and uniform weighting meet}

For the data fidelity terms $h_j$ we propose the use of a non-trivial invertible preconditioning matrix $\bm{U}_j$ which has links to the standard weighting schemes.
The weighting is used to mitigate the effects produced by the sampling strategy \citep{Briggs1995, Yatawatta2014} and serves as an important pre-processing step for the \sw{clean} family of algorithms.
We aim to incorporate the sampling density information into the \ac{pd} algorithmic structure, through $\bm{U}_j$, while solving the same problems defined in (\ref{split-min-problem}) and (\ref{split-min-dual-problem}). 
This does not change the overall results due to the convergence guarantees of the convex optimisation methods and increases the speed of convergence, as will be shown through simulations.

With this aim, we employ a diagonal preconditioning matrix $\bm{U}_j$, for each visibility block $\bs{y}_j$.
The matrix $\bm{U}_j$ accounts for the sampling density similarly to the uniform weighting. 
It contains on the diagonal the inverse of the sampling density in the vicinity of each associated visibility point.
This has the benefit of allowing for a facile computation of its inverse which is important to the computational complexity of the resulting strategy.
Other types of preconditioning could also be supported.

To give further insight into the behaviour of this preconditioning strategy, consider the problem (\ref{split-min-problem}) written in an equivalent formulation
\begin{equation}%
	\min_{\bs{x}} f(\bs{x}) + \gamma \sum_{i=1}^b l_i(\bm{W}_i^\dagger\bm{\Psi}^\dagger_i\bs{x}) + \sum_{j=1}^d \tilde{h}_j(\bm{G}_j\bm{M}_j\bm{F}\bm{Z}\bs{x}),
	\label{eqv-split-min-problem}
\end{equation}
by introducing the natural weighting matrix $\bm{\Theta}_j$ in the definition of the function $\tilde{h}_j(\bs{z}) = \iota_{\tilde{\mc{E}}_j}  (\bs{z}), \tilde{\mc{E}}_j = \{ \bs{z} \in \bb{C}^{M_j}: \| \bm{\Theta}_j\bs{z} - \bs{y}_j \|_2 \leq \epsilon_j \}$.
Now, the convex set associated with $\tilde{h}_j$ becomes the ellipsoid $\tilde{\mc{E}}$ associated with the natural weight matrix $\bm{\Theta}_j$.
\bc This does not change the definition of the minimisation problems but changes significantly how the problem is approached algorithmically.
It changes the manner in which the data fidelity constraint is enforced to make it similar to the way the generalised proximity operator is used in the algorithm. As such, it allows us to provide an intuitive link between the whitening matrices $\bm{\Theta}_j$ and the preconditioning matrices $\bm{U}_j$ by highlighting that they enter the algorithmic structure through a similar mechanism.

Thus, \ec based on the definition of the proximity operator (\ref{generalised-proximity-operator}) and by performing the variable change $\bs{s}=\bm{\Theta}_j\bs{z}$ and  $\bar{\bs{s}}=\bm{\Theta}_j\bar{\bs{z}}$, we can write $\prox^{\bm{U}_j}_{\tilde{h}_j} (\bs{z})$ as
\begin{equation}
	\begin{aligned}
	& \prox^{\bm{U}_j}_{\tilde{h}_j} \big(\bm{\Theta}_j^{-1}\bs{s}\big) = \bm{\Theta}_j^{-1} \argmin_{\bar{\bs{s}}} \tilde{h}_j\big(\bm{\Theta}_j^{-1}\bar{\bs{s}}\big) + \\
	& \qquad \qquad \qquad \qquad +  \frac{1}{2} \big( \bs{s} - \bar{\bs{s}}\big)^\dagger \bm{\Theta}_j^{-1^\dagger} \bm{U}_j \bm{\Theta}_j^{-1} \big(\bs{s} - \bar{\bs{s}}\big).
	\end{aligned}
	\label{elip-proj-3}
\end{equation}
Since both $\bm{\Theta}_j$ and $\bm{U}_j$ are diagonal matrices and since $\tilde{h}_j\big(\bm{\Theta}_j^{-1}\bs{s}\big) = h_j(\bs{s})$, (\ref{elip-proj-3}) becomes
\begin{equation}
	\begin{aligned}
	& \prox^{\bm{U}_j}_{\tilde{h}_j} \big(\bm{\Theta}_j^{-1}\bs{s}\big) =  \\
	& \qquad \qquad \quad \bm{\Theta}_j^{-1} \argmin_{\bar{\bs{s}}} h_j (\bar{\bs{s}}) +  \frac{1}{2} \big( \bs{s} - \bar{\bs{s}}\big)^\dagger \bm{D} \big(\bs{s} - \bar{\bs{s}}\big),
	\end{aligned}
	\label{elip-proj-4}
\end{equation}
with diagonal elements $d_{e, e} = \sigma_e^2 \op{D}_e(\bm{U}_j)$. 
The operator $\op{D}_e$ selects the $e$th diagonal element from $\bm{U}_j$.
Since they affect the data fidelity term $h_j$ in a similar way, this provides an intuitive link between the natural weighting matrix $\bm{\Theta}_j$ and the preconditioning matrix $\bm{U}_j$, which is based on the inverse of the sampling density.
A large value for $d_{e, e}$ corresponds to either a low sample density for the frequency vicinity of the given measurement $e$ or a large noise variance for the same measurement.
Low values $d_{e, e}$ correspond to less noisy measurements or a high sampling density.
Since sampling the same $u$--$v$ region multiple times can be seen as lowering the noise by averaging the data, the similitude between the effect of the noise on the measurement and the sampling density is immediate.

Let us emphasise again that only the natural weighting
performed through $\bm{\Theta}_j$ is reflected back into the definition
of the minimisation problem through the application of $\bm{\Theta}_j^{-1}$ in (\ref{elip-proj-4}).  In contrast, the preconditioning matrix is only an internal algorithmic flexibility to solve the very same problem.
Thus, such an approach can be seen to incorporate all the benefits from both natural and uniform weighting in \sw{clean} terms. On one hand it optimises resolution by accounting for the correct noise statistics, leveraging natural weighting in the definition of the minimisation problem for image reconstruction. On the other hand it optimises sensitivity by enabling accelerated convergence through a preconditioning strategy incorporating sampling density information \`a la uniform weighting.

\subsection{Convergence requirements}
The variables $\bs{x}^{(t)}$, $\bs{v}_j^{(t)}$ and $\bs{u}_i^{(t)}$, $\forall i, j$, are guaranteed to converge to the solution of the PD problem (\ref{split-min-problem})--(\ref{split-min-dual-problem}) for an adequately chosen set of configuration parameters,  $\tau$, $\zeta$ and $\eta$.
The convergence conditions \citep[Lemma 4.3]{Pesquet2015} can be stated explicitly for Algorithm \ref{alg-primal-dual} as%
\begin{equation}%
	\begin{aligned}
		&\left\| \begin{bmatrix}
			\zeta \bm{I} & \bm{0} \\
			\bm{0} & \eta \bm{U} \\
		\end{bmatrix}^{\frac{1}{2}}
		\begin{bmatrix}
			\bm{W}^\dagger\bm{\Psi}^\dagger \\
			\bm{\Phi} \\
		\end{bmatrix}
		\begin{bmatrix}
%			\tau \bm{I} & \bm{0} \\
%			\bm{0} & \tau \bm{I} \\
			\bc \tau \bm{I} \ec
		\end{bmatrix}^{\frac{1}{2}} \right\|_{\rm{S}}^2 \leq\\
		 & \qquad \qquad \qquad \qquad \leq \tau \zeta \left \| \bm{W}^\dagger\bm{\Psi}^\dagger \right \|_{\rm{S}}^2 + \tau \eta \left \| \bm{U}^{\frac{1}{2}} \bm{\Phi} \right \|_{\rm{S}}^2 <1,
	\end{aligned}
	\label{convergence-req-explicit-pd}
\end{equation}%
with the use of the triangle and Cauchy-Schwarz inequalities and with the diagonal matrices $\bm{I}$ of a proper dimension.
The matrix $\bm{U}$ represents a diagonal concatenation of all the preconditioning matrices $\bm{U}_j$ associated with the differently split operators and data blocks.
A relaxation with the factor $0 < \lambda \leq 1$ of the updates is also permitted.
The additional parameter $\gamma > 0$ imposes that $\kappa > 0$ as well.
For the randomised setup, the probabilities with which the active sets $\mc{P}$ and $\mc{D}$ are generated have to be nonzero and the activated variables need to be drawn in an independent and identical manner along the iterations.

The general framework of the \ac{pd} with forward-backward iterations approach and its mathematical analysis are presented by \cite{Pesquet2015}.

\subsection{Computational complexity}
The complexity and parallelised and distributed implementation details follow closely the study from \cite{Onose2016}.
The only difference is the introduction of the preconditioning matrix and the need for the iterative computation of the resulting proximity operator.
The complexity class of Algorithm \ref{alg-fb-proj} is $\mc{O}(M_j)$ \bc per data block $j$. \ec
The computations involving the projection are to be performed in a distributed fashion similarly to the computations involving the data fidelity terms.
The convergence speed of Algorithm \ref{alg-fb-proj} is linked to the conditioning number of the preconditioning matrix and may slow down for ill-conditioned matrices.
In such case, Algorithm $4$ proposed by \cite{Dai2006} or faster proximal gradient methods \citep{Tseng2008} become preferable.
Empirical evidence however suggests that the accuracy of the projection can be lowered by reducing the number of iteration performed without damaging the convergence speed of the whole algorithm.
The algorithm is resilient to errors in the computations and in practice as little as $1$ iteration can be enough to achieve a significant acceleration.
This can serve to control the added complexity due to the sub-iterative computation of the preconditioned proximity operator.
\bc Comparing the added total computational complexity of the preconditioning, which is $\mc{O}(M)$ per sub-iteration, with that of the basic non-preconditioned \ac{pd} algorithm, which is of the order $\mc{O}(nN \log nN) + \mc{O}(dN) + \mc{O}(MN)$ per iteration, it is evident that the added cost due to the preconditioning in \ac{ppd} is negligible when the number of sub-iterations is kept small. \ec

For more details regarding the complexity, randomisation and general structure of the \ac{pd} algorithm solving (\ref{split-min-problem}) and (\ref{split-min-dual-problem}) we direct the reader to \cite{Onose2016}.

\section{Simulations and results}
\label{sec-sims}

\begin{samepage}
\begin{figure*}
	\centering
	\begin{minipage}{.30\linewidth}
	\centering
  		\includegraphics[trim={0px 0px 0px 0px}, clip, height=4.6cm]{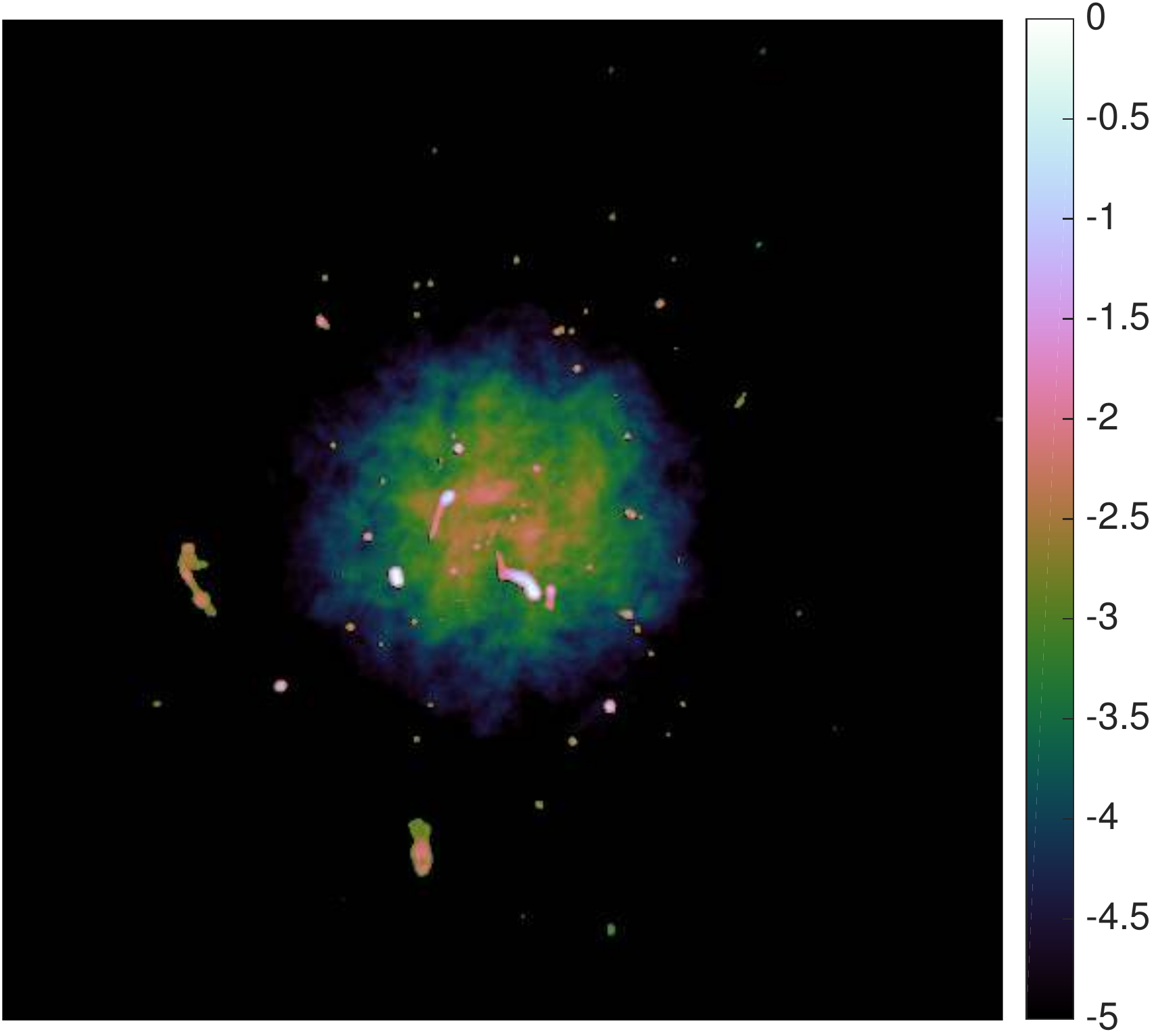}
	\end{minipage}%
	\begin{minipage}{.60\linewidth}
  	\centering
  		\includegraphics[trim={0px 0px 0px 0px}, clip, height=4.6cm]{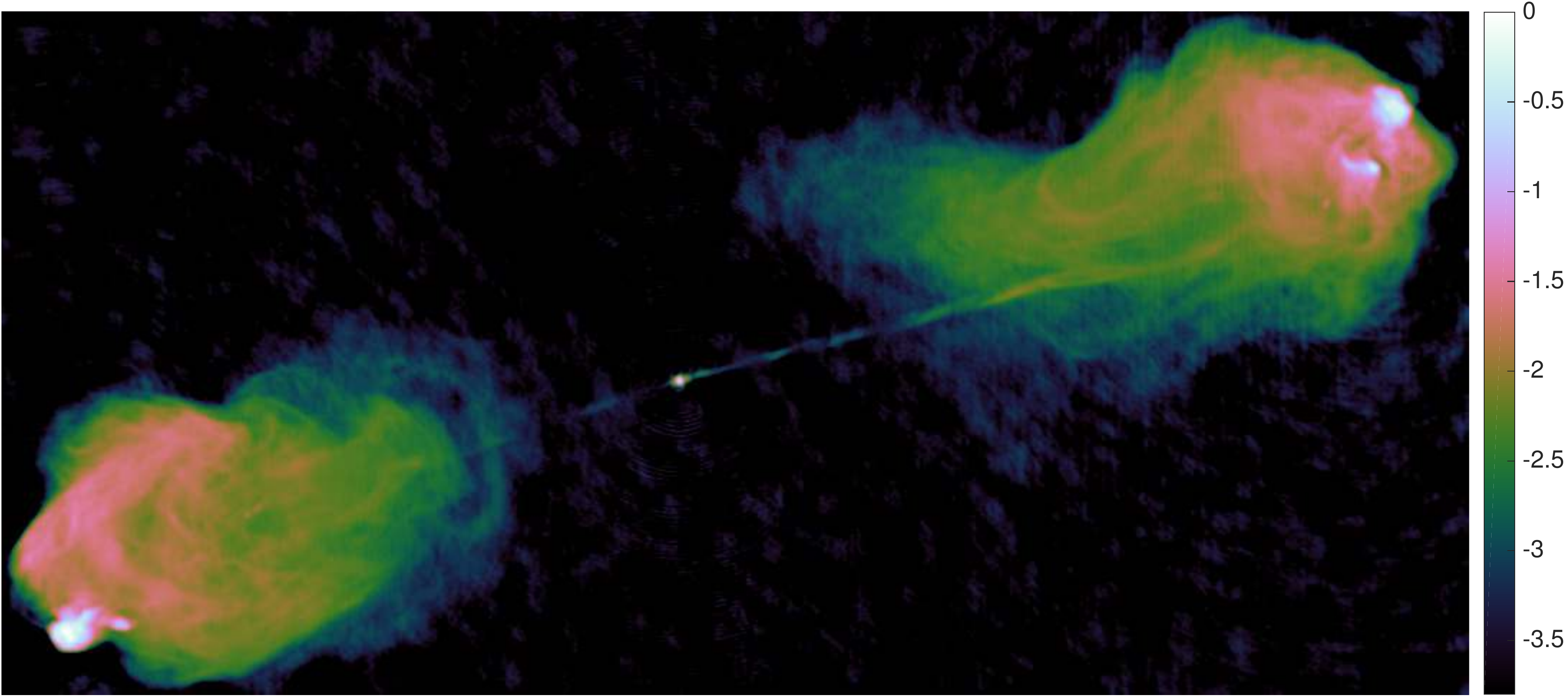}
	\end{minipage}
	\caption[]{The test images, a $512 \times 512$ galaxy cluster image and a $477 \times 1024$ image of Cygnus A, all shown in $\log_{10}$ scale.\footnotemark}
	\label{fig-test-images}
	\afterpage{\footnotetext{\bc We display $\log_{10} \bs{z}$ where $\bs{z}$ is the current image of interest.\ec}}
\end{figure*}

\begin{figure*}
	\centering
	\begin{minipage}{.33\linewidth}
	        \centering
  		\includegraphics[trim={0px 0px 0px 0px}, clip, height=5.3cm]{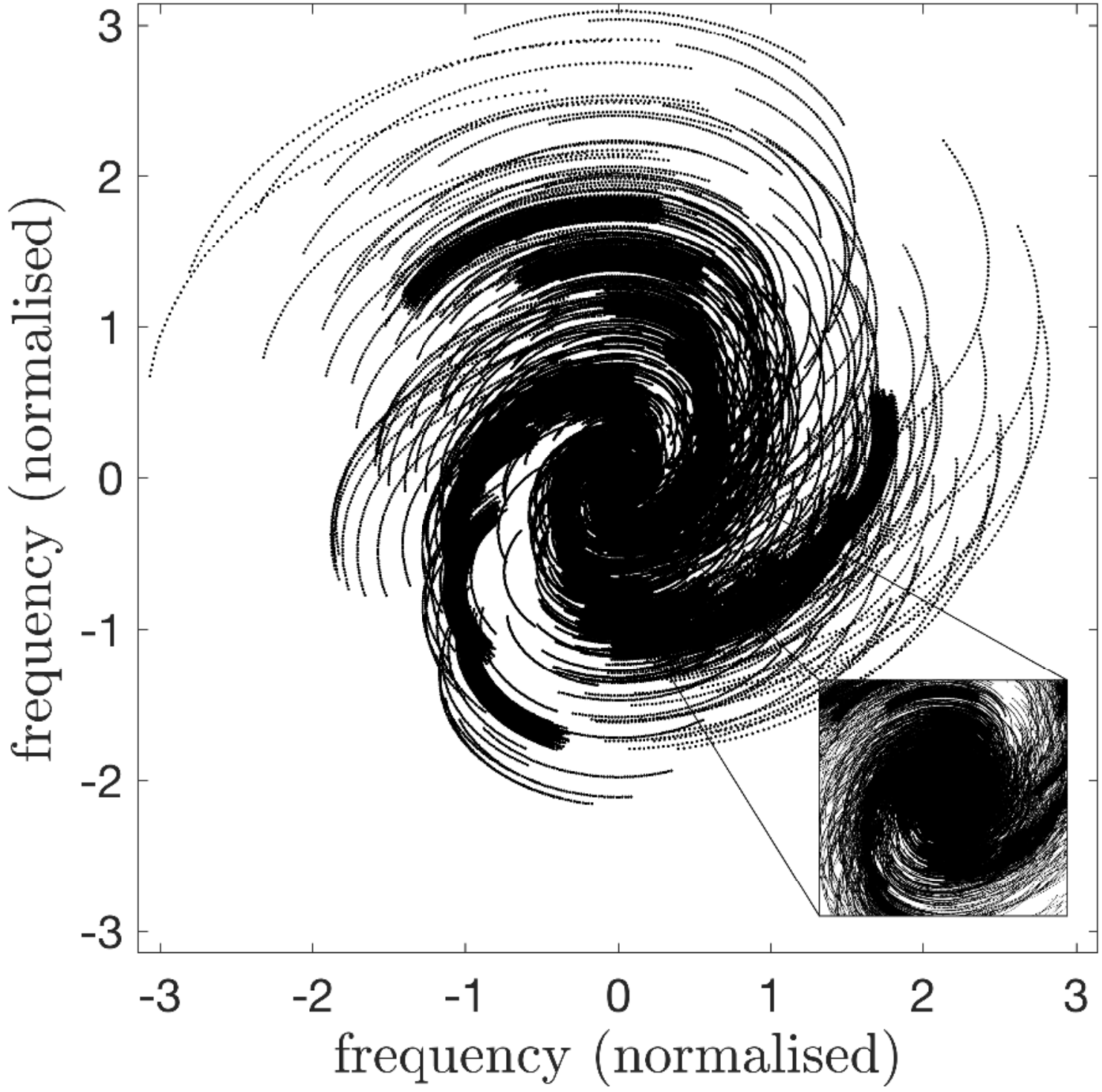}
	\end{minipage}
	\begin{minipage}{.33\linewidth}
	        \centering
  		\includegraphics[trim={0px 0px 0px 0px}, clip, height=5.3cm]{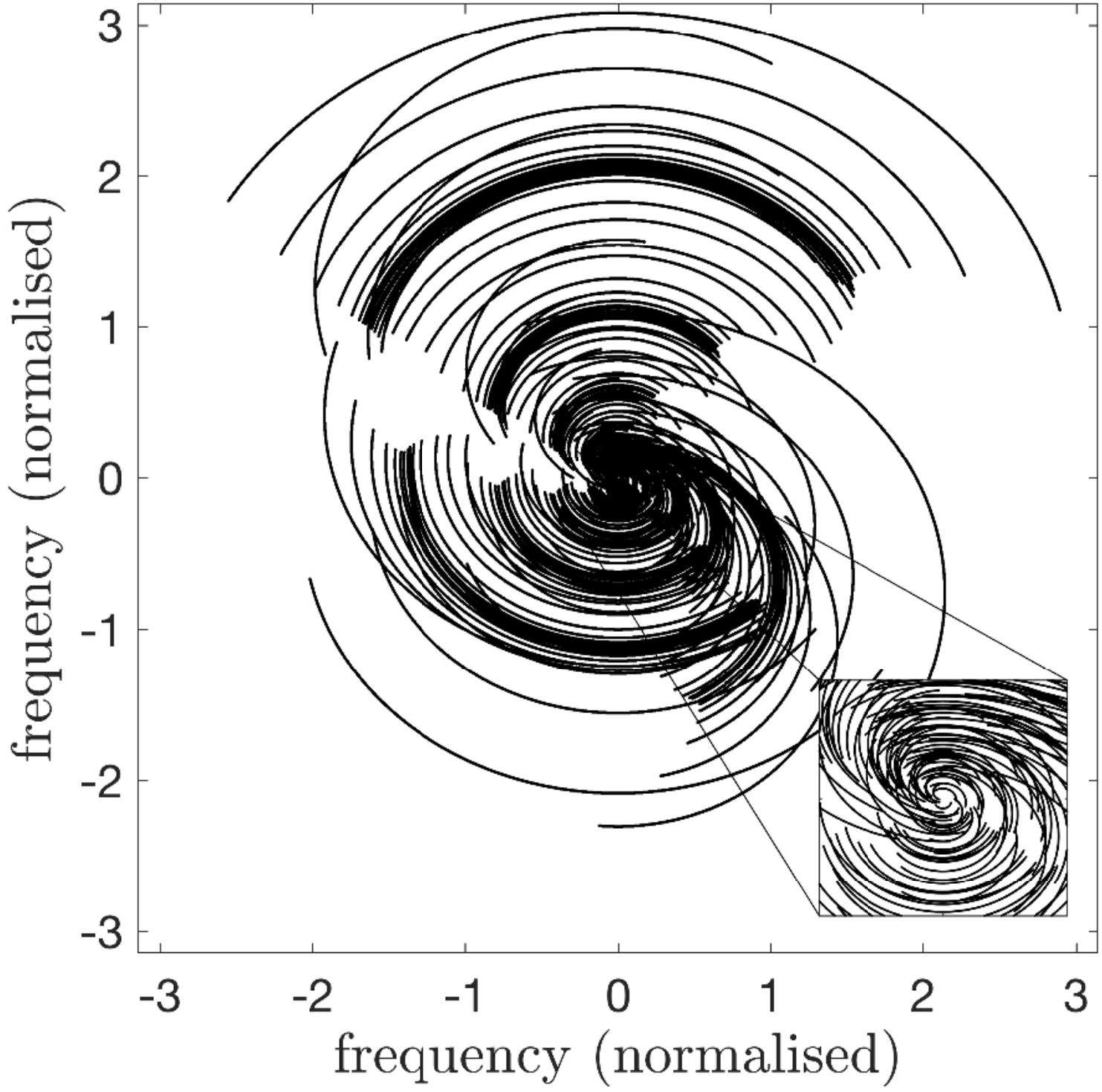}
	\end{minipage}
	\begin{minipage}{.33\linewidth}
	        \centering
  		\includegraphics[trim={0px 0px 0px 0px}, clip, height=5.3cm]{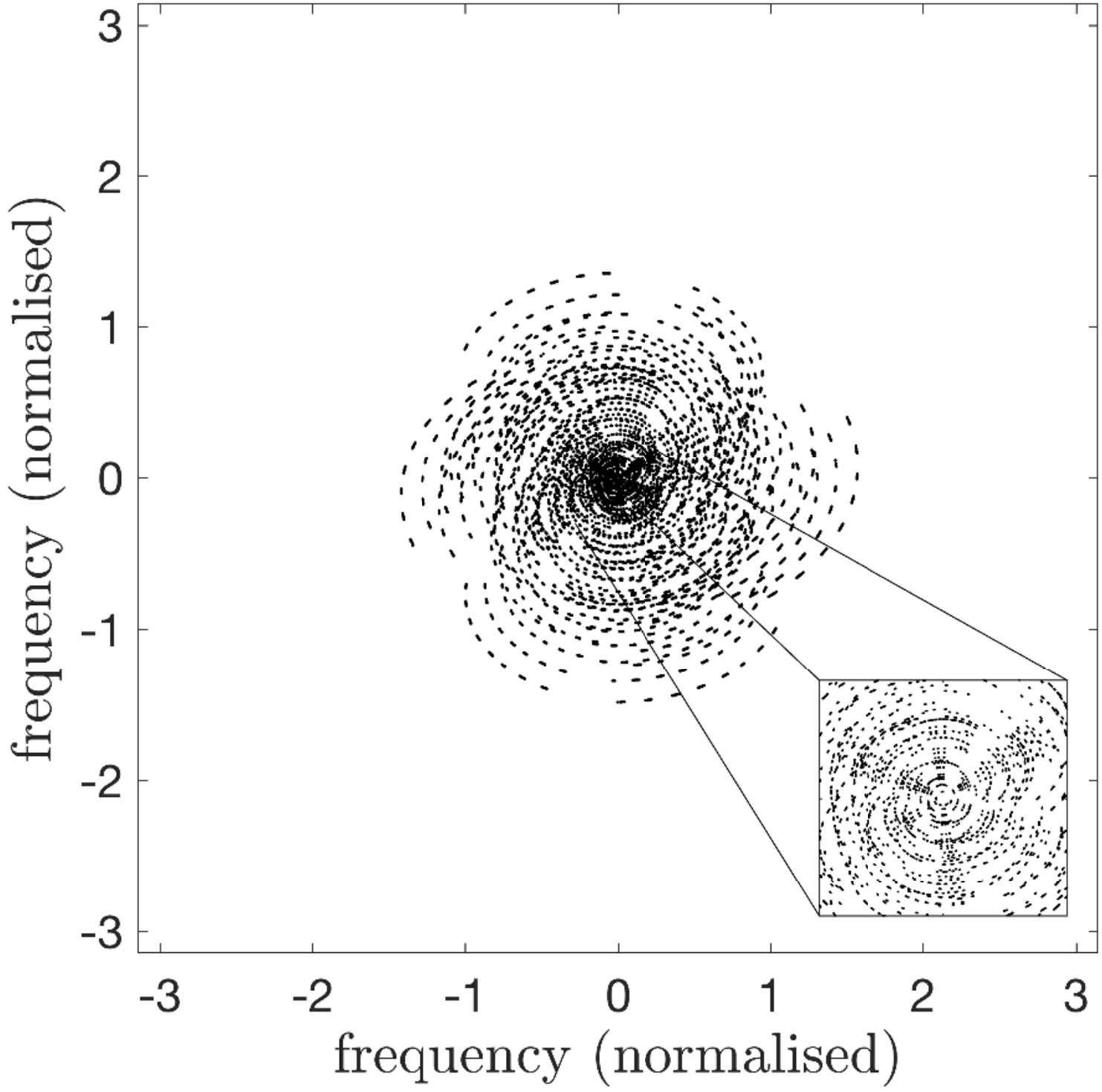}
	\end{minipage}
  	
	\caption{From left to right, the \ac{ska} coverage containing $1~447~950$ $u$--$v$ points, the \ac{vla} coverage containing $894~240$ $u$--$v$ points and the coverage of the real \ac{vla} observations containing $307~780$ $u$--$v$ points. \bc All frequencies are normalised with the largest corresponding baseline and rescaled to the interval $[-\pi, \pi]$ to produce the coverages presented.\ec}
	\label{fig-coverage-example}
\end{figure*}
\end{samepage}

We study the acceleration for different sampling strategies of the $u$--$v$ space.
To judge the efficacy of the acceleration, we compare the preconditioned algorithm \ac{ppd} against the non-preconditioned \ac{pd} and \ac{admm} algorithms \citep{Onose2016}, solving the same minimisation problem.
We also compare the reconstruction quality and acceleration using real interferometric measurement of the 3C129 radio galaxy.
In this case, we showcase the reconstruction in comparison with \sw{clean}, as implemented by the \sw{wsclean} package \citep{Offringa2014}.
We provide reconstruction for multi-scale \sw{clean}, denoted as \ac{ms-clean}.
We do not study the distribution and randomisation, an extensive study being performed by \cite{Onose2016}.

We work with pre-calibrated measurements, for both simulated and real data.
We assume the absence of \ac{dde}s and a small field of view such that the measurement operator is a Fourier operator.
We used an oversampled factor $n=4$ and a matrix $\bm{G}$ that performs an interpolation of the frequency data using $8 \times 8$ \bc Kaiser-Bessel \ec interpolation kernels \citep{Fessler2003} to average nearby uniformly distributed frequency.
The diagonal preconditioning matrix $\bm{U}$ contains the inverse of the sampling density as diagonal elements.

Thus, we begin by performing synthetic tests with the $u$--$v$ space sampled using a zero-mean, generalised Gaussian distribution \citep{Novey2010} with shape parameter $\beta$.
This allows us to have control of the sampling densities and see how the preconditioning is able to accelerate the convergence speed for various sampling patterns.
We also use realistic simulations of \ac{vla} and \ac{ska} coverages and we study, through simulations, the behaviour of the algorithms.
The $u$--$v$ coverages used are included in Figure \ref{fig-coverage-example}.
For all these tests we use two test images to generate the visibilities, namely a $477 \times 1024$ image of the Cygnus A radio galaxy and a $512 \times 512$ simulated image of a galaxy cluster with faint extended emission, respectively.
The galaxy cluster image was produced using the \sw{faraday} tool \citep{Murgia2004}.
The two images are presented in Figure \ref{fig-test-images}.
The simulated visibilities are corrupted by zero-mean complex \bc independent \ec Gaussian noise.
We run simulations for two noise levels, to produce an input signal to noise ratio $\rm{iSNR} = 30~\rm{dB}$ and $\rm{iSNR} =  50~\rm{dB}$ on the visibilities, respectively. \bc This is accomplished by choosing the appropriate noise power relative to the power of the simulated, noise free, signal. \ec
\bc In this case, the resulting noise statistics
are used to generate the weight matrix $\bm{\Theta}$. \ec

For the comparison with \sw{clean} we rely on observations of the 3C129 radio galaxy: right ascension 04h 45m 31.695s, declination 44$^\circ$ 55' 19.95'', J2000.
The observations were performed using the \ac{vla} for two 50 MHz channels centred at 4.59 and 4.89 GHz on the 25th of July 1994 in configuration B and 3rd of November 1994 in configuration C, respectively.
The calibration and flagging for radio frequency interference have been performed in \cite{Pratley2016} according to the \sw{casa} manual.
We additionally remove approximatively $20~000$ visibility points that contained large noise outliers, probably visibilities affected by radio frequency interference or poorly calibrated.
The remaining data consist of $307~780$ visibilities.
The normalised $u$--$v$ coverage is also included in Figure \ref{fig-coverage-example}.
All reconstructions are performed at twice the resolution of the telescope array.
This is necessary to avoid tension between the band limitation of the reconstructed image and the positivity constraint introduced by our approach.

For the synthetic tests, we assess the reconstruction performance in terms of the signal to noise ratio,
\begin{equation}
	{\rm SNR} = 20 \log_{10} \left( \frac{\|\bs{x}^{\circ} \|_2}{\|\bs{x}^{\circ} \! - \bs{x}^{(t)}\|_2}\right),
\end{equation}
where $\bs{x}^{\circ}$ is the original image and $\bs{x}^{(t)}$ is the reconstructed estimate of the original.
For the real data reconstructions, since we do not have access to the ground truth, we report the dynamic range obtained for the reconstruction,
\begin{equation}
	{\rm DR} = \frac{\sqrt{N}\|\bm{\Phi}\|_{\rm{S}}^2}{\|\bm{\Phi}^\dagger(\bs{y} - \bm{\Phi}\bs{x}^{\bc (t) \ec})\|_2} \max_{e} {x^{\bc (t) \ec}_{e}}.
\end{equation}

\subsection{Choice of parameters}

The \ac{ppd} algorithms converge given that (\ref{convergence-req-explicit-pd}) is satisfied.
To ensure this we set $\bc \zeta \ec = \sfrac{1}{\| \bm{\Psi} \bm{W} \|_{\rm{S}}^2}$, $\bc \eta \ec = \sfrac{1}{\| \bm{U}^{\frac{1}{2}}\bm{\Phi} \|_{\rm{S}}^2}$ and $\tau=0.49$.
The relaxation parameter is set to 1.
For the \ac{admm} and \ac{pd} algorithms we set the parameters as recommended by \cite{Onose2016}.
\bc We do not use randomisation, all data and all sparsity priors are used at each iteration. \ec
We use the \ac{sara} collection of wavelets \citep{Carrillo2012}, namely a concatenation of a Dirac basis with the first eight Daubechies wavelets, as sparsity prior.
For the simulations, we set the normalised soft-threshold values $\kappa = 10^{-4}$ for all three methods, \ac{ppd}, \ac{pd} and \ac{admm}.
We run \ac{ppd} for a number of sub-iteration $n_{\rm itr} \in \{1,5,50\}$.
In all tests we impose that the square of the global bound $\epsilon^2$ is 2 standard deviations above the mean of the $\chi^2$ distribution associated with the noise \citep{Onose2016}.

For the real data reconstruction we set $\kappa = 10^{-6}$, since the recovered image has the brightest pixel on the order of $10^{-2}$.
In this case we also perform $10$ re-weighting steps, one every $1024$ iterations, according to Algorithm $\ref{reweighted-pd}$.
We start with $\omega^{(0)} = 10^{-2}$ and set $\omega^{(k)} = 0.25^k \omega^{(0)}$ for each step $k$.
In this case the global bound $\epsilon^2$ is set to be $1.05$ times mean of the $\chi^2$ distribution associated with the thermal noise affecting the visibilities.
Such a bound was observed to provide good reconstruction results.
\ac{ms-clean} was run using the \sw{wsclean} software package, \bc version $2.2.1$\ec, with both uniform and natural weighting. \bc For both weighting schemes, we use 6 scales, $\{0,16,24,32,48,64\}$. We set the major loop gain to $\gamma_M=0.6$ and the minor loop gain to $\gamma_m=0.08$. The stopping threshold is set to 2 standard deviations above the automatically estimated noise level on the different scales.
The uniform weighting test reached the stopping threshold.
The natural weighting test was stopped after $35~000$ iterations since, for a larger number of iterations, the method was only accumulating spurious components without improving the solution. \ec

\subsection{Simulations}

\begin{figure*}
	\centering
	\begin{minipage}{.33\linewidth}
  		\includegraphics[trim={0px 0px 0px 0px}, clip, height=2.3cm]{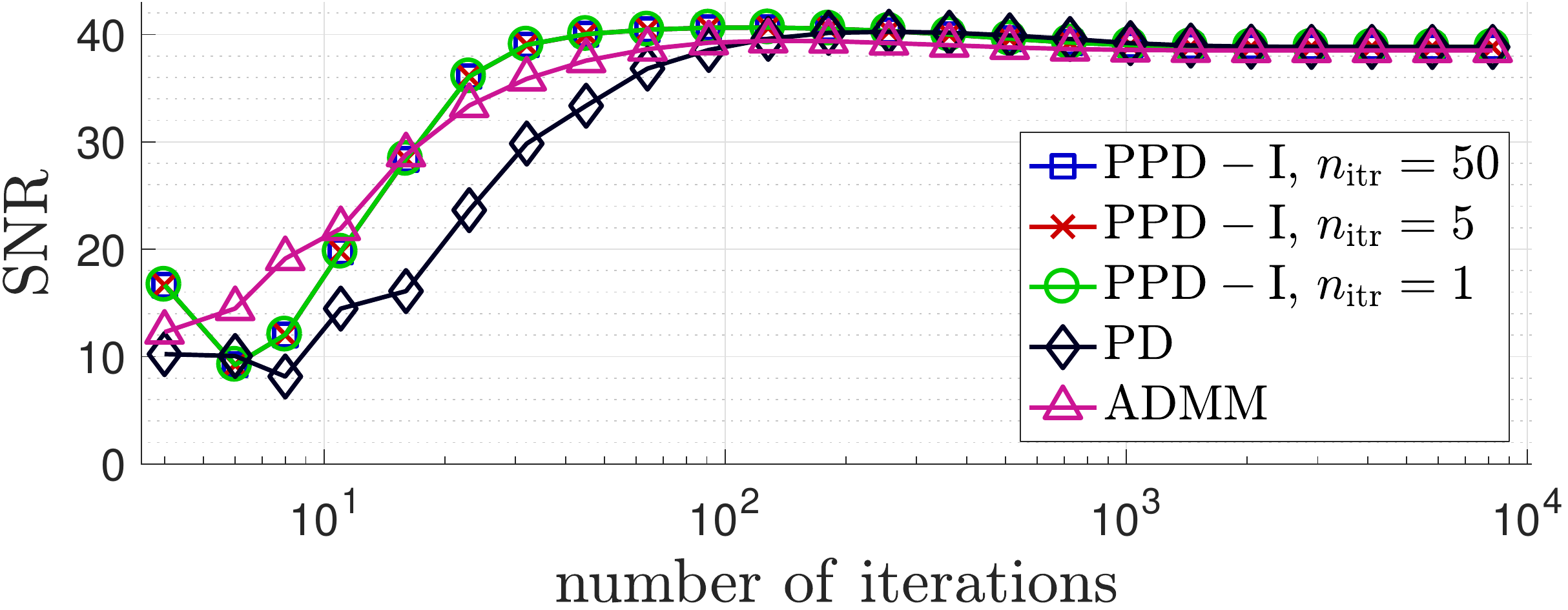}
	\end{minipage}%
	\begin{minipage}{.14\linewidth}
  		\includegraphics[trim={0px 0px 0px 0px}, clip, height=2.3cm]{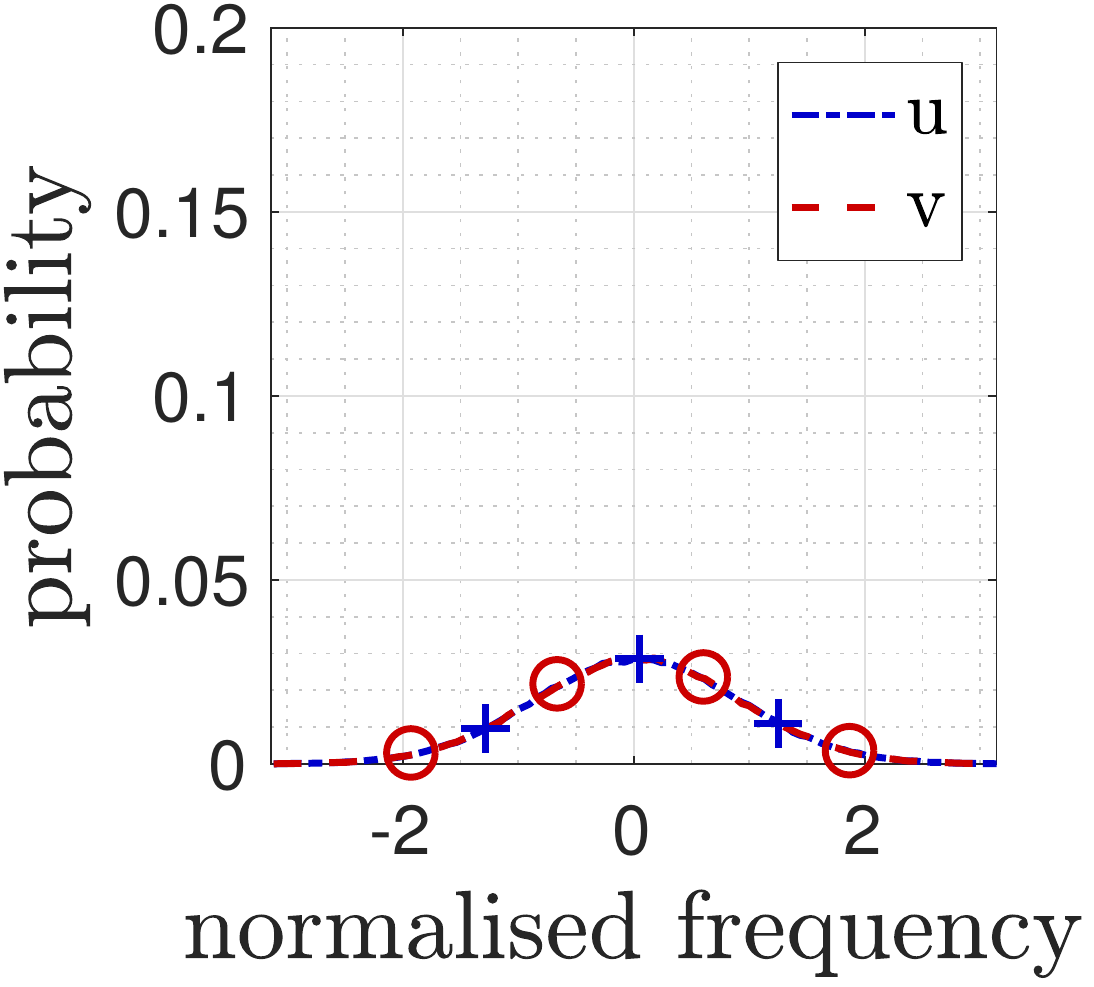}
	\end{minipage}%
	\hspace{.05\linewidth}
	\begin{minipage}{.33\linewidth}
  		\includegraphics[trim={0px 0px 0px 0px}, clip, height=2.3cm]{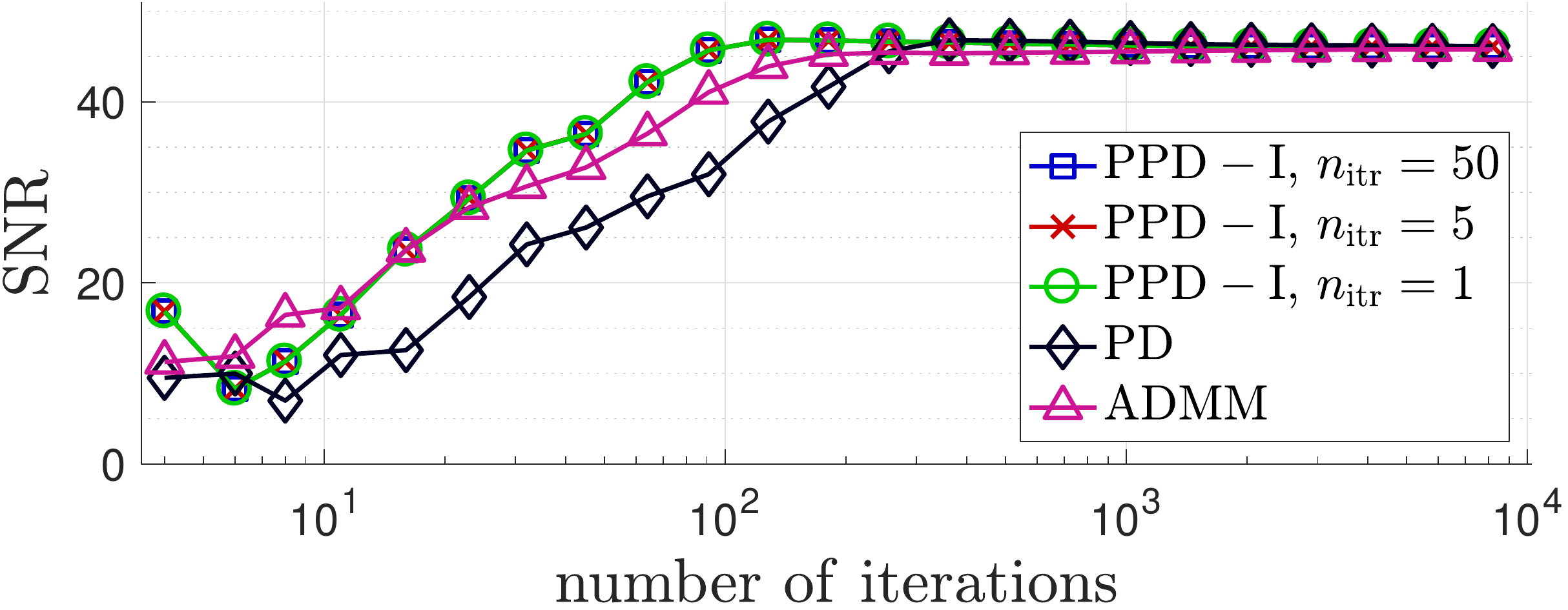}
	\end{minipage}%
	\begin{minipage}{.14\linewidth}
  		\includegraphics[trim={0px 0px 0px 0px}, clip, height=2.3cm]{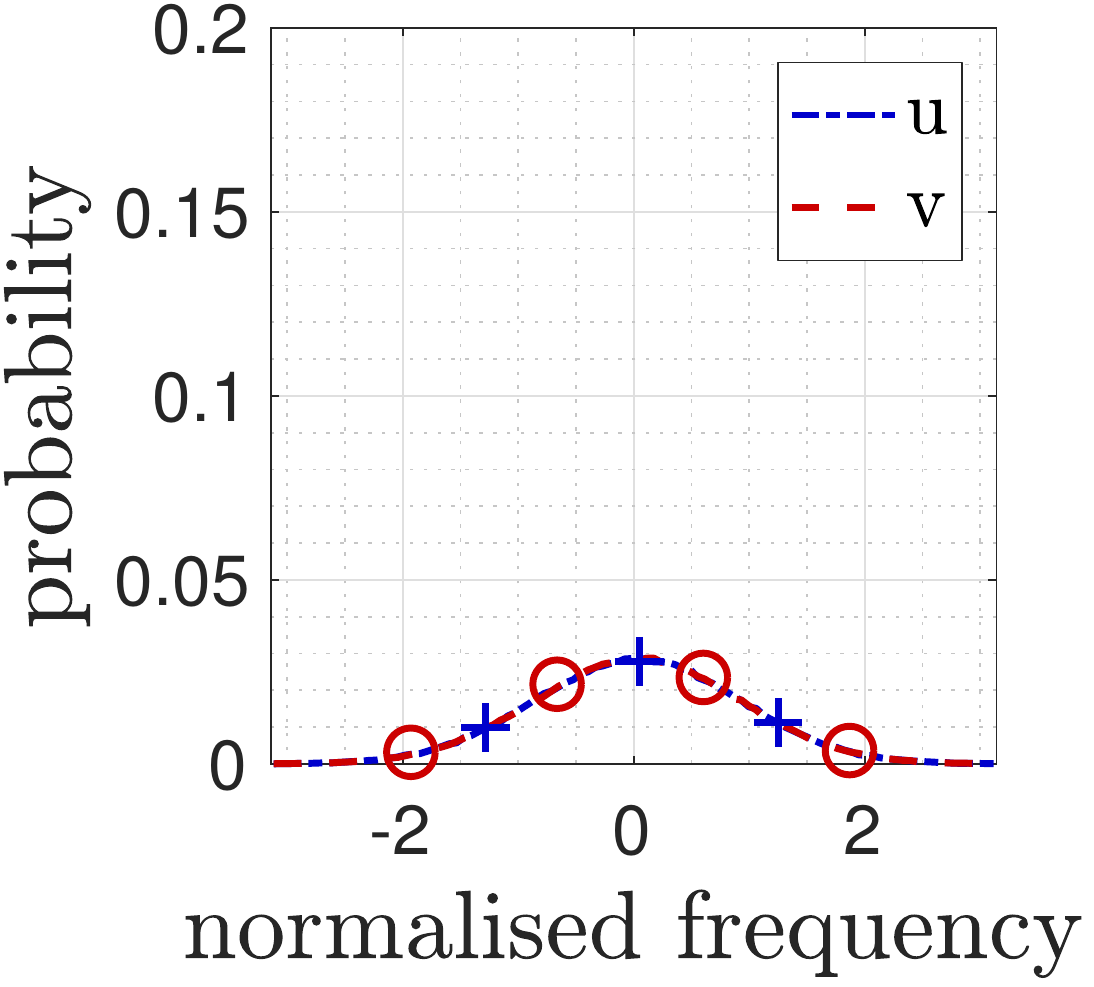}
	\end{minipage}

	\vspace{2pt}
	\begin{minipage}{.33\linewidth}
  		\includegraphics[trim={0px 0px 0px 0px}, clip, height=2.3cm]{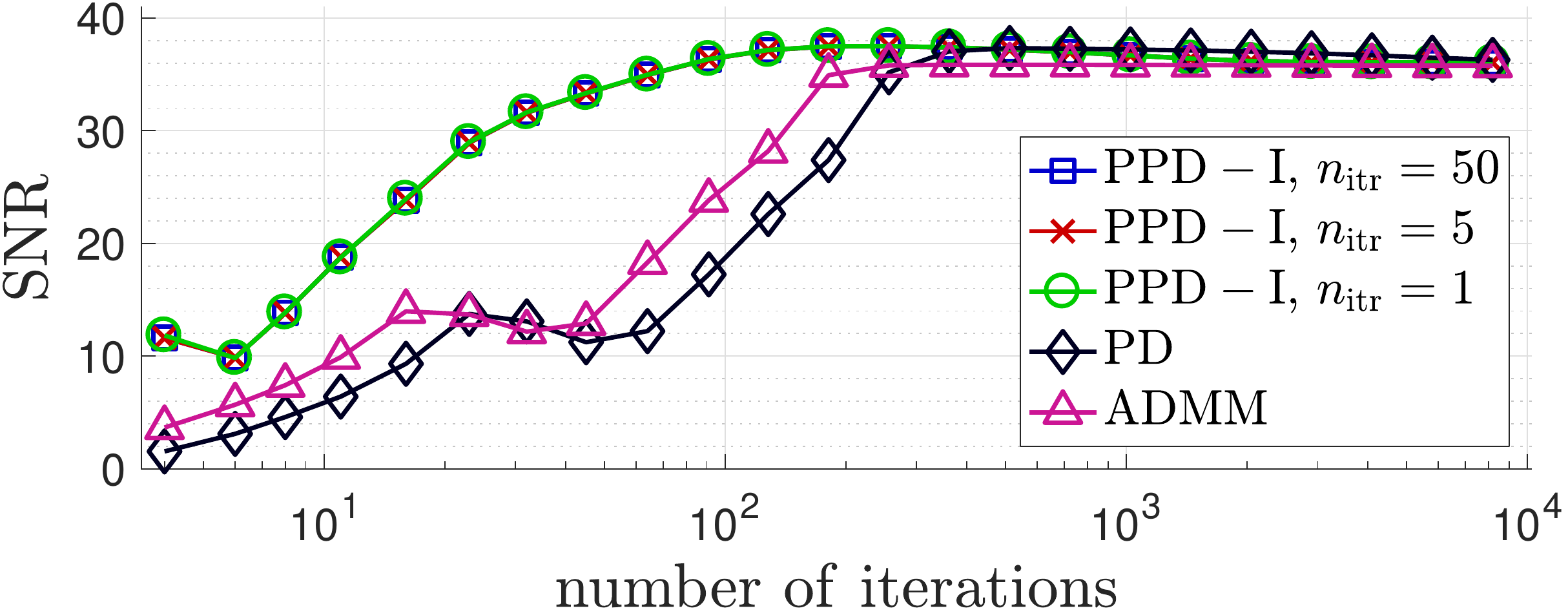}
	\end{minipage}%
	\begin{minipage}{.14\linewidth}
  		\includegraphics[trim={0px 0px 0px 0px}, clip, height=2.3cm]{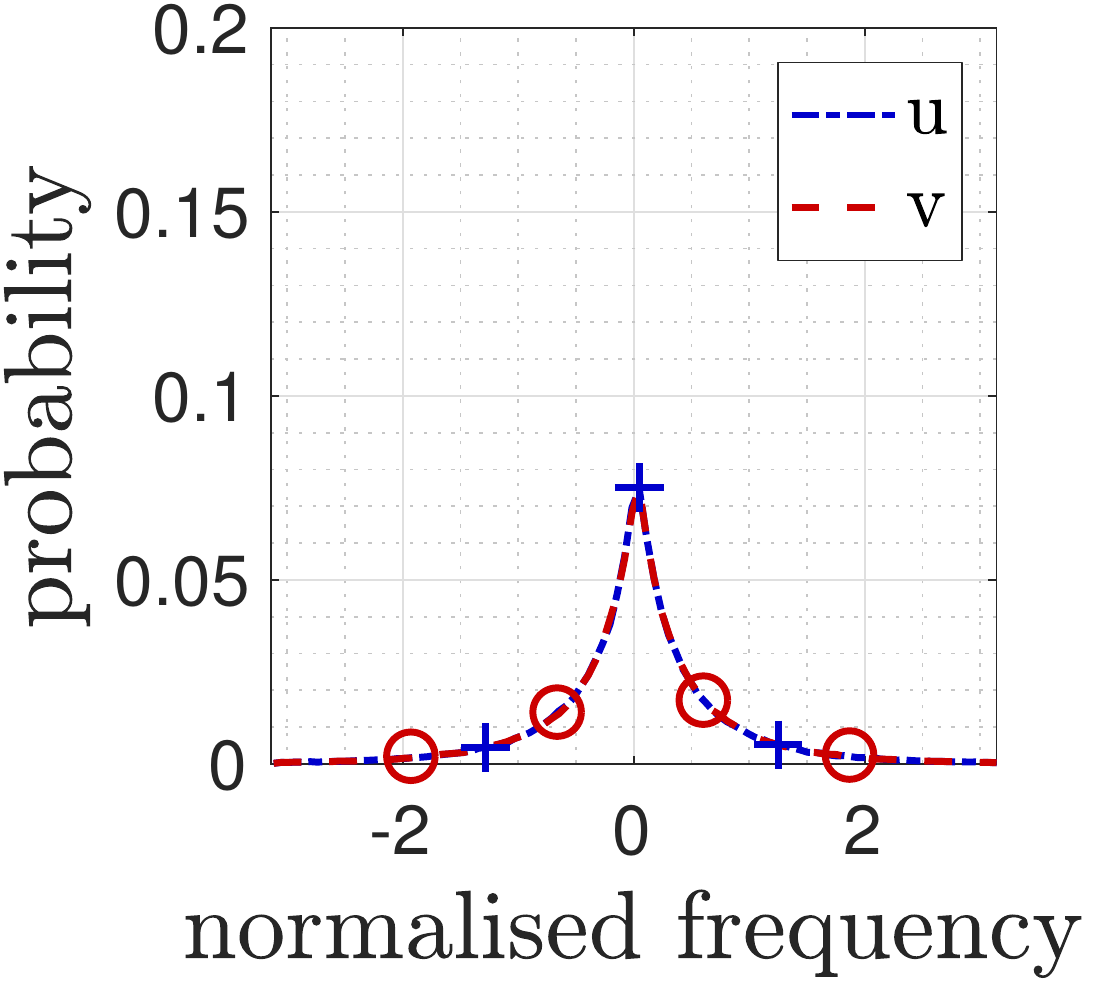}
	\end{minipage}%
	\hspace{.05\linewidth}
	\begin{minipage}{.33\linewidth}
  		\includegraphics[trim={0px 0px 0px 0px}, clip, height=2.3cm]{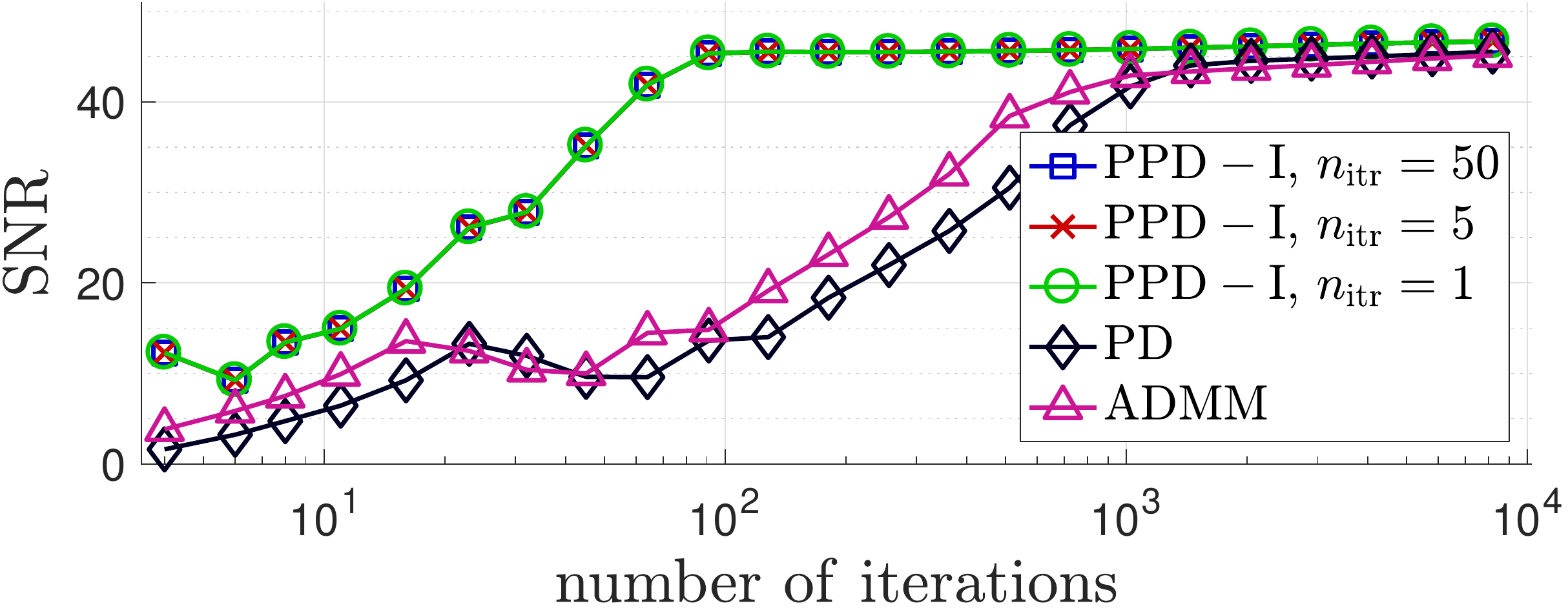}
	\end{minipage}%
	\begin{minipage}{.14\linewidth}
  		\includegraphics[trim={0px 0px 0px 0px}, clip, height=2.3cm]{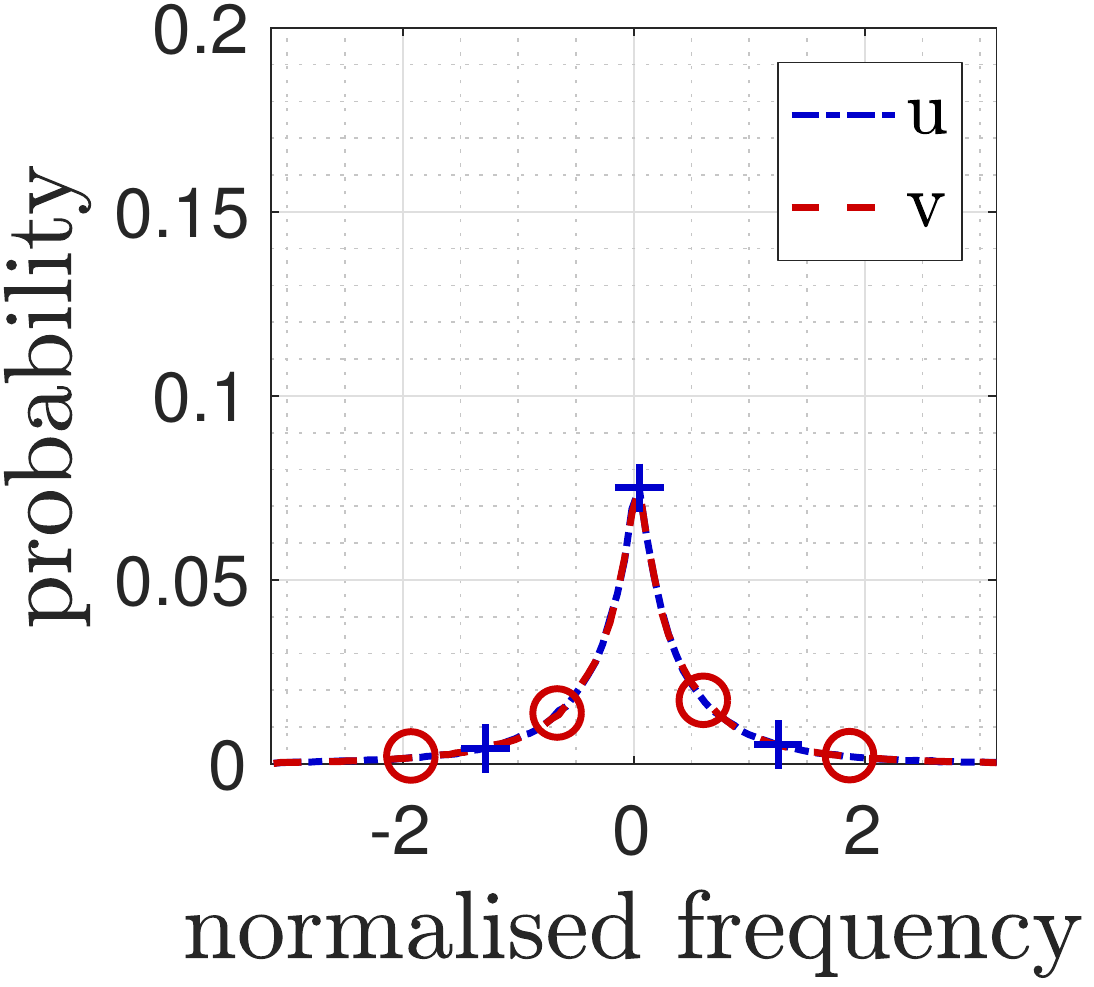}
	\end{minipage}

	\vspace{2pt}
	\begin{minipage}{.33\linewidth}
  		\includegraphics[trim={0px 0px 0px 0px}, clip, height=2.3cm]{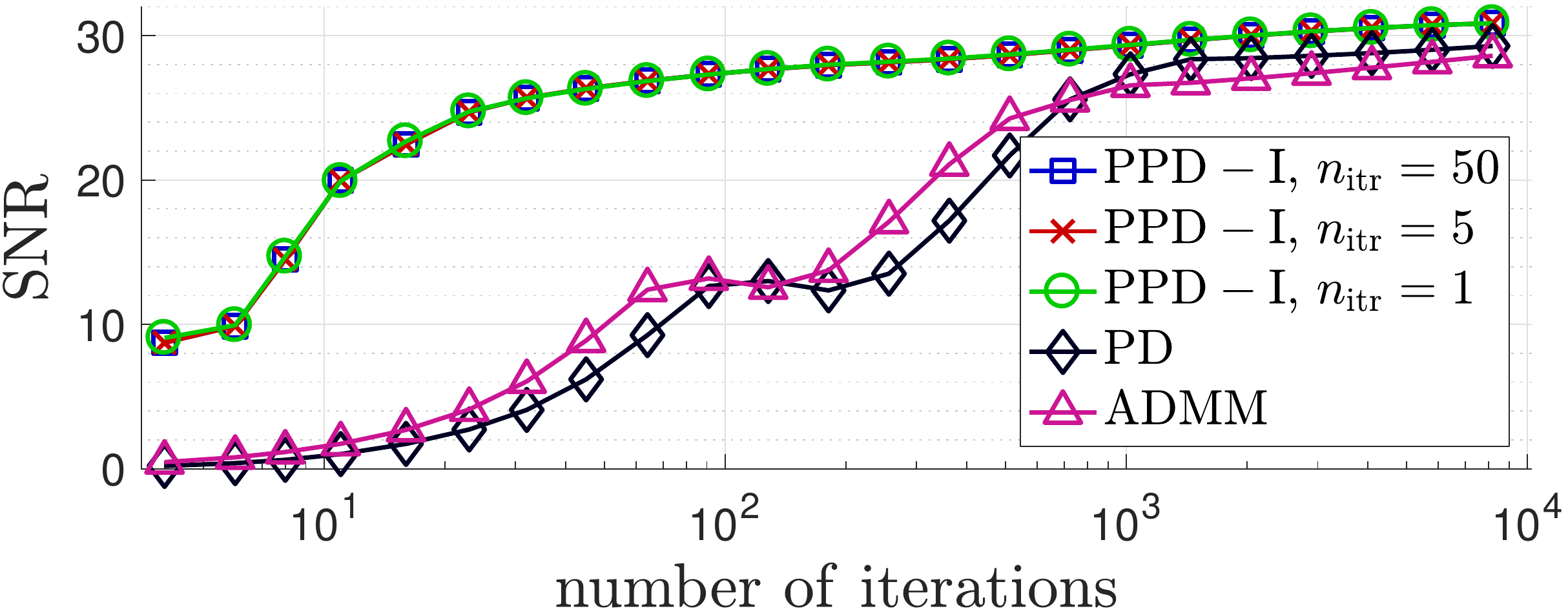}
	\end{minipage}%
	\begin{minipage}{.14\linewidth}
  		\includegraphics[trim={0px 0px 0px 0px}, clip, height=2.3cm]{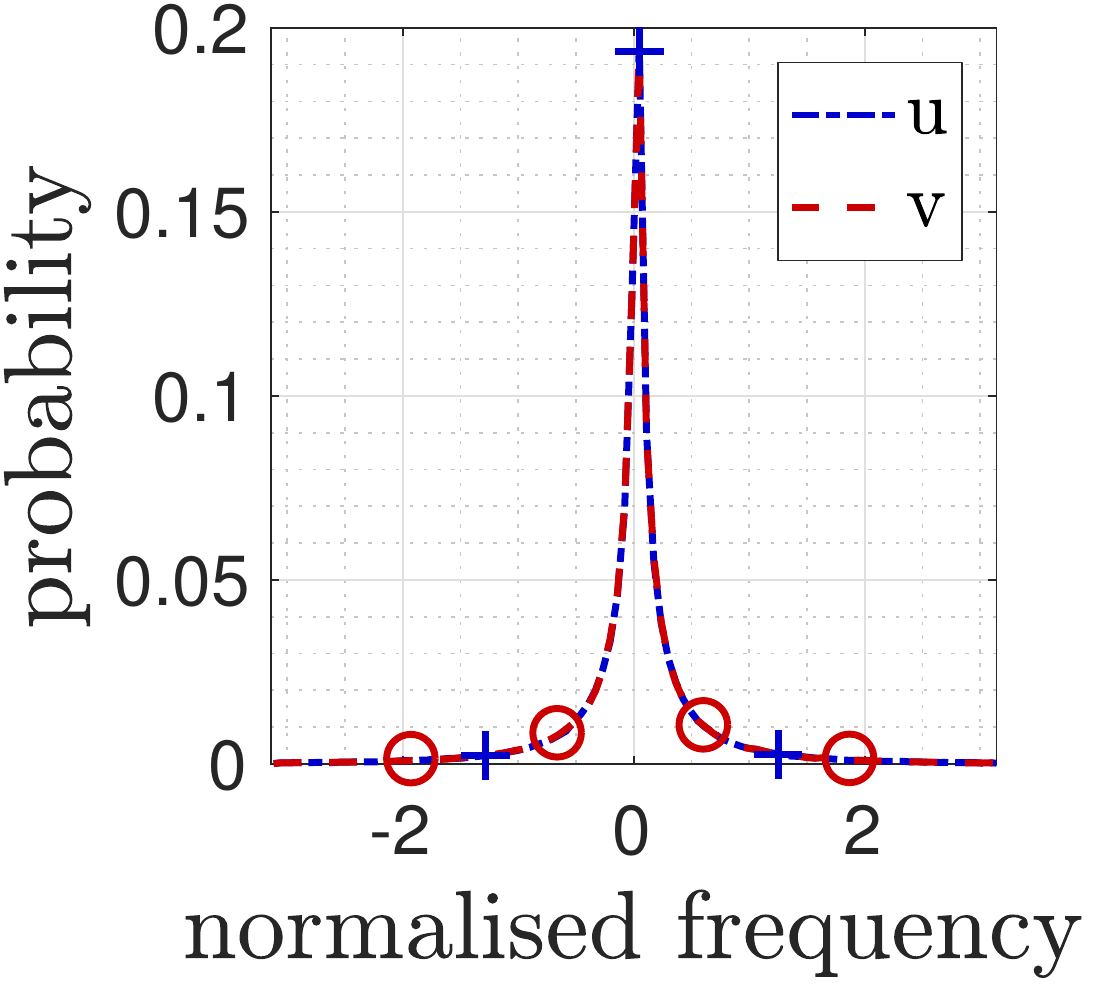}
	\end{minipage}%
	\hspace{.05\linewidth}
	\begin{minipage}{.33\linewidth}
  		\includegraphics[trim={0px 0px 0px 0px}, clip, height=2.3cm]{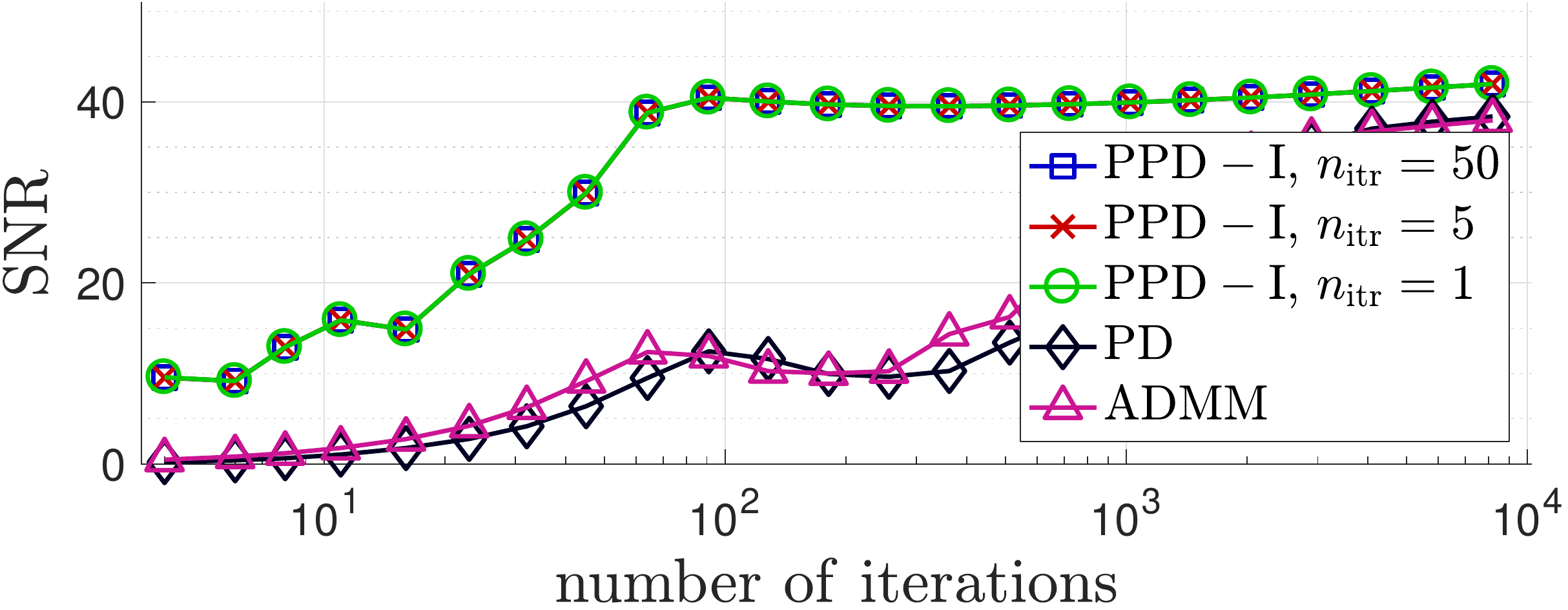}
	\end{minipage}%
	\begin{minipage}{.14\linewidth}
  		\includegraphics[trim={0px 0px 0px 0px}, clip, height=2.3cm]{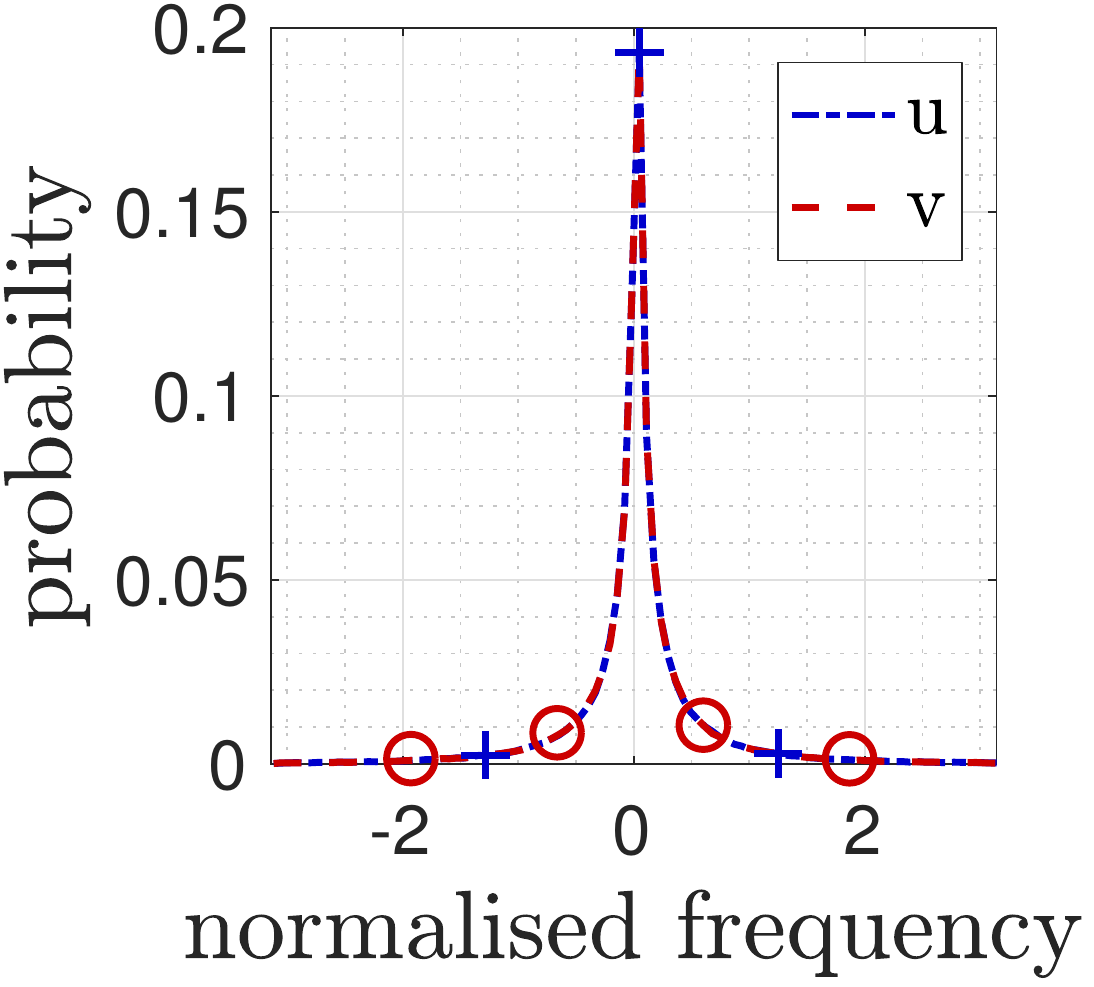}
	\end{minipage}

	\caption{Evolution of the $\rm SNR$ for the \ac{ppd}, \ac{pd} and \ac{admm} algorithms for the reconstruction of the galaxy cluster test image with a $u$--$v$ coverage randomly generated such that the sampling follows a \ac{ggd} with shape parameter $\beta$, from top to bottom, $2$, $0.5$ and $0.25$, respectively. The shape of the distribution of the $u$ and $v$ normalised coordinates is presented next to the graph portraying the evolution of the $\rm SNR$.
The visibilities are corrupted by Gaussian noise to produce a 30$\rm dB$  $\rm iSNR$ for the figures on the right and a 50$\rm dB$ $\rm iSNR$ for the figures on the left. The number of sub-iteration $n_{\rm itr}$ performed by \ac{ppd} to estimate the ellipsoid projection is also reported.}
	\label{ggd-gc}
\end{figure*}

\begin{figure*}
	\centering
	\begin{minipage}{.33\linewidth}
  		\includegraphics[trim={0px 0px 0px 0px}, clip, height=2.3cm]{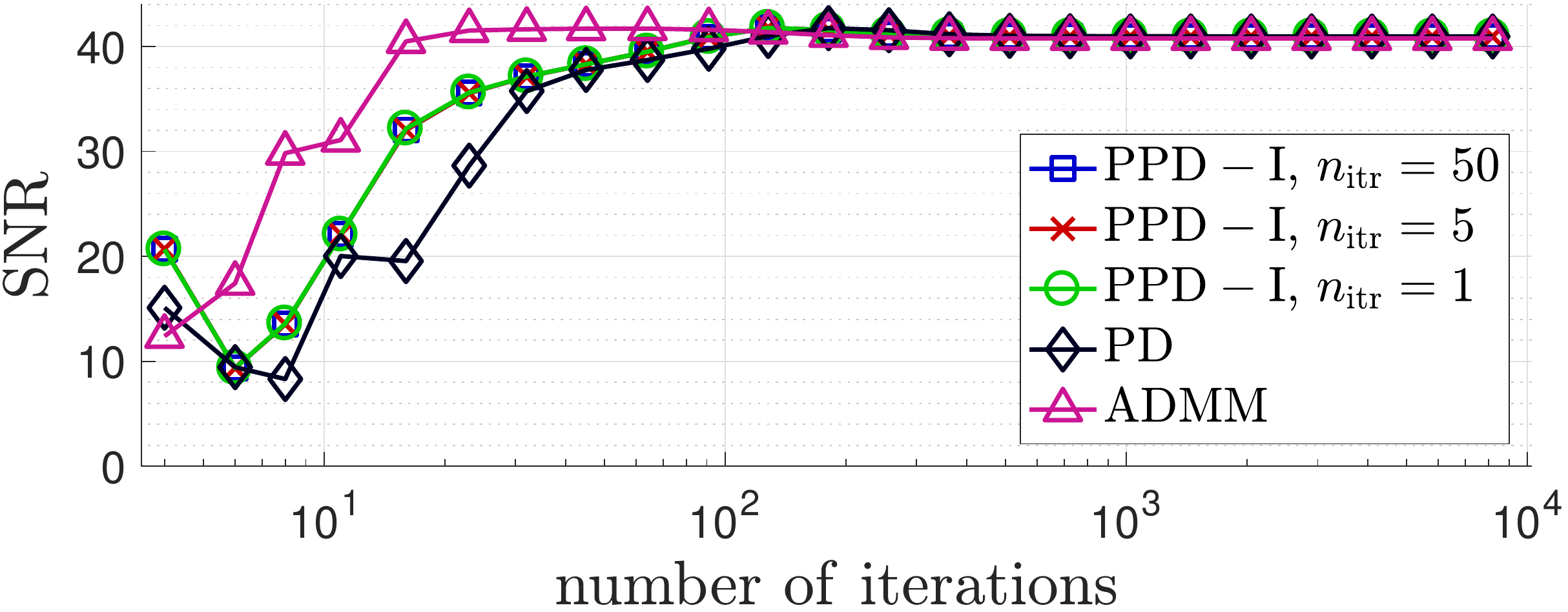}
	\end{minipage}%
	\begin{minipage}{.14\linewidth}
  		\includegraphics[trim={0px 0px 0px 0px}, clip, height=2.3cm]{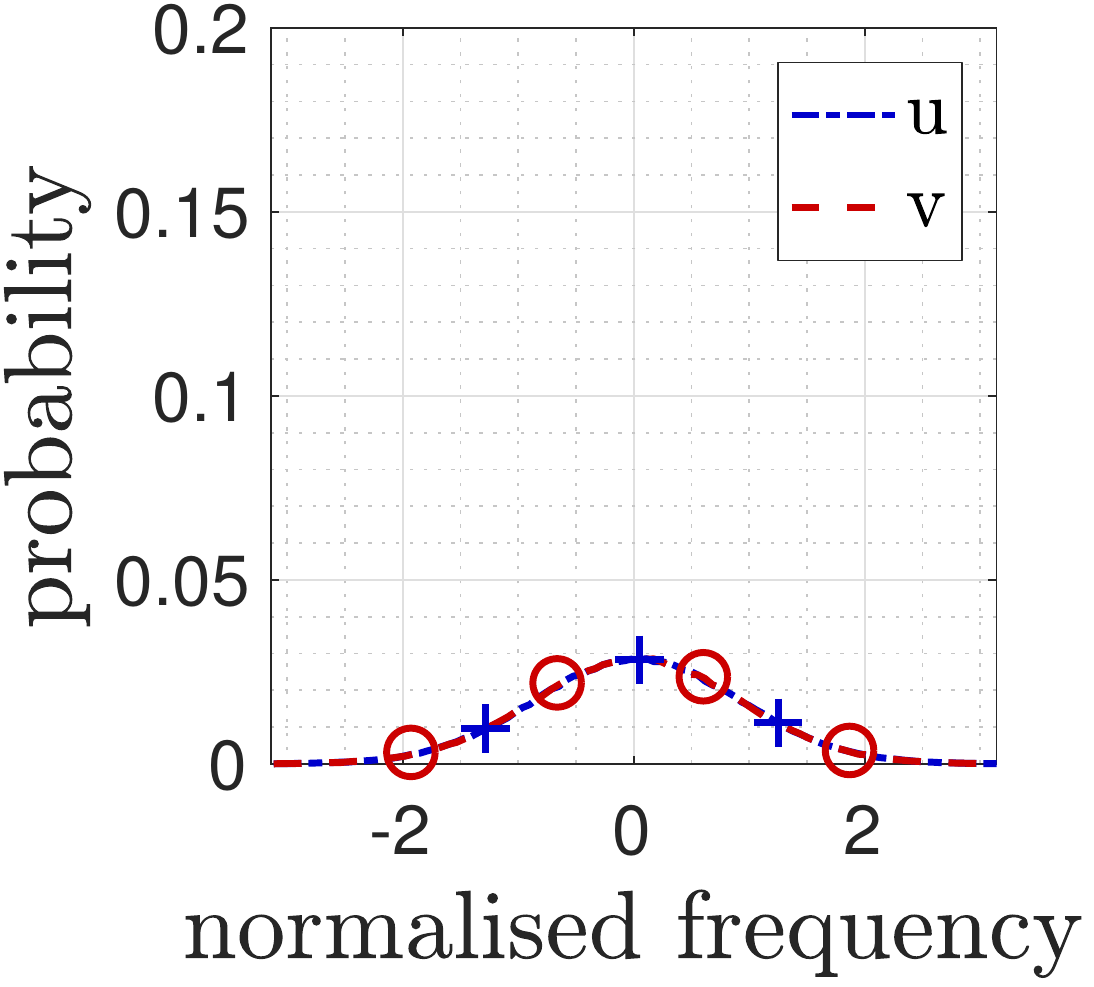}
	\end{minipage}%
	\hspace{.05\linewidth}
	\begin{minipage}{.33\linewidth}
  		\includegraphics[trim={0px 0px 0px 0px}, clip, height=2.3cm]{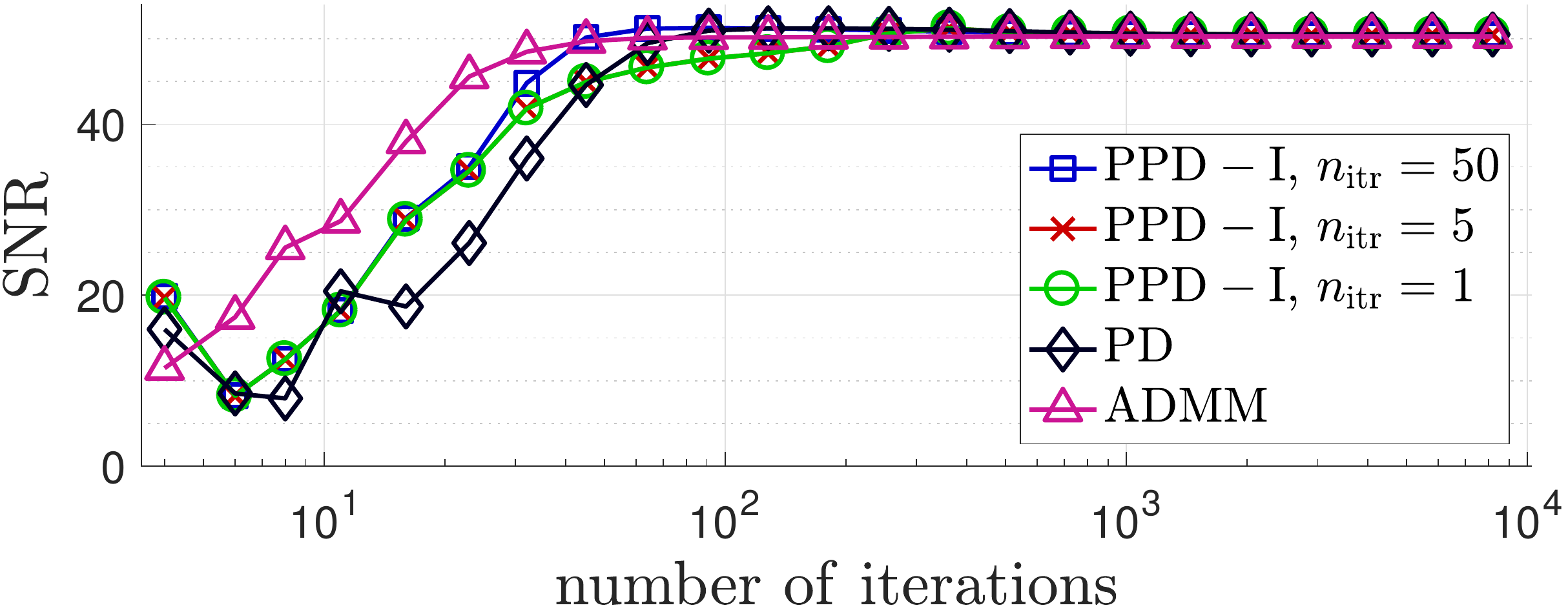}
	\end{minipage}%
	\begin{minipage}{.14\linewidth}
  		\includegraphics[trim={0px 0px 0px 0px}, clip, height=2.3cm]{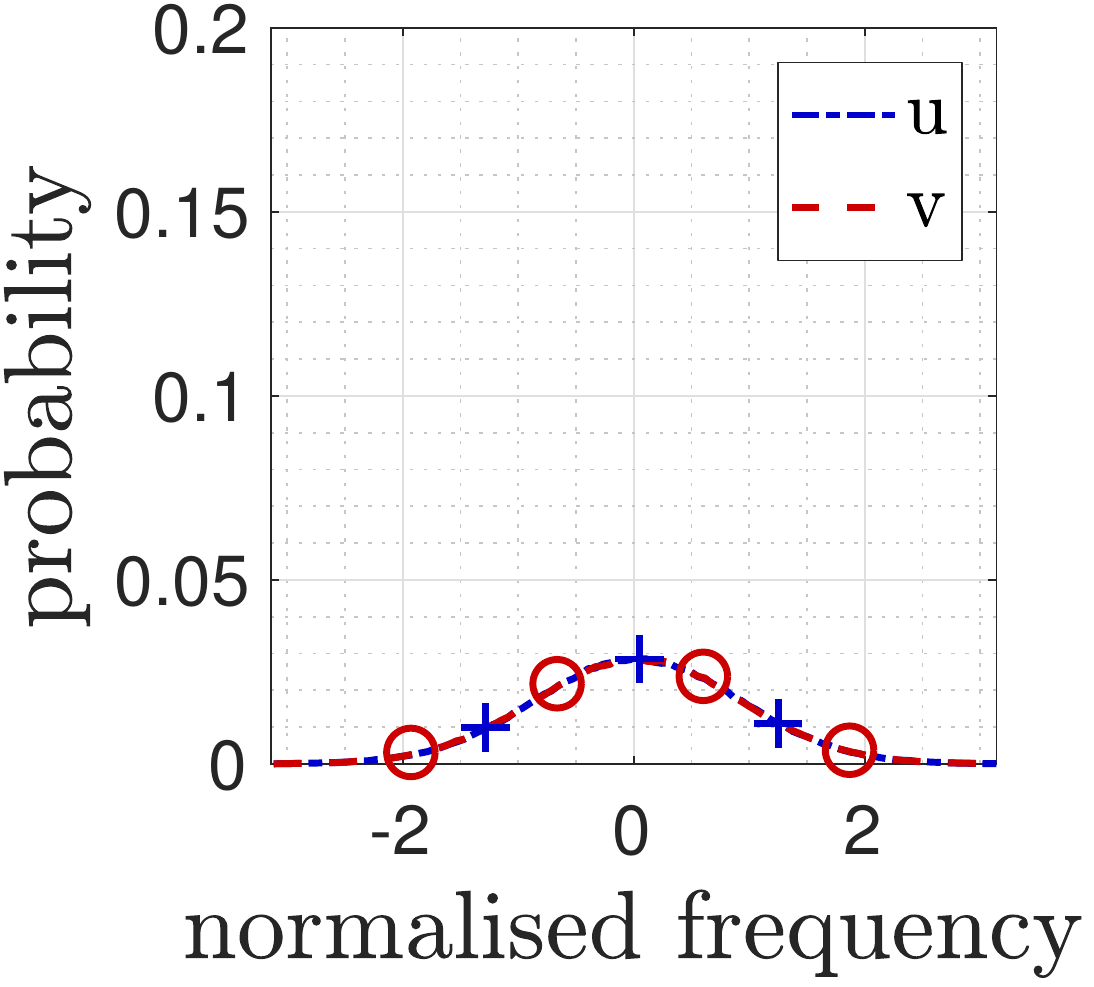}
	\end{minipage}

	\vspace{2pt}
	\begin{minipage}{.33\linewidth}
  		\includegraphics[trim={0px 0px 0px 0px}, clip, height=2.3cm]{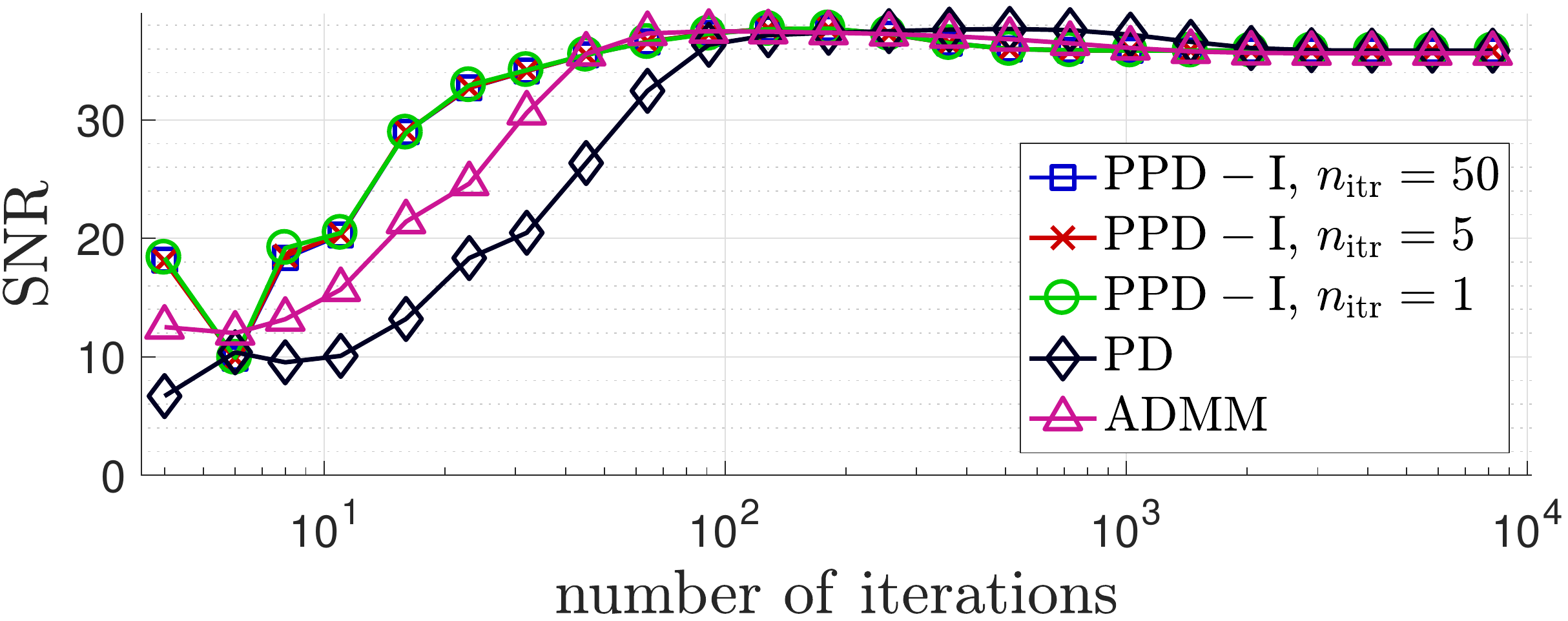}
	\end{minipage}%
	\begin{minipage}{.14\linewidth}
  		\includegraphics[trim={0px 0px 0px 0px}, clip, height=2.3cm]{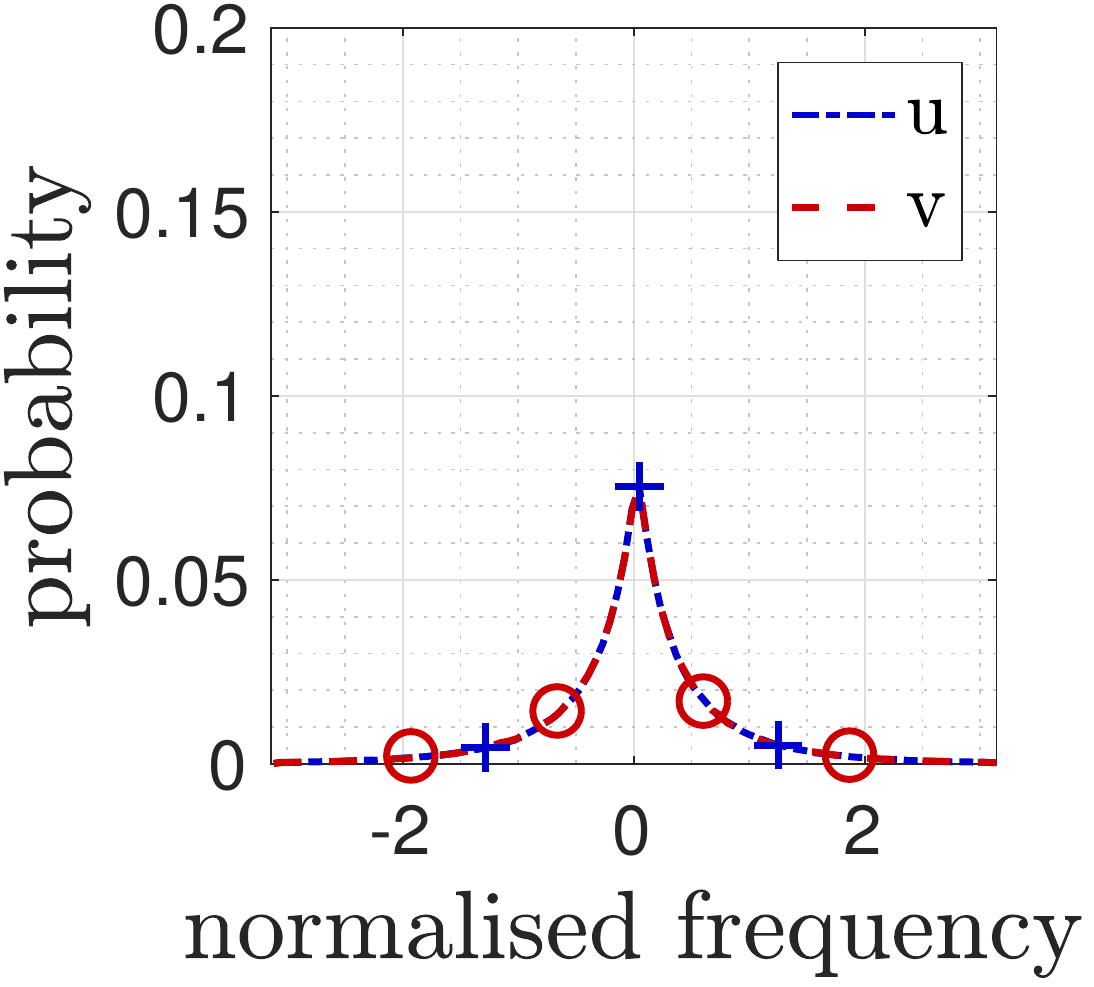}
	\end{minipage}%
	\hspace{.05\linewidth}
	\begin{minipage}{.33\linewidth}
  		\includegraphics[trim={0px 0px 0px 0px}, clip, height=2.3cm]{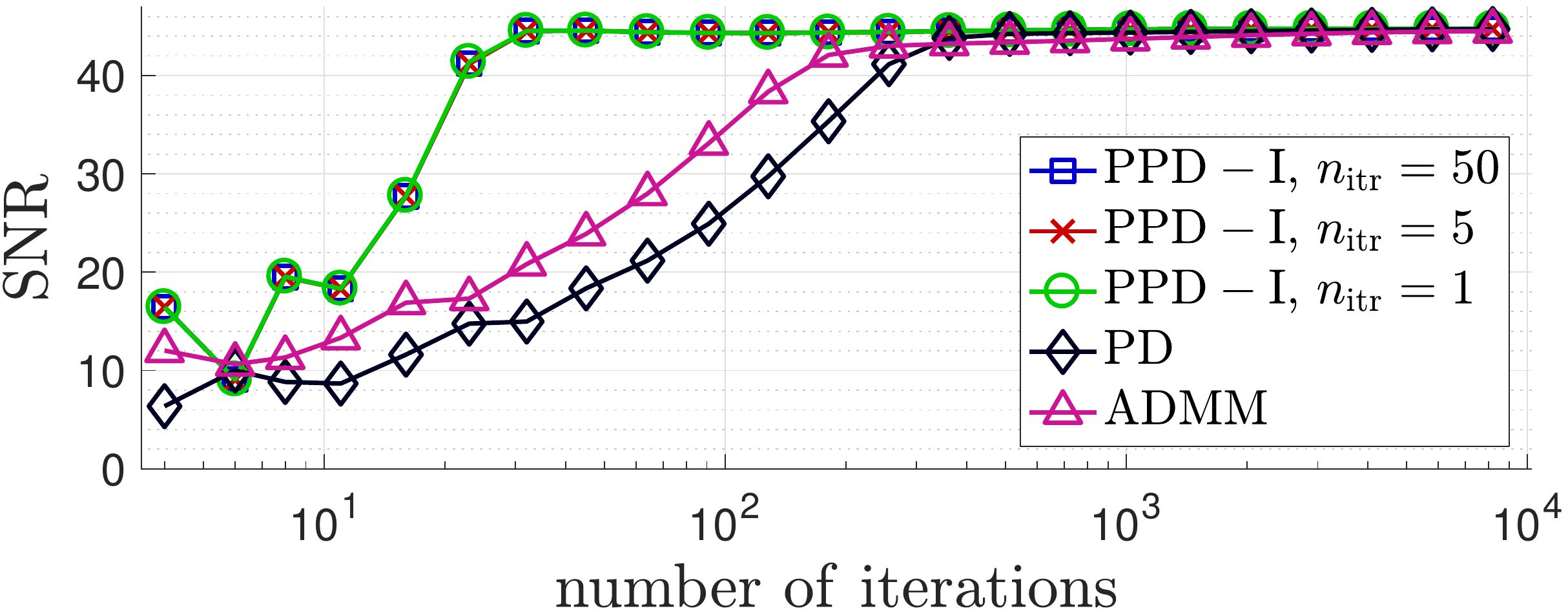}
	\end{minipage}%
	\begin{minipage}{.14\linewidth}
  		\includegraphics[trim={0px 0px 0px 0px}, clip, height=2.3cm]{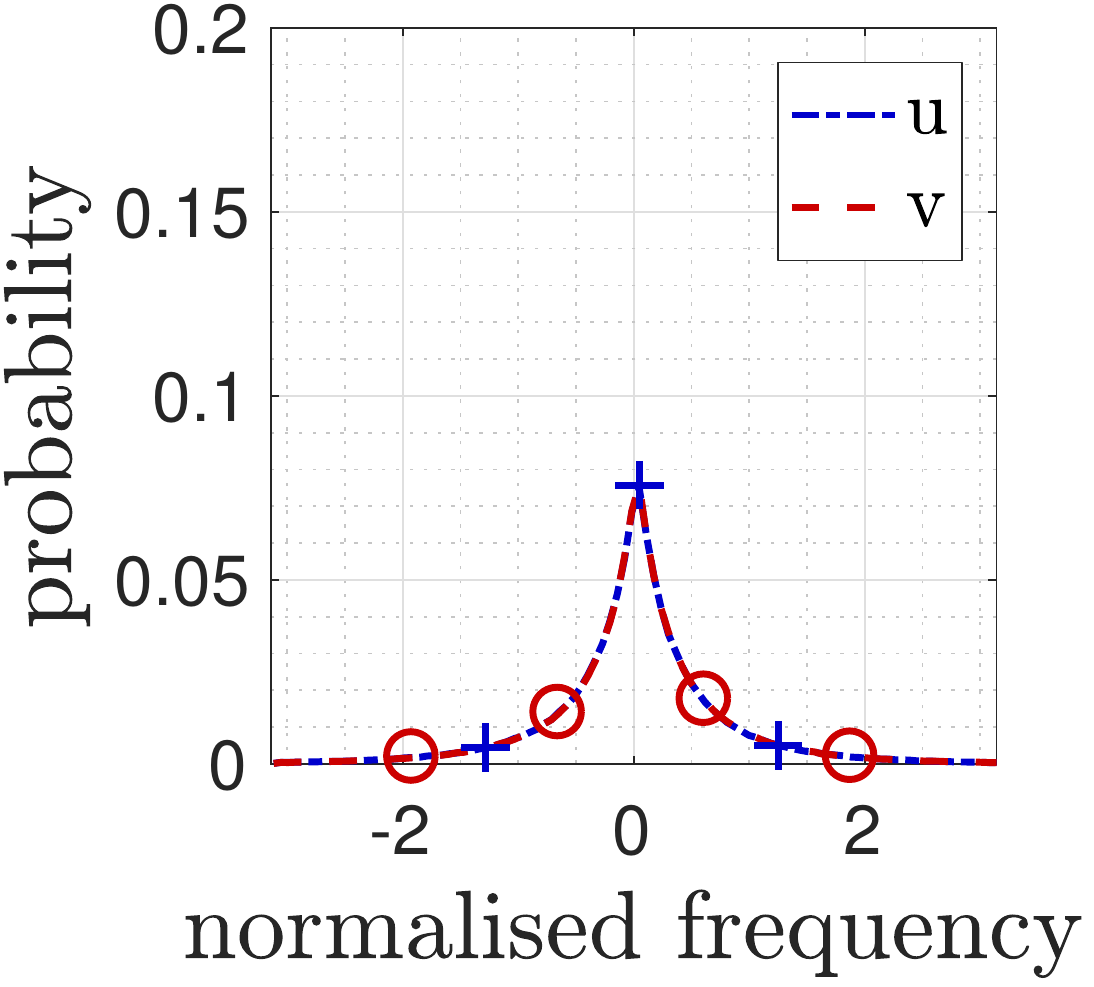}
	\end{minipage}

	\vspace{2pt}
	\begin{minipage}{.33\linewidth}
  		\includegraphics[trim={0px 0px 0px 0px}, clip, height=2.3cm]{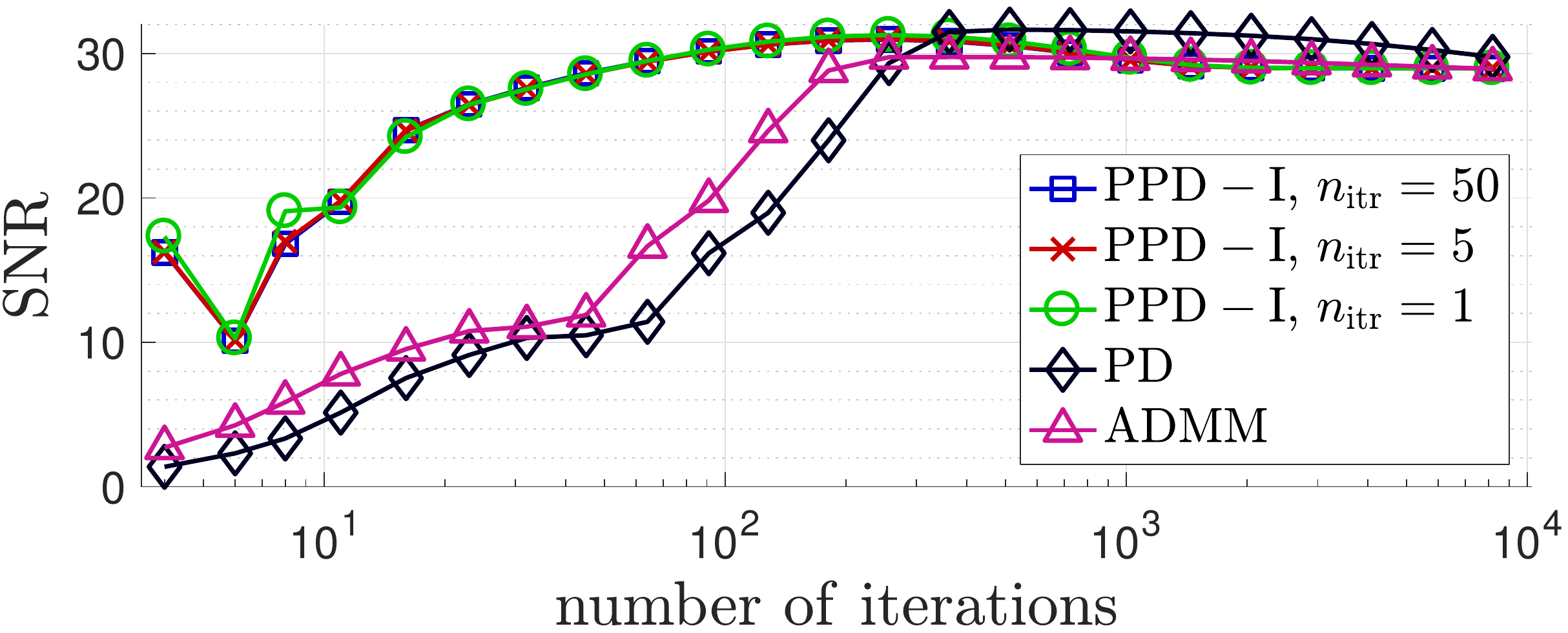}
	\end{minipage}%
	\begin{minipage}{.14\linewidth}
  		\includegraphics[trim={0px 0px 0px 0px}, clip, height=2.3cm]{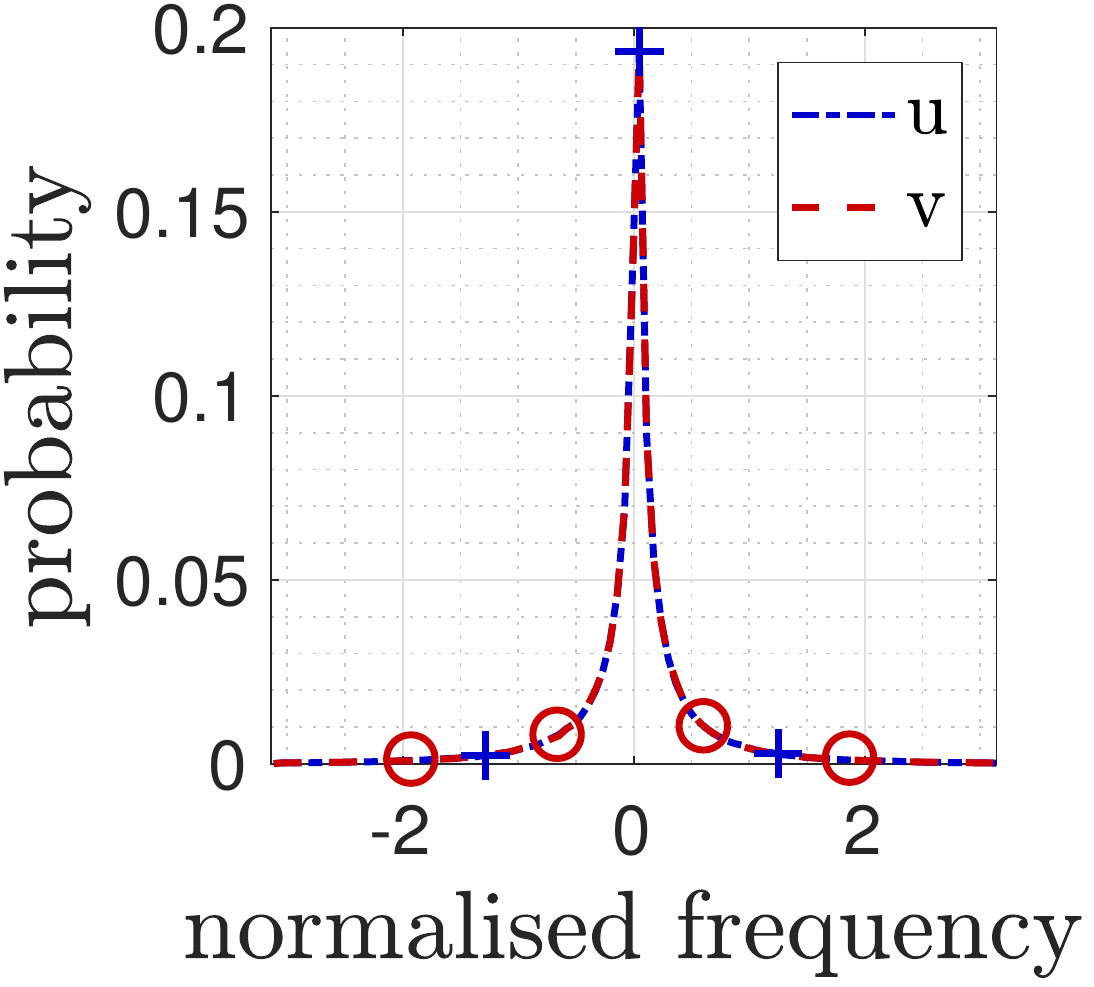}
	\end{minipage}%
	\hspace{.05\linewidth}
	\begin{minipage}{.33\linewidth}
  		\includegraphics[trim={0px 0px 0px 0px}, clip, height=2.3cm]{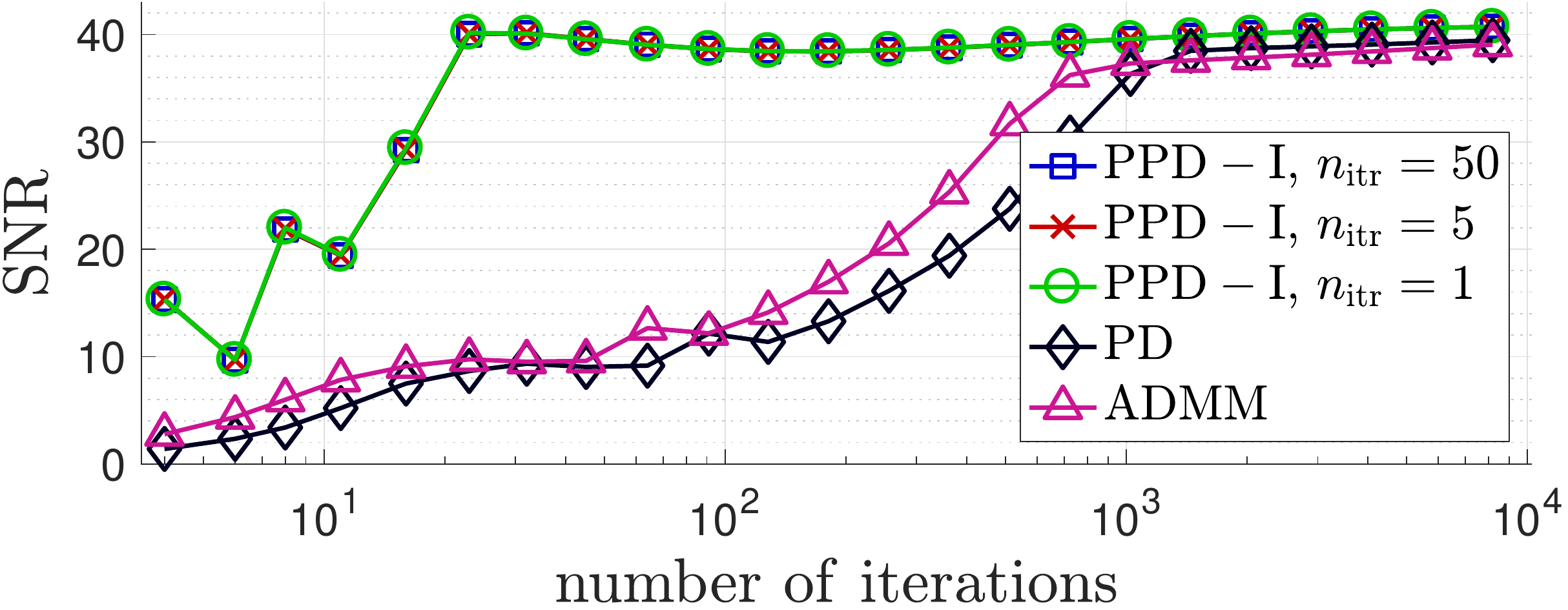}
	\end{minipage}%
	\begin{minipage}{.14\linewidth}
  		\includegraphics[trim={0px 0px 0px 0px}, clip, height=2.3cm]{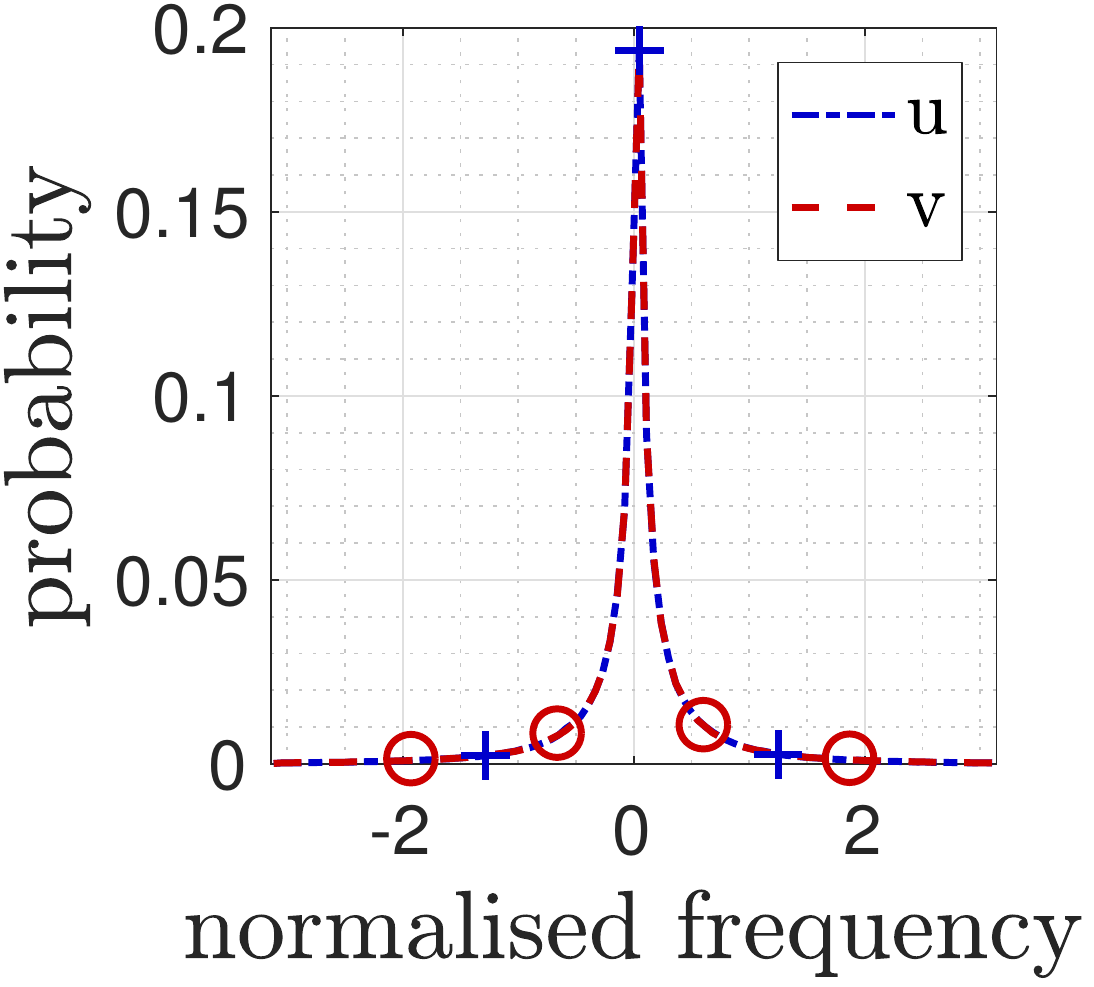}
	\end{minipage}

	\caption{Evolution of the $\rm SNR$ for the \ac{ppd}, \ac{pd} and \ac{admm} algorithms for the reconstruction of the Cygnus A test image with a $u$--$v$ coverage randomly generated such that the sampling follows a \ac{ggd} with shape parameter $\beta$, from top to bottom, $2$, $0.5$ and $0.25$, respectively. The shape of the distribution of the $u$ and $v$ normalised coordinates is presented next to the graph portraying the evolution of the $\rm SNR$.
The visibilities are corrupted by Gaussian noise to produce a 30$\rm dB$ $\rm iSNR$ for the figures on the right and a 50$\rm dB$ $\rm iSNR$ for the figures on the left. The number of sub-iteration $n_{\rm itr}$ performed by \ac{ppd} to estimate the ellipsoid projection is also reported.}
	\label{ggd-ca}
\end{figure*}

To study the behaviour of \ac{ppd} across a broad range of $u$--$v$ sampling strategies, we use coverages with the sampled $u$--$v$ points distributed according to a generalised Gaussian distribution with the shape parameter $\beta$.
We study the acceleration for the reconstruction of the galaxy cluster test image in Figure \ref{ggd-gc} and for the reconstruction of the Cygnus A test image in Figure \ref{ggd-ca}.
Here, we report the evolution of the ${\rm SNR}$ as a function of the number of iterations.
In both cases we have performed tests for two levels of input noise, $30~\rm{dB}$ and $50~\rm{dB}$.
For all test cases we provide the distribution of the normalised $u$ and $v$ coordinates to showcase the link between the convergence speed and sampling pattern.

For sampling strategies that are farther away from uniform, the preconditioning strategy improves the convergence rate dramatically in all test cases.
For a Gaussian sampling, when $\beta=2$, the converge speed of the \ac{ppd} is similar to that of \ac{pd} and \ac{admm}.
A decrease in $\beta$ does not affect \ac{ppd} greatly. It maintains almost the same convergence speed throughout all the test cases.
In the extreme case when $\beta=0.25$, the density of measurements is much greater in the centre of the $u$--$v$ space and \ac{ppd} becomes one order of magnitude faster than \ac{pd} and \ac{admm}.
In all test cases, the \ac{ppd} algorithm remains robust to an inexact computation of the ellipsoid projection.
In practice there is little difference between performing $1$ sub-iteration and performing as many as $50$.
Due to this, its complexity per iteration is marginally larger than that of \ac{pd}.
This, coupled with the improved convergence rate, makes \ac{ppd} much more suitable for the large-scale problems arising in \ac{ri}.
Comparing the two input noise regimes, for lower input noise, the gap between \ac{ppd} and \ac{pd} becomes larger.
For less noisy data, the sampling density becomes the most important factor that limits the convergence speed.
This is due to the high frequency data having lower power than the low frequency data.
For large noise, the high frequency visibilities are below the noise level and the effective coverage can be considered to be truncated at the point where the data are overwhelmed by the noise.
For the low noise setup, the algorithms can improve the reconstruction and achieve a higher $\rm SNR$ but the coverage becomes more important for the convergence speed because the effective useful visibilities cover a wider range of frequencies in the $u$--$v$ space.

\begin{figure*}
	\centering
	
	\begin{minipage}{.33\linewidth}
  		\includegraphics[trim={0px 0px 0px 0px}, clip, height=2.3cm]{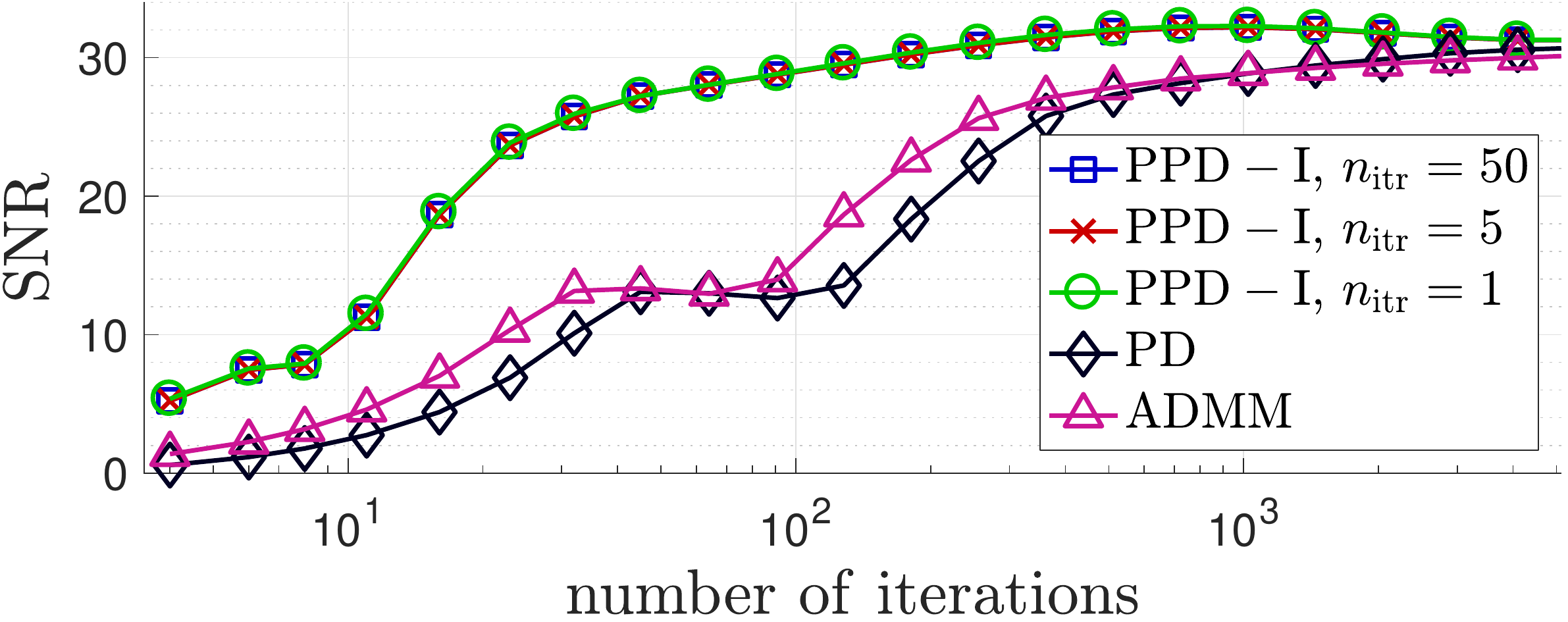}
	\end{minipage}%
	\begin{minipage}{.14\linewidth}
  		\includegraphics[trim={0px 0px 0px 0px}, clip, height=2.3cm]{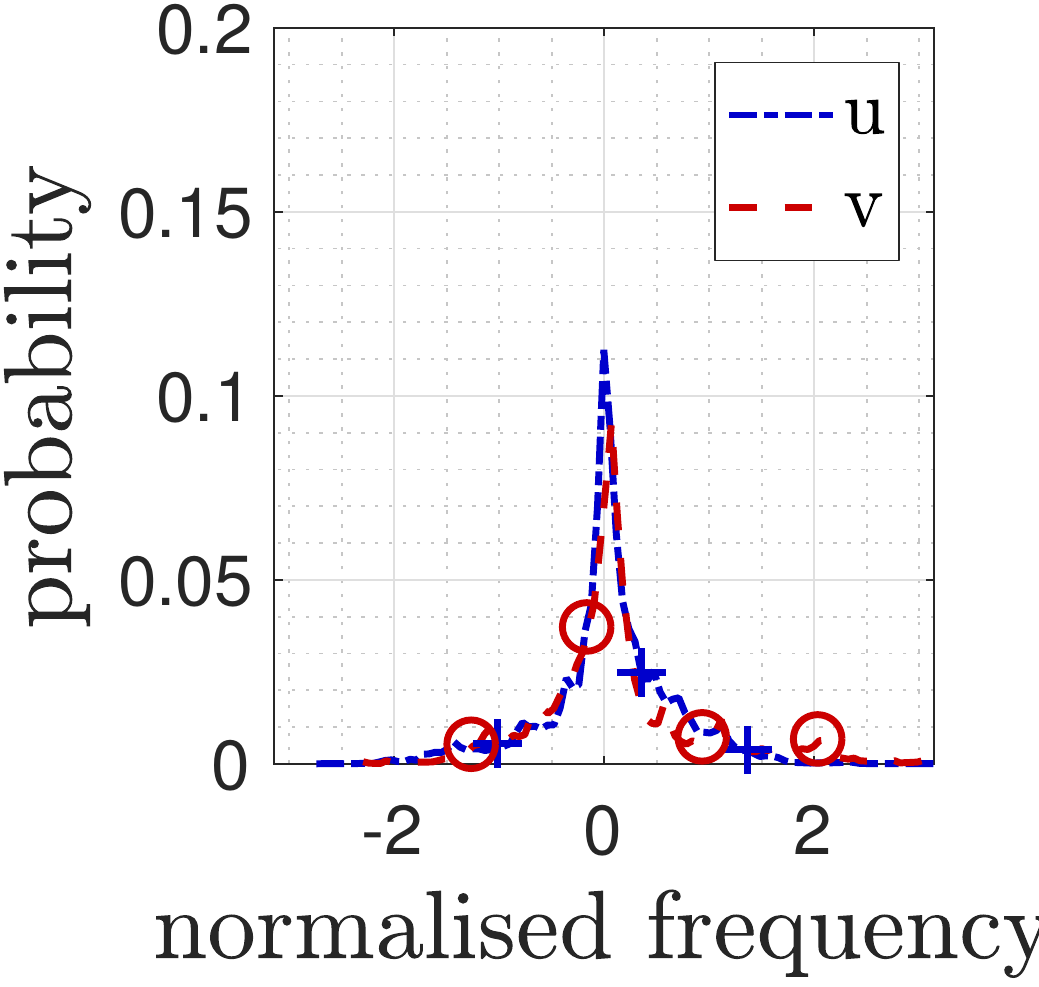}
	\end{minipage}%
	\hspace{.05\linewidth}
	\begin{minipage}{.33\linewidth}
  		\includegraphics[trim={0px 0px 0px 0px}, clip, height=2.3cm]{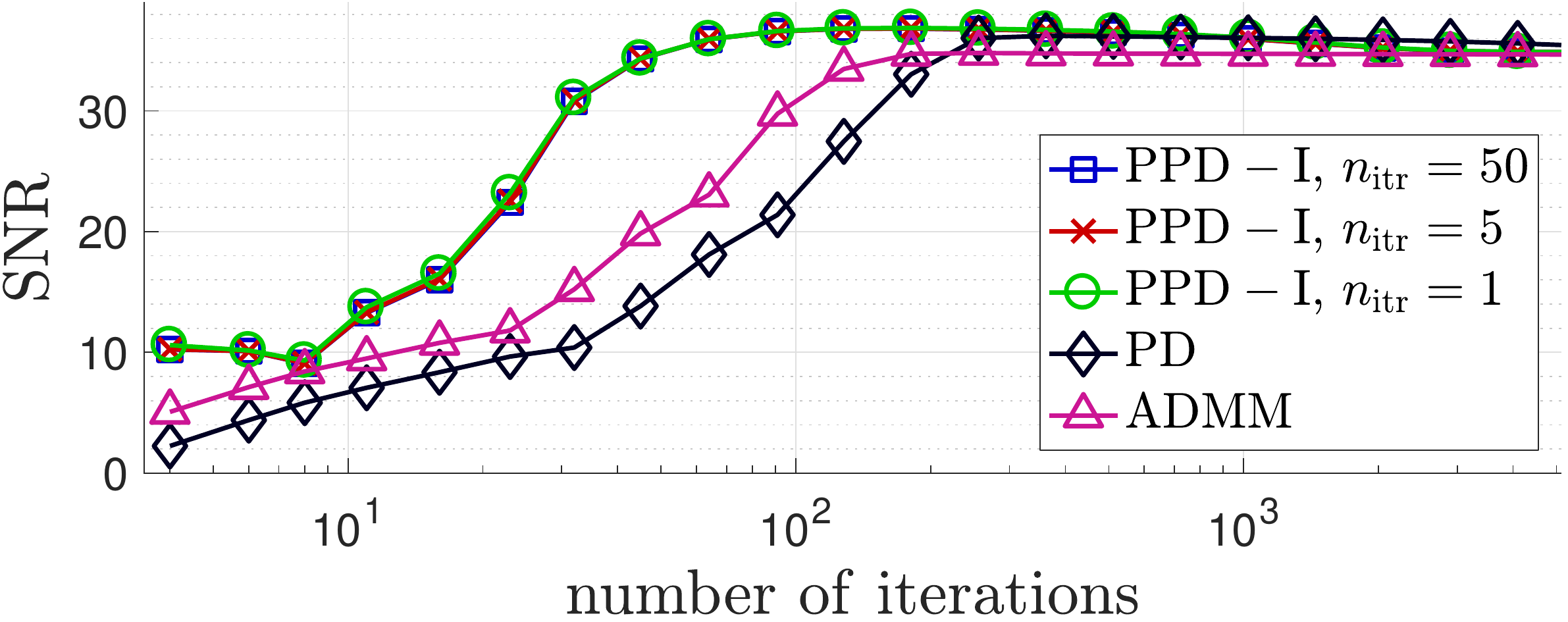}
	\end{minipage}%
	\begin{minipage}{.14\linewidth}
  		\includegraphics[trim={0px 0px 0px 0px}, clip, height=2.3cm]{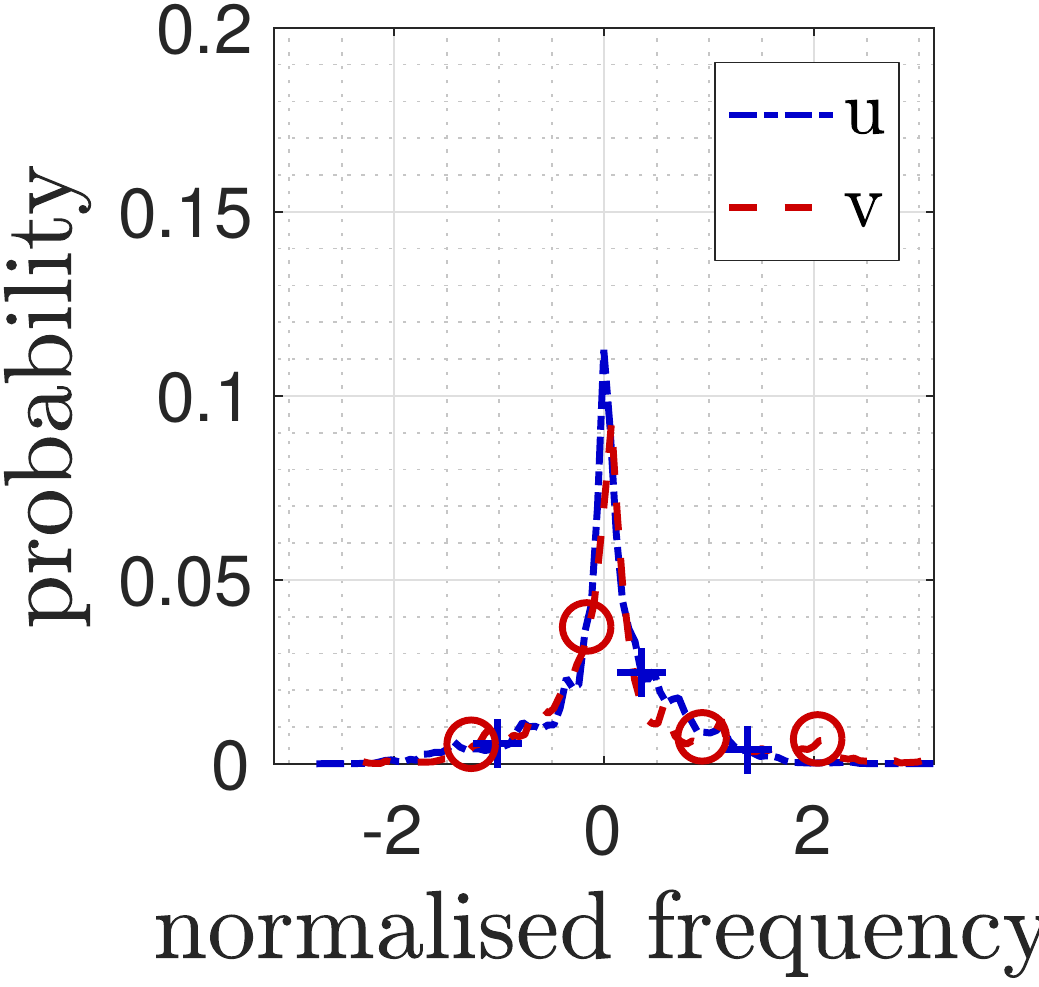}
	\end{minipage}
	
	\vspace{2pt}
	\begin{minipage}{.33\linewidth}
  		\includegraphics[trim={0px 0px 0px 0px}, clip, height=2.3cm]{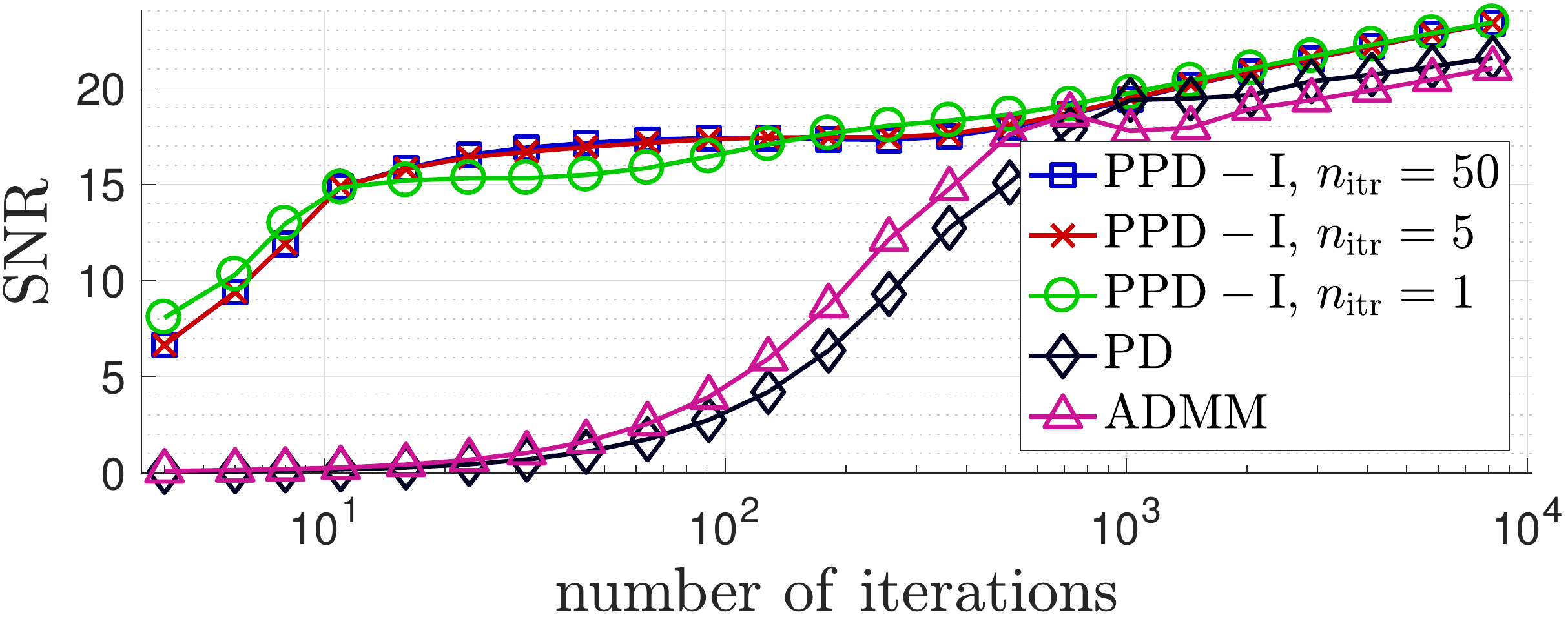}
	\end{minipage}%
	\begin{minipage}{.14\linewidth}
  		\includegraphics[trim={0px 0px 0px 0px}, clip, height=2.3cm]{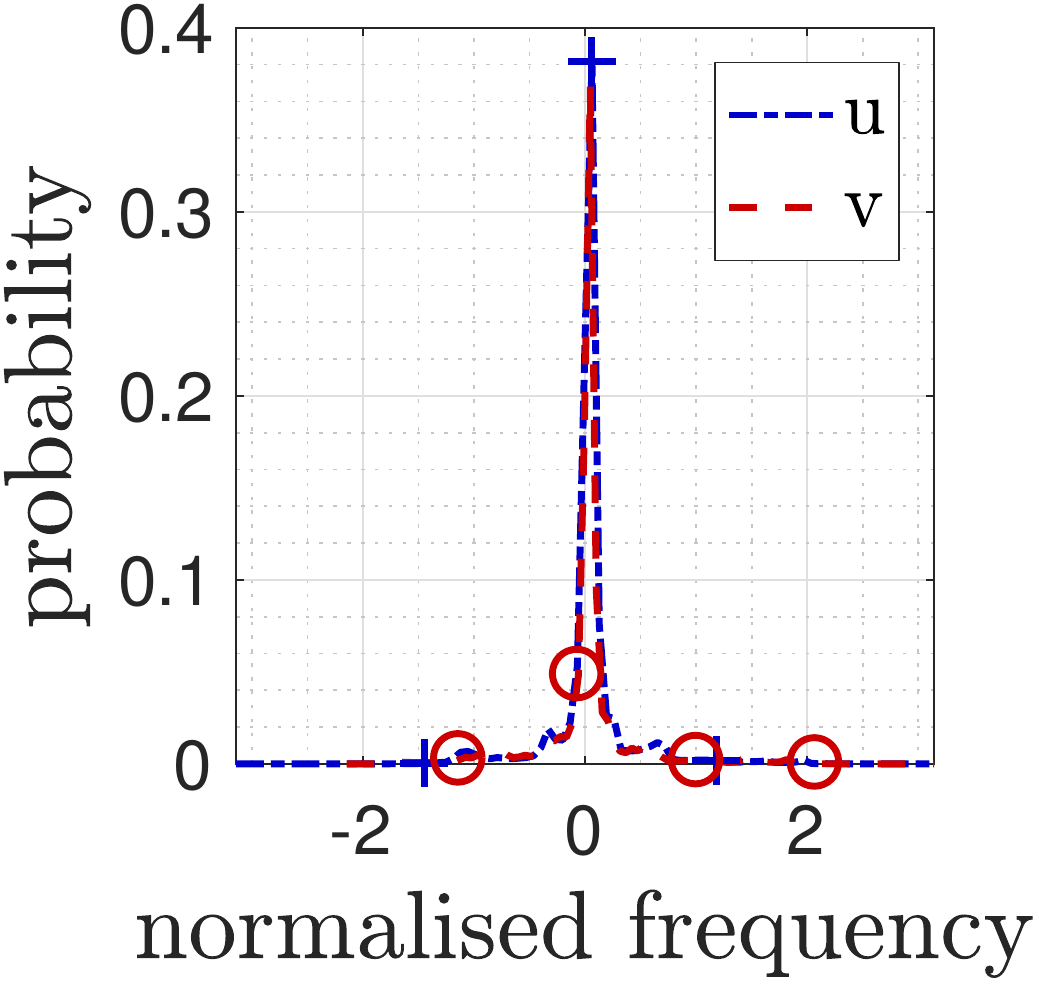}
	\end{minipage}%
	\hspace{.05\linewidth}
	\begin{minipage}{.33\linewidth}
  		\includegraphics[trim={0px 0px 0px 0px}, clip, height=2.3cm]{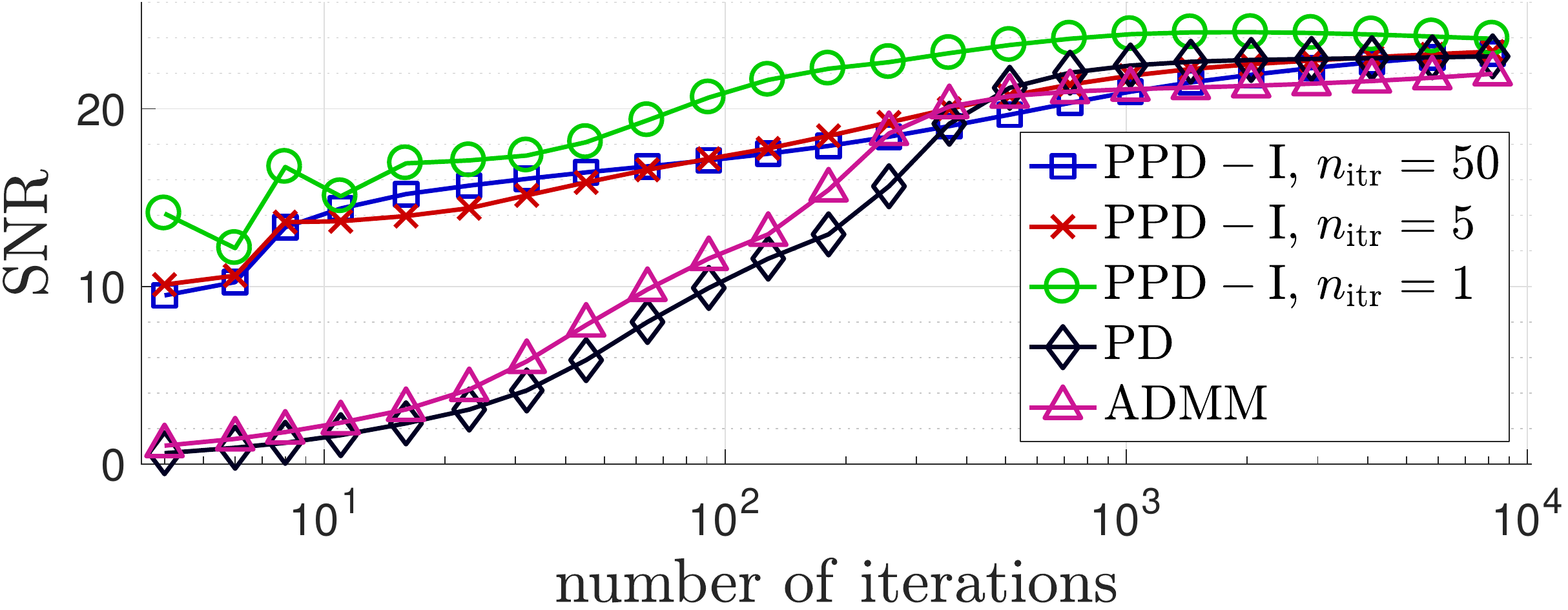}
	\end{minipage}%
	\begin{minipage}{.14\linewidth}
  		\includegraphics[trim={0px 0px 0px 0px}, clip, height=2.3cm]{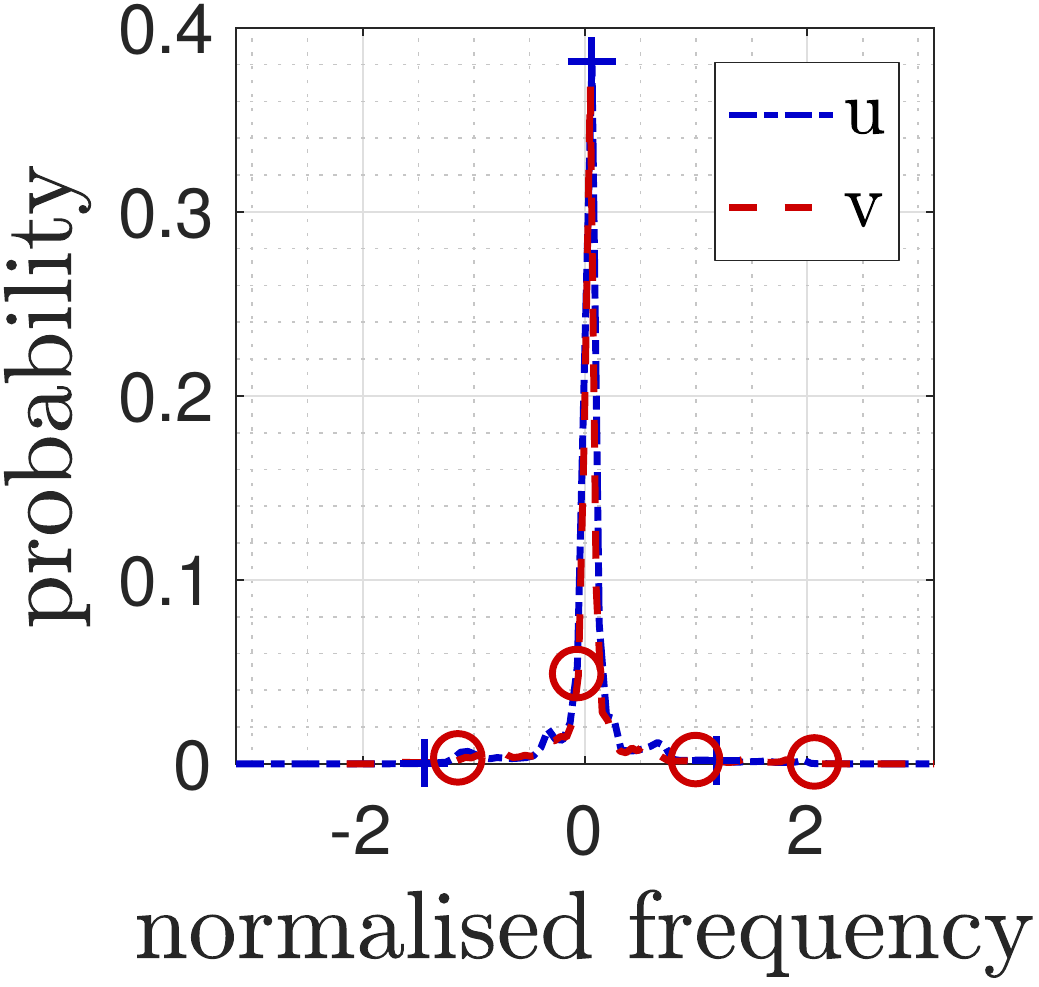}
	\end{minipage}

	\caption{Evolution of the $\rm SNR$ for the \ac{ppd}, \ac{pd} and \ac{admm} algorithms for the reconstruction of the (left) galaxy cluster and (right) Cygnus A test image with a realistic $u$--$v$ coverage corresponding to (top) \ac{vla} and (bottom) \ac{ska}.
The shape of the distribution of the $u$ and $v$ normalised coordinates is presented next to the graph portraying the evolution of the $\rm SNR$.
The visibilities are corrupted by Gaussian noise to produce a 30$\rm dB$ $\rm iSNR$.
The number of sub-iteration $n_{\rm itr}$ performed by \ac{ppd} to estimate the ellipsoid projection is also reported.}
	\label{real-cov}
\end{figure*}

To further validate the behaviour of the algorithms, we also study them for the reconstruction of the two test images using simulated, but realistic \ac{ska} and \ac{vla} coverages.
The evolution of the ${\rm SNR}$ a s function of iteration number for these test cases is presented in Figure \ref{real-cov}.
In all tests \ac{ppd} maintains a similar level of acceleration as observed before, for the generalised Gaussian distributed $u$--$v$ coverages.
For the \ac{ska} coverages, where the conditioning number of the preconditioning matrix is large, the number of sub-iterations begins to affect the evolution of \ac{ppd}.
Especially of the Cygnus A image it seems that using only one sub-iteration is actually faster.
This behaviour is probably due to the fact that the preconditioning matrix is not optimal. 
Performing only one sub-iteration can be understood as projection onto a \bc slightly \ec different ellipsoid.

\begin{figure}
	\centering
	\begin{minipage}{.99\linewidth}
	        \centering
	        \includemedia[noplaybutton, attachfiles, addresource=sim-figs/vla-gc.mp4, activate=onclick,
	        flashvars={
source=sim-figs/vla-gc.mp4
&autoPlay=true
&loop=false
}]{\includegraphics[trim={0px 0px 0px 0px}, clip, height=3.6cm]{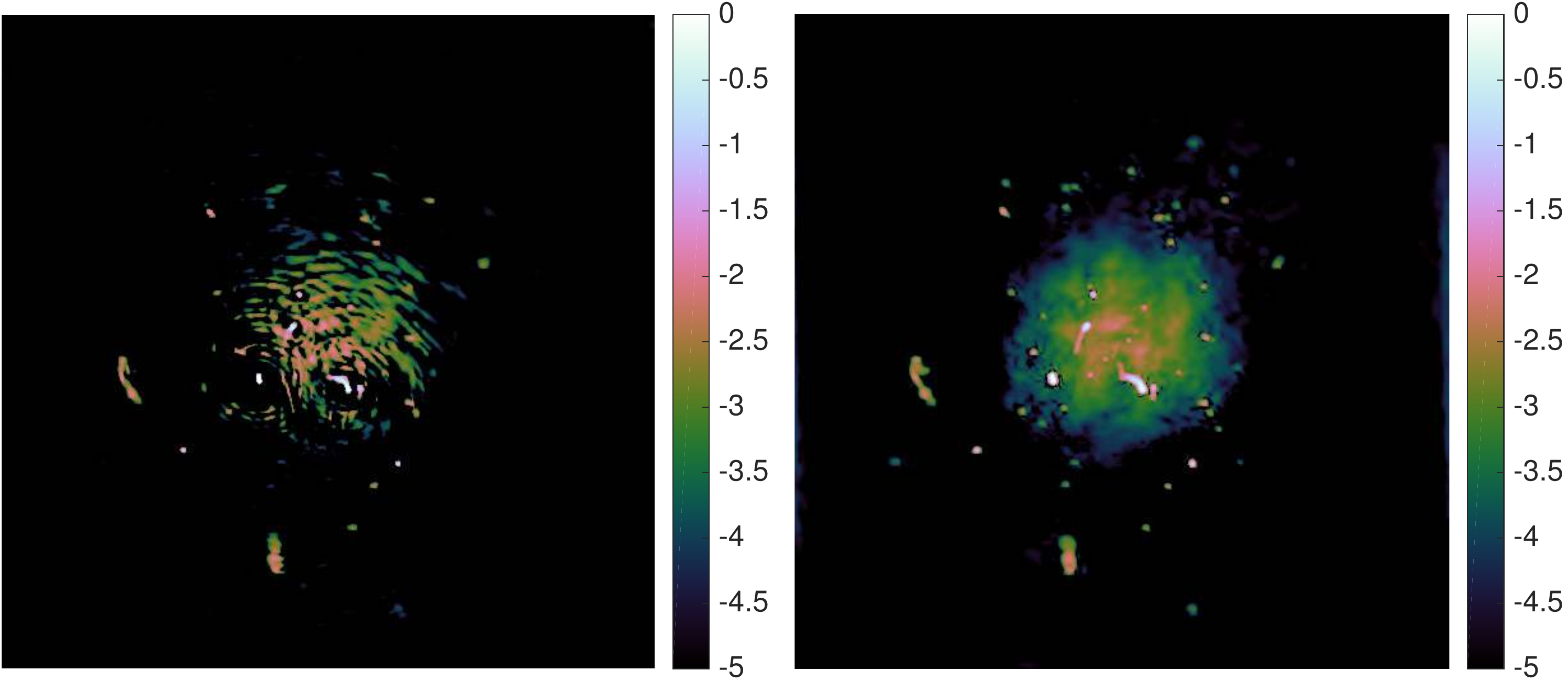}}{VPlayer9.swf}
	\end{minipage}

	\caption{The reconstructed images for \ac{ppd}, with the number of sub-iteration $n_{\rm itr} = 1$, and \ac{pd} at iteration $99$ for the galaxy cluster image with \ac{vla} coverage, corresponding to the tests presented in top, left graph from Figure \ref{real-cov}. The figure contains an animation with the solutions obtained during the first $2048$ iterations. The animation is only supported when the \ac{pdf} file is opened using Adobe Acrobat Reader.}
	\label{real-cov-gc-movie}
\end{figure}

\begin{figure}
	\centering
	\begin{minipage}{.99\linewidth}
	        \centering
	        \includemedia[noplaybutton, attachfiles, addresource=sim-figs/ska-ca.mp4, activate=onclick,
	        flashvars={
source=sim-figs/ska-ca.mp4
&autoPlay=true
&loop=false
}]{\includegraphics[trim={0px 0px 0px 0px}, clip, height=7.2cm]{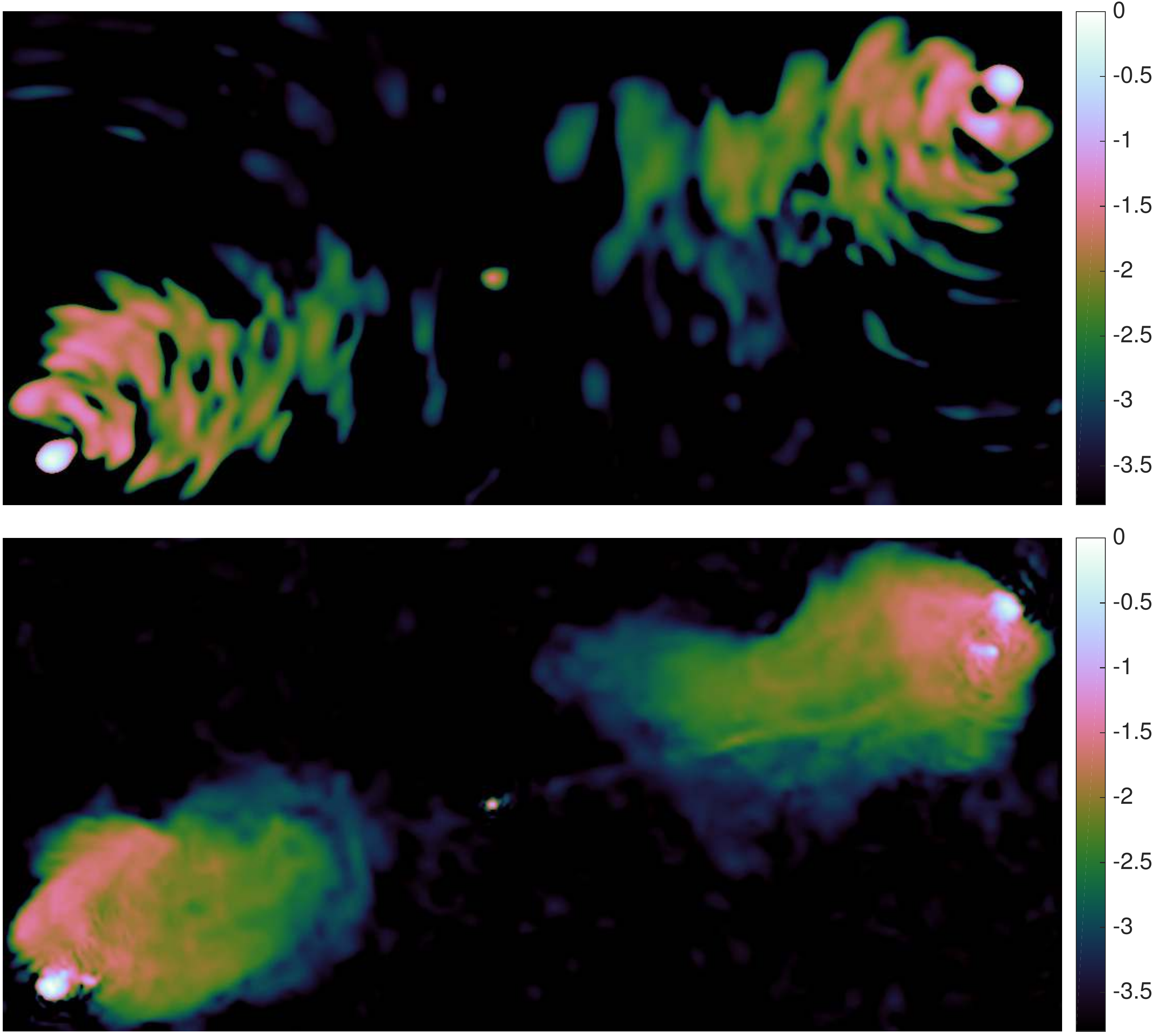}}{VPlayer9.swf}
	\end{minipage}

	\caption{The reconstructed images for \ac{ppd}, with the number of sub-iteration $n_{\rm itr} = 1$, and \ac{pd} at iteration $99$ the Cygnus A image with the \ac{ska} coverage, corresponding to the tests presented in bottom, right graph from Figure \ref{real-cov}. The figure contains an animation with the solutions obtained during the first $2048$ iterations. The animation is only supported when the \ac{pdf} file is opened using Adobe Acrobat Reader.}
	\label{real-cov-ca-movie}
\end{figure}

Figures \ref{real-cov-gc-movie} and \ref{real-cov-ca-movie} contain the reconstructed images for \ac{ppd} and \ac{pd} at iteration $99$ for the galaxy cluster image with \ac{vla} coverage and the Cygnus A image with the \ac{ska} coverage, respectively.
The reconstruction quality achieved by \ac{ppd} at this iteration is evident.
Such a reconstruction is possible with \ac{pd} only by performing approximatively $10$ times more iterations.
The figures also contain embedded an animation that cycles through the iterations and shows the solution estimates at each iteration.\footnote{The animation is only supported when the \ac{pdf} file is opened using Adobe Acrobat Reader, \url{https://get.adobe.com/reader/}}
The evolution of \ac{ppd} resembles a behaviour that is associated with the uniform weighting used for \sw{clean} while the evolution of \ac{pd} resembles that associated with natural weighting. 
Both methods however converge towards the same global solution, the solution of the natural weighted data.
The sampling density information is only incorporated into \ac{ppd} to accelerate the convergence speed.

\subsection{Real data reconstruction}

\begin{figure*}
	\centering
	\begin{minipage}{.33\linewidth}
	        \centering
  		\includegraphics[trim={0px 0px 0px 0px}, clip, height=5cm]{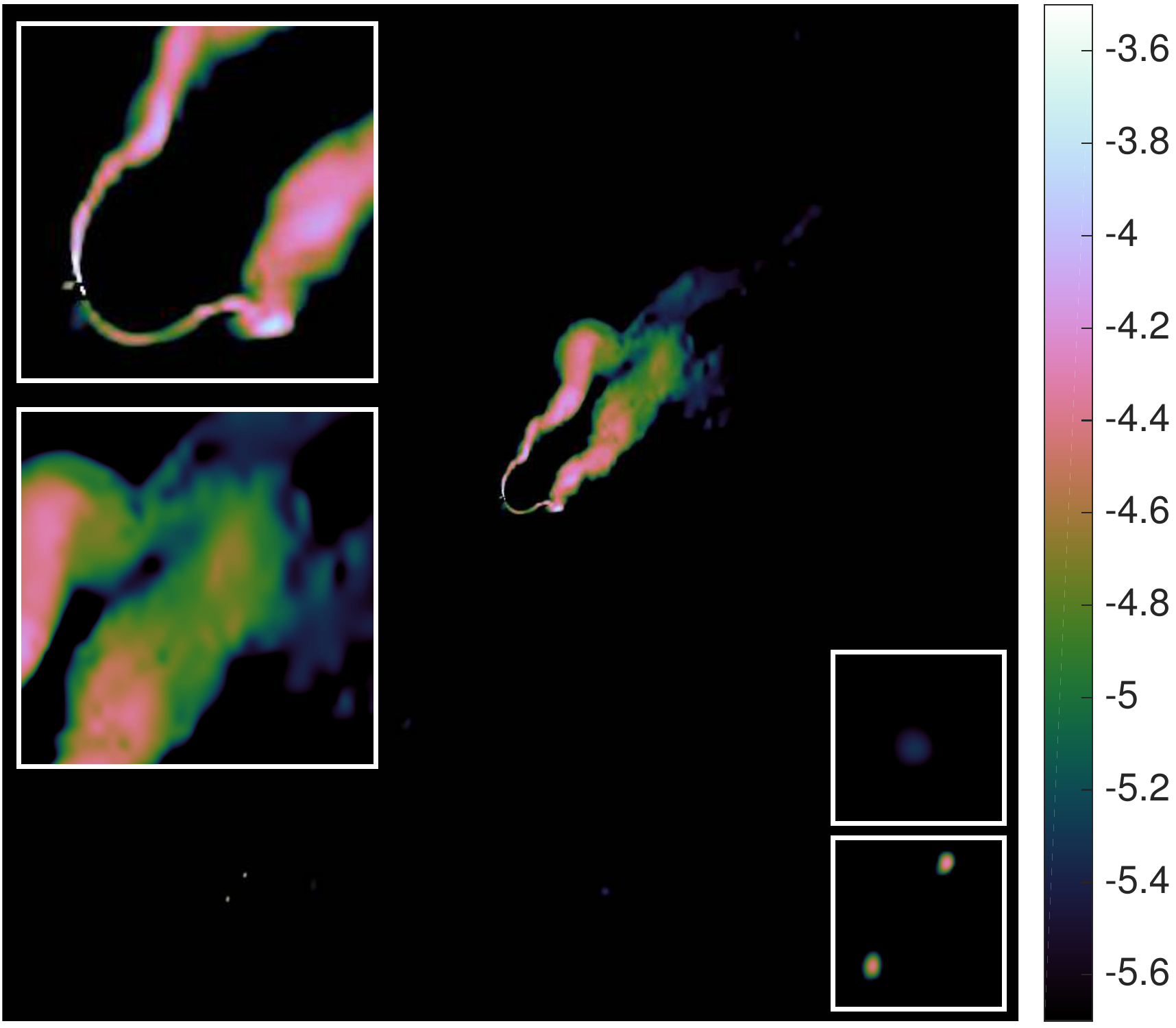}
	\end{minipage}
	\begin{minipage}{.33\linewidth}
	        \centering
  		\includegraphics[trim={0px 0px 0px 0px}, clip, height=5cm]{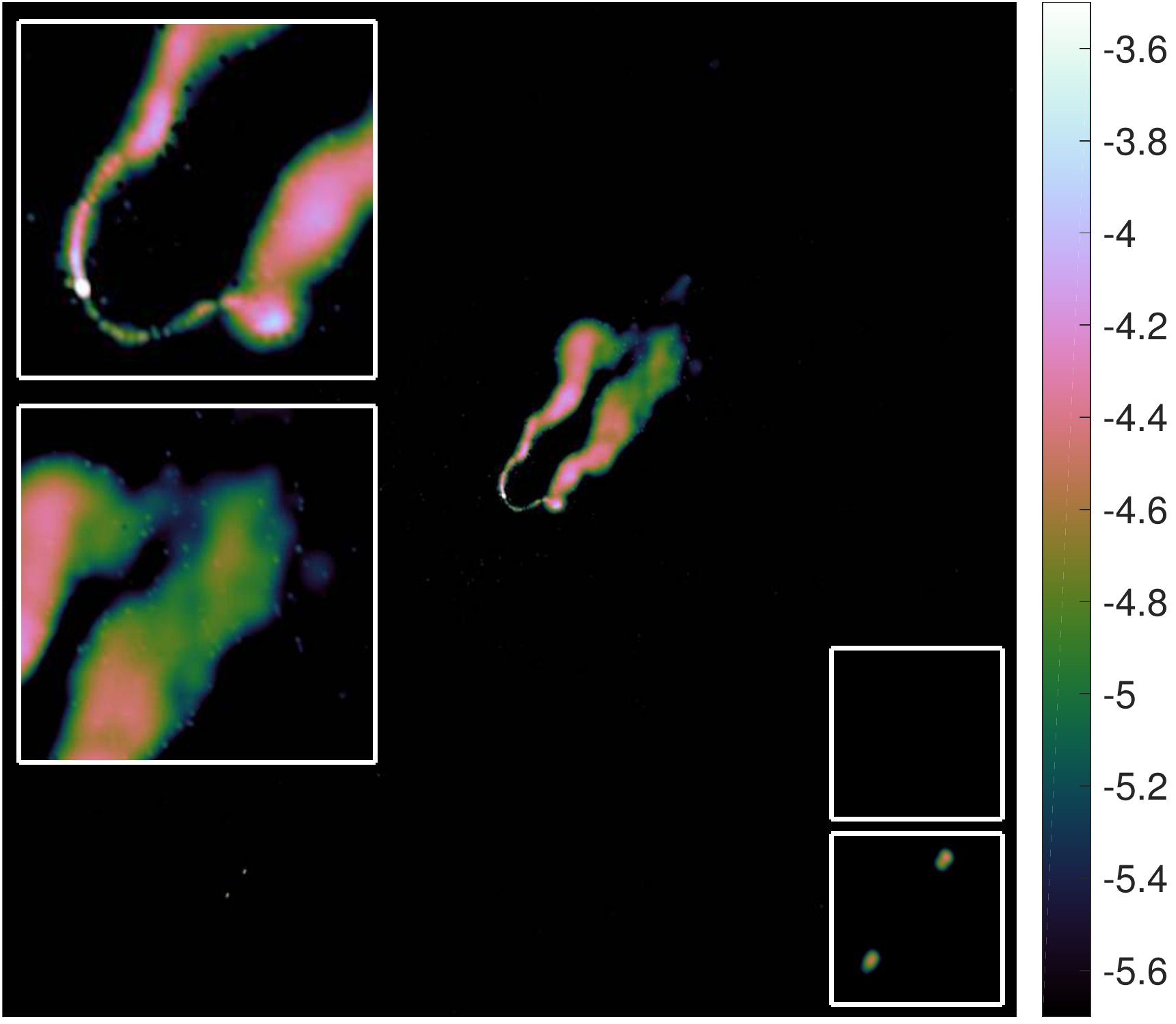}
	\end{minipage}
	\begin{minipage}{.33\linewidth}
	        \centering
  		\includegraphics[trim={0px 0px 0px 0px}, clip, height=5cm]{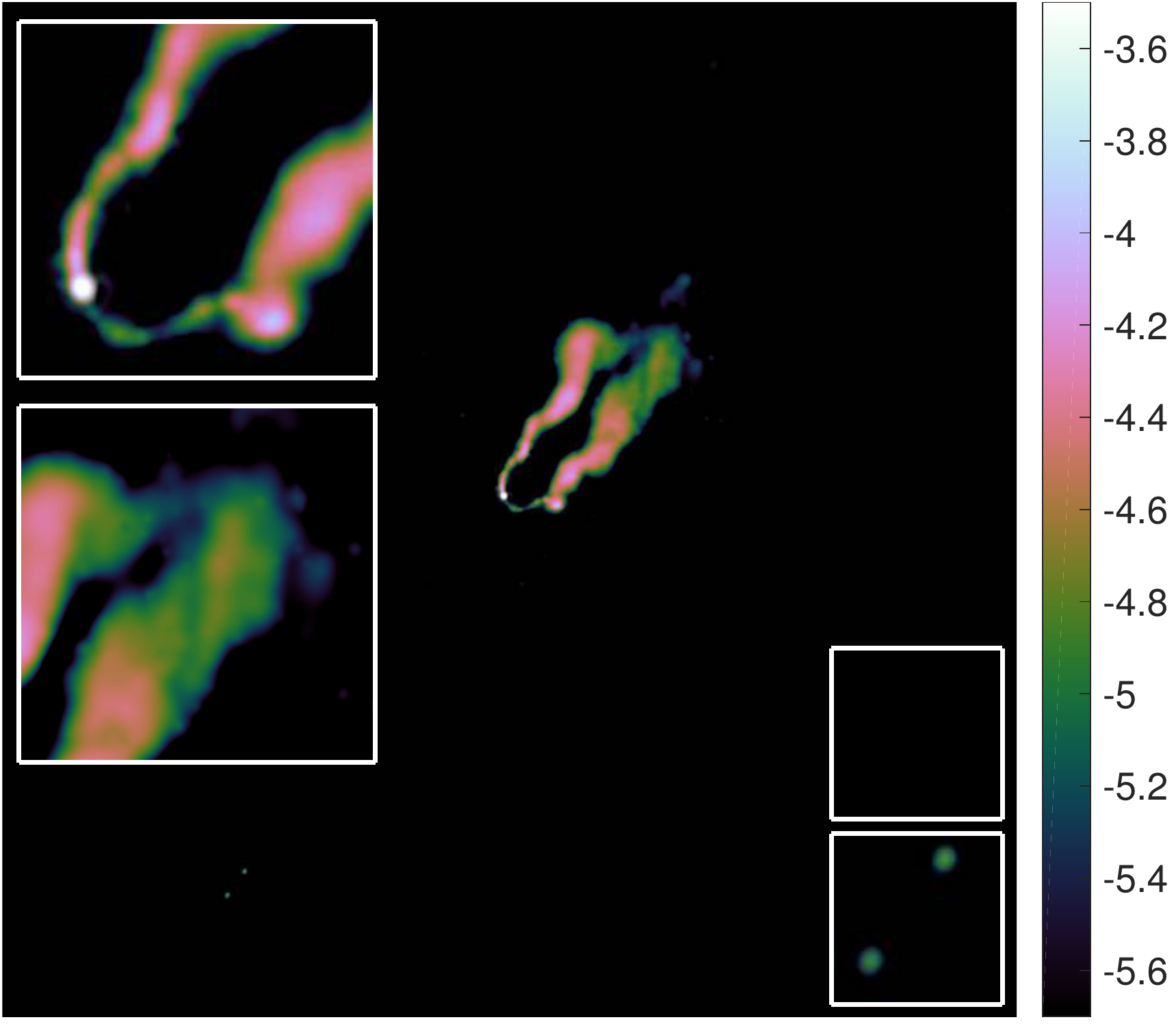}
	\end{minipage}

	\begin{minipage}{.33\linewidth}
	        \centering
  		\phantom{\includegraphics[trim={0px 0px 0px 0px}, clip, height=5cm]{sim-figs/ppd-3c129.pdf}}
		\hspace{10pt}
	\end{minipage}
	\begin{minipage}{.33\linewidth}
	        \centering
  		\includegraphics[trim={0px 0px 0px 0px}, clip, height=5cm]{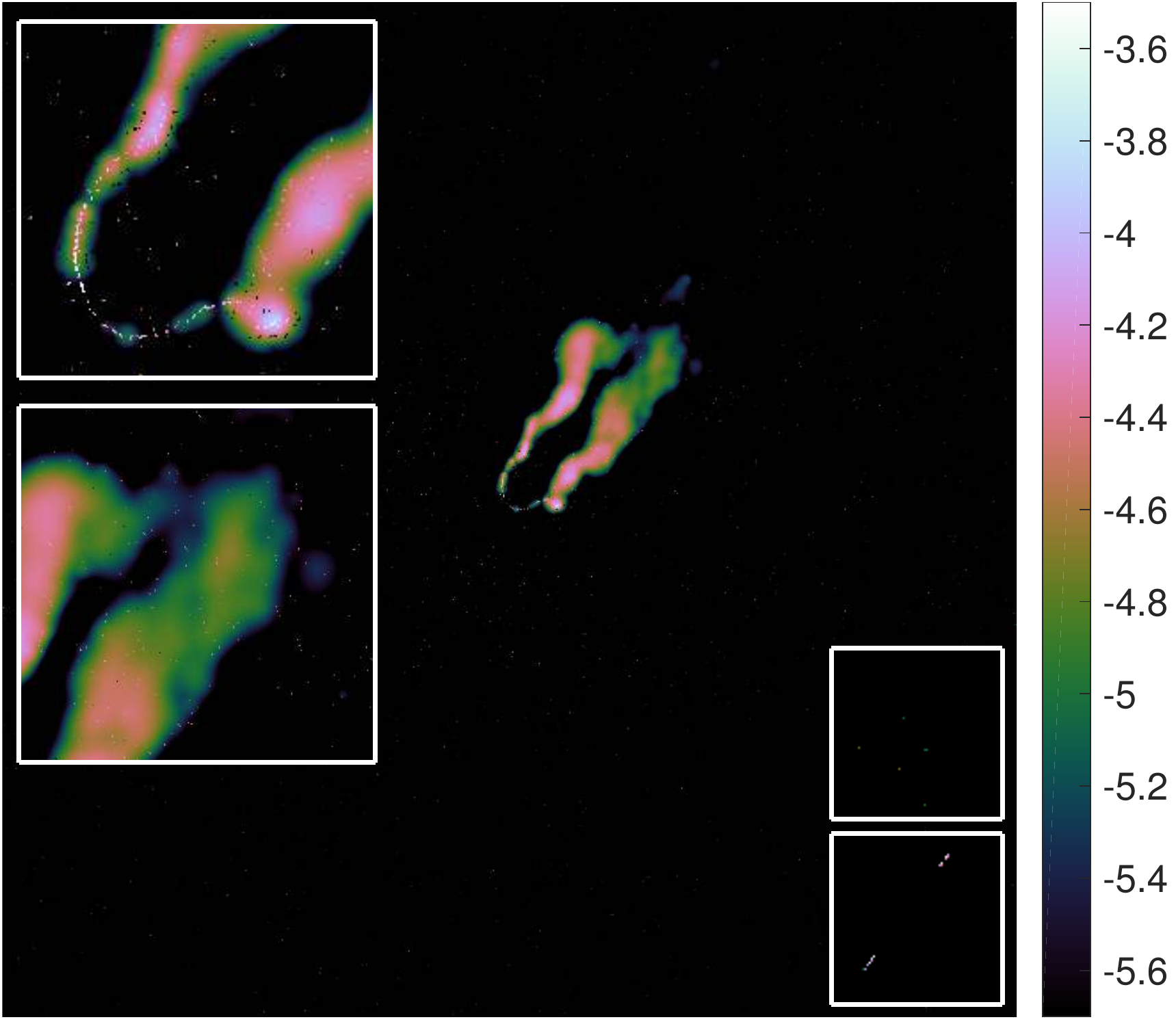}
	\end{minipage}
	\begin{minipage}{.33\linewidth}
	        \centering
  		\includegraphics[trim={0px 0px 0px 0px}, clip, height=5cm]{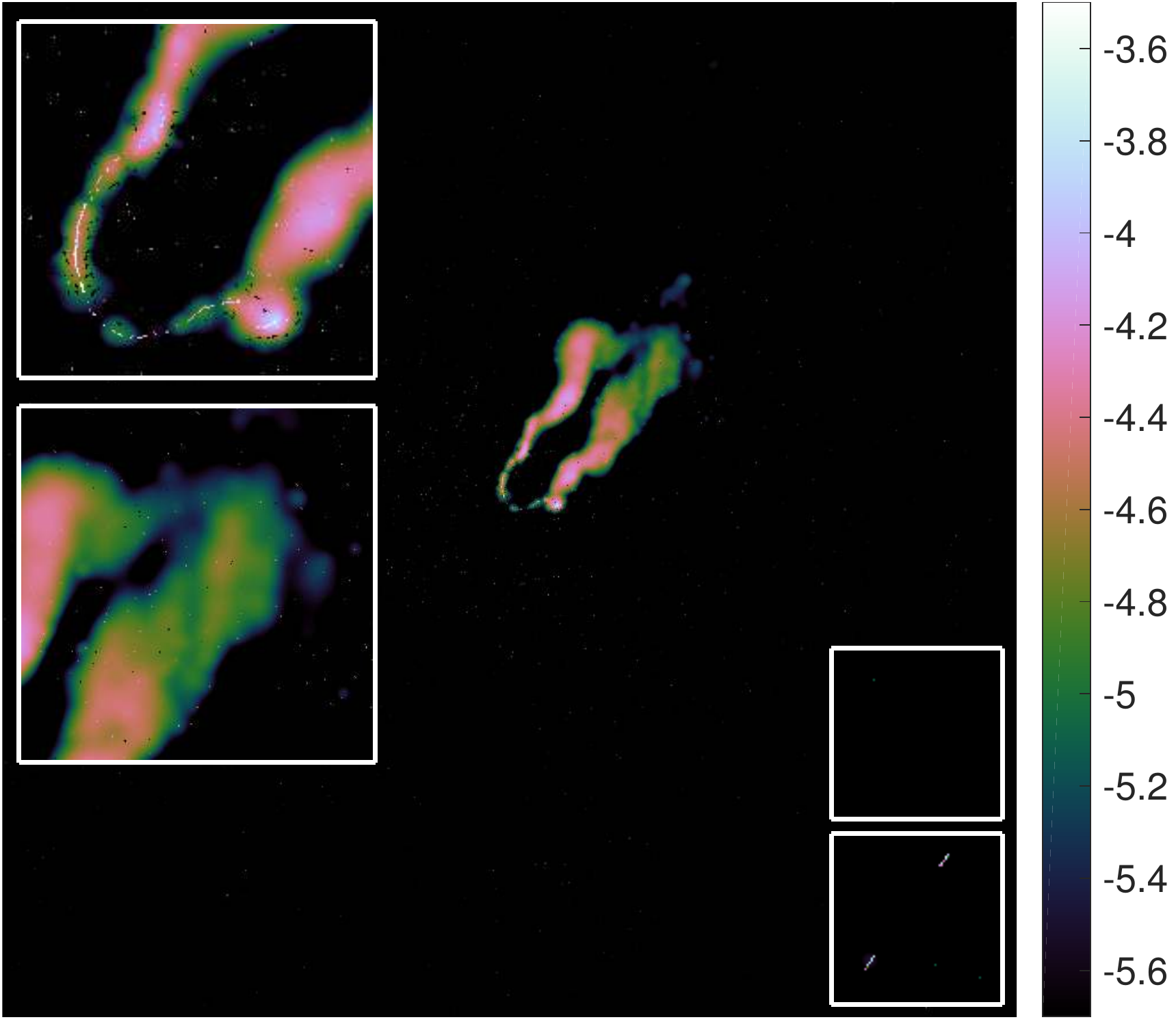}
	\end{minipage}

	\begin{minipage}{.33\linewidth}
	        \centering
  		\includegraphics[trim={0px 0px 0px 0px}, clip, height=5.37cm]{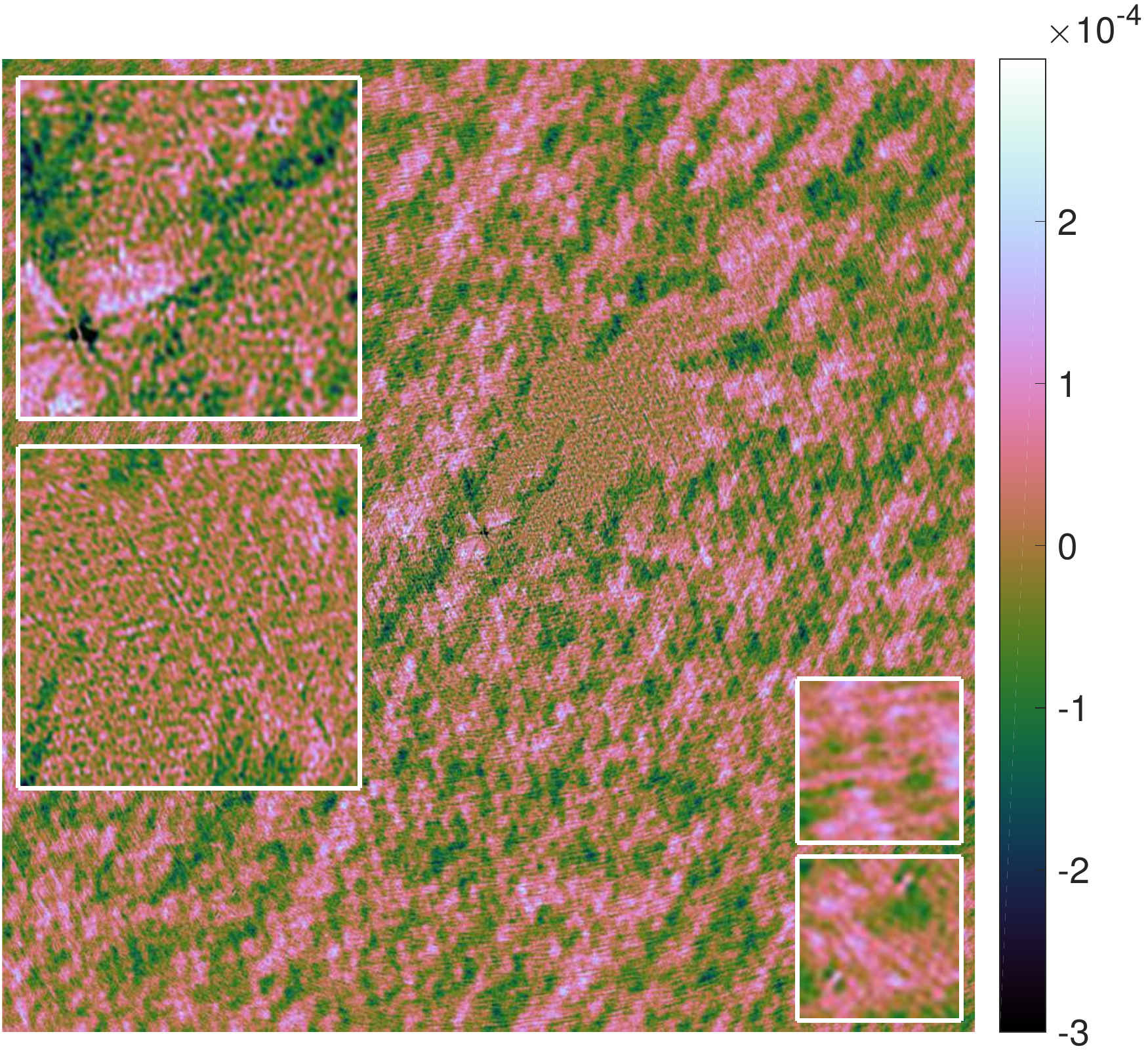}
	\end{minipage}
	\begin{minipage}{.33\linewidth}
	        \centering
  		\includegraphics[trim={0px 0px 0px 0px}, clip, height=5.37cm]{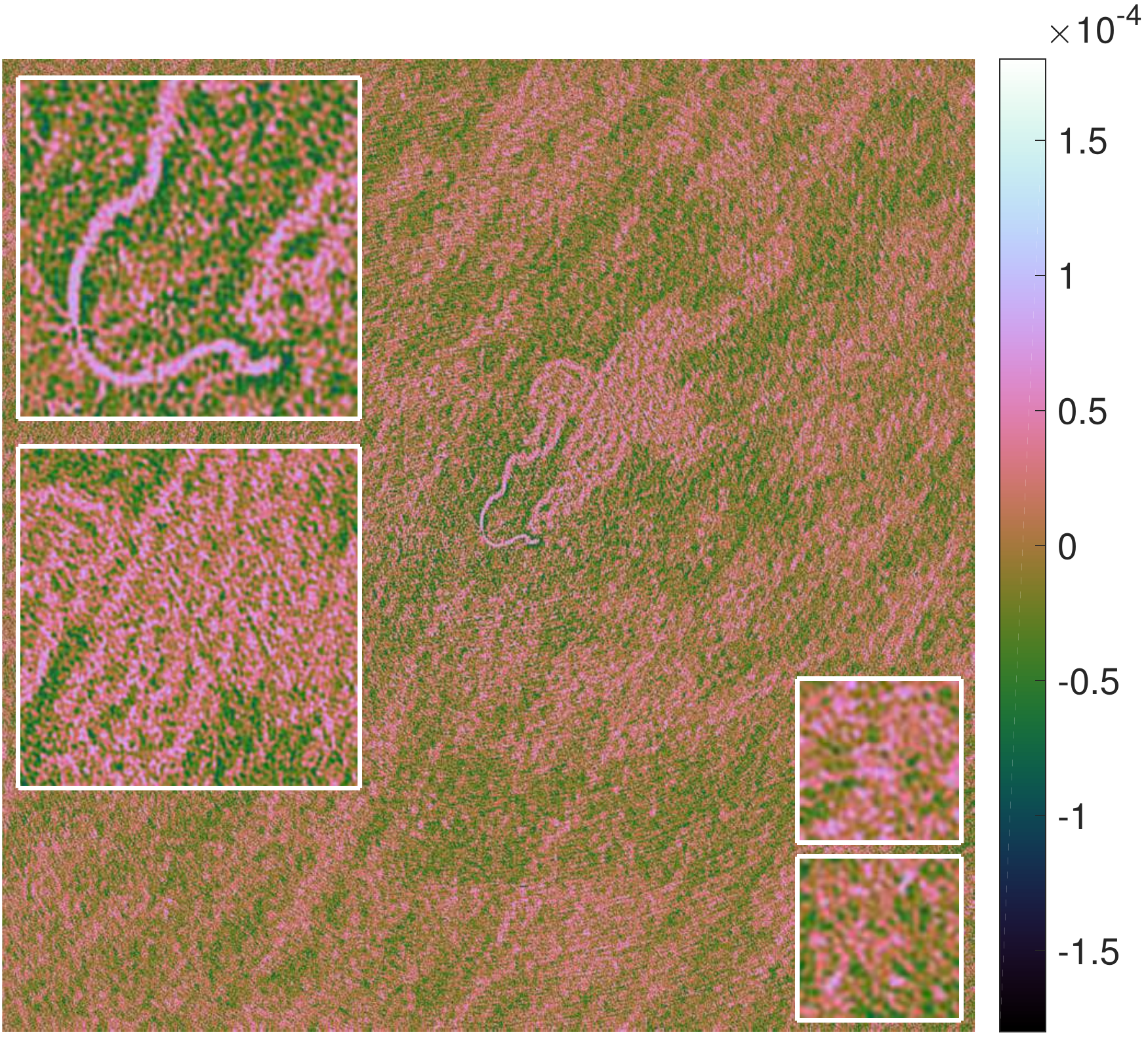}
	\end{minipage}
	\begin{minipage}{.33\linewidth}
	        \centering
  		\includegraphics[trim={0px 0px 0px 0px}, clip, height=5.37cm]{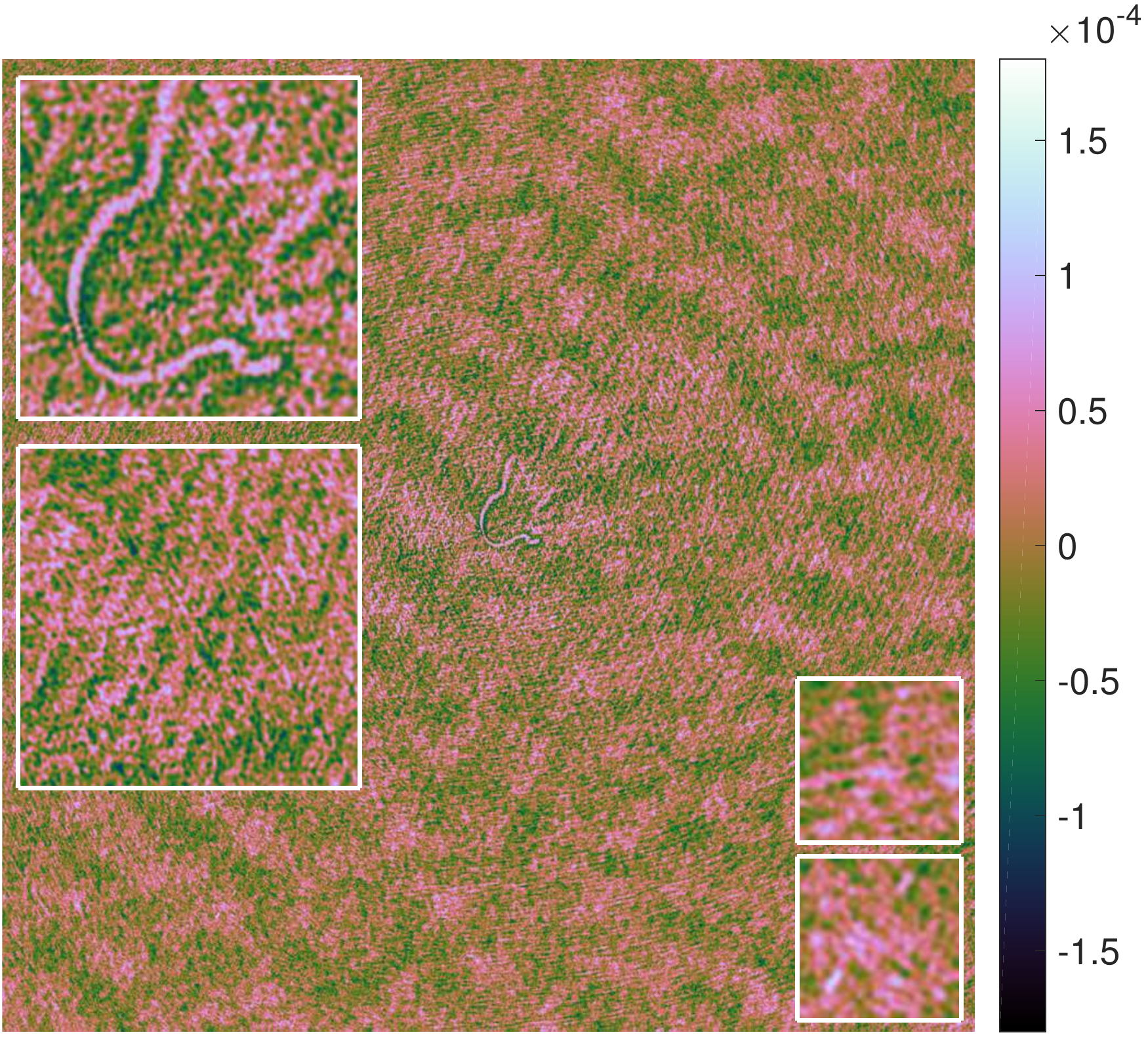}
	\end{minipage}
  	
	\caption{Reconstruction of the 3C129 radio galaxy from $307~780$ visibilities acquired using the \ac{vla}. The resolution of the images is twice the resolution of the telescope. The images from left to right correspond to the \ac{ppd} algorithm with $n_{\rm itr} = 5$, \ac{ms-clean} with uniform weighting and \ac{ms-clean} with natural weighting, respectively. From top to bottom the images are the $\log_{10}$ scale reconstructed image, the $\log_{10}$ \sw{clean} model image and the linear scale residual image.
\bc Each residual is computed as $\bm{\Phi}^\dagger \bs{y} - \bm{\Phi}^\dagger \bm{\Phi} \bs{x}^{(t)}$ normalised such that the associated point spread function has a maximum value of $1$. \ec
For \sw{clean}, the reconstructed image is produced by convolving the model image with the normalised \bc \sw{clean} beam. \ec
Both the \sw{clean} model and reconstructed images have negative components which are not displayed.
The \ac{ppd} reconstruction does not require any post processing. It does not produce a model image that needs to be convolved with the \sw{clean} beam, this space being left blank for \ac{ppd}.}
	\label{real-data}
\end{figure*}

\begin{figure}
	\centering
	\begin{minipage}{.65\linewidth}	
		\hspace{-10pt}
  		\includegraphics[trim={0px 0px 0px 0px}, clip, height=2.42cm]{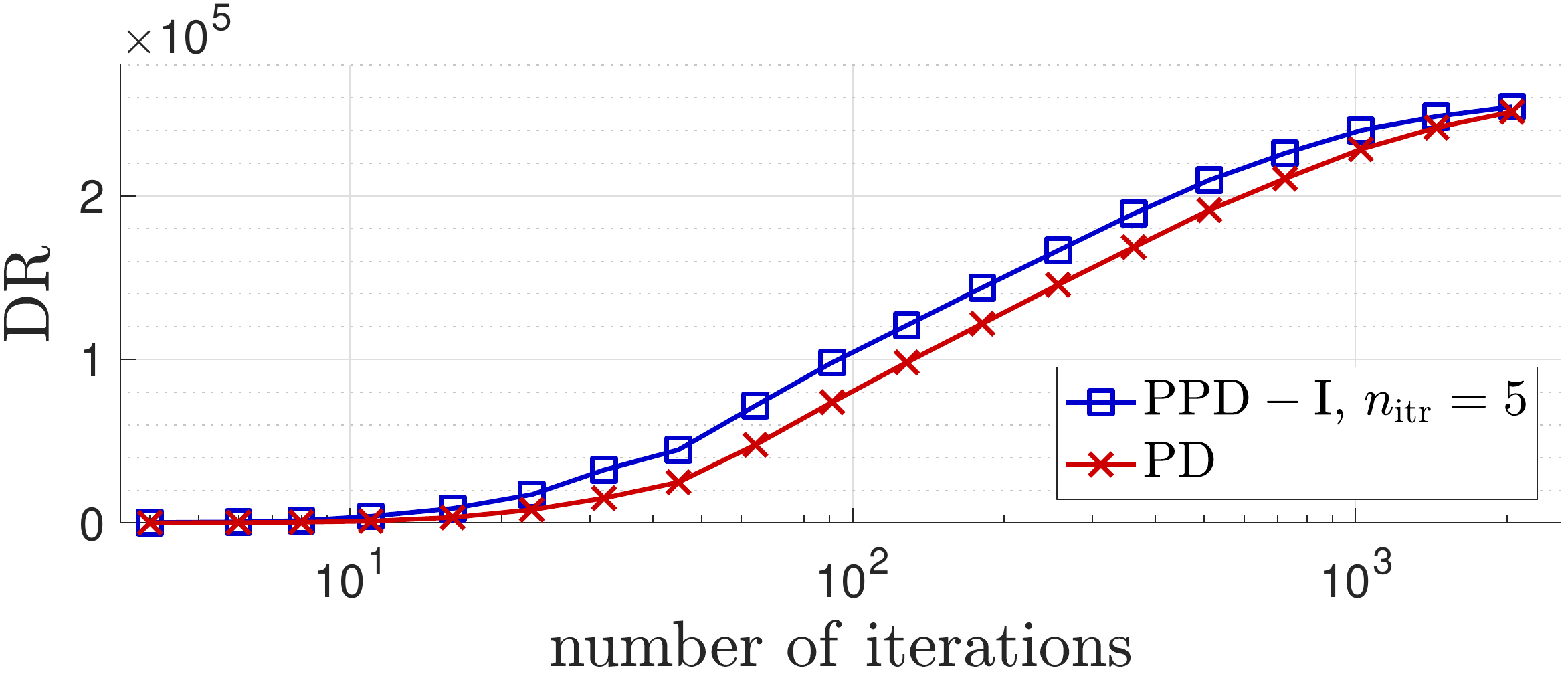}
	\end{minipage}%
	\begin{minipage}{.25\linewidth}
		\vspace{4pt}
  		\includegraphics[trim={0px 0px 0px 0px}, clip, height=2.3cm]{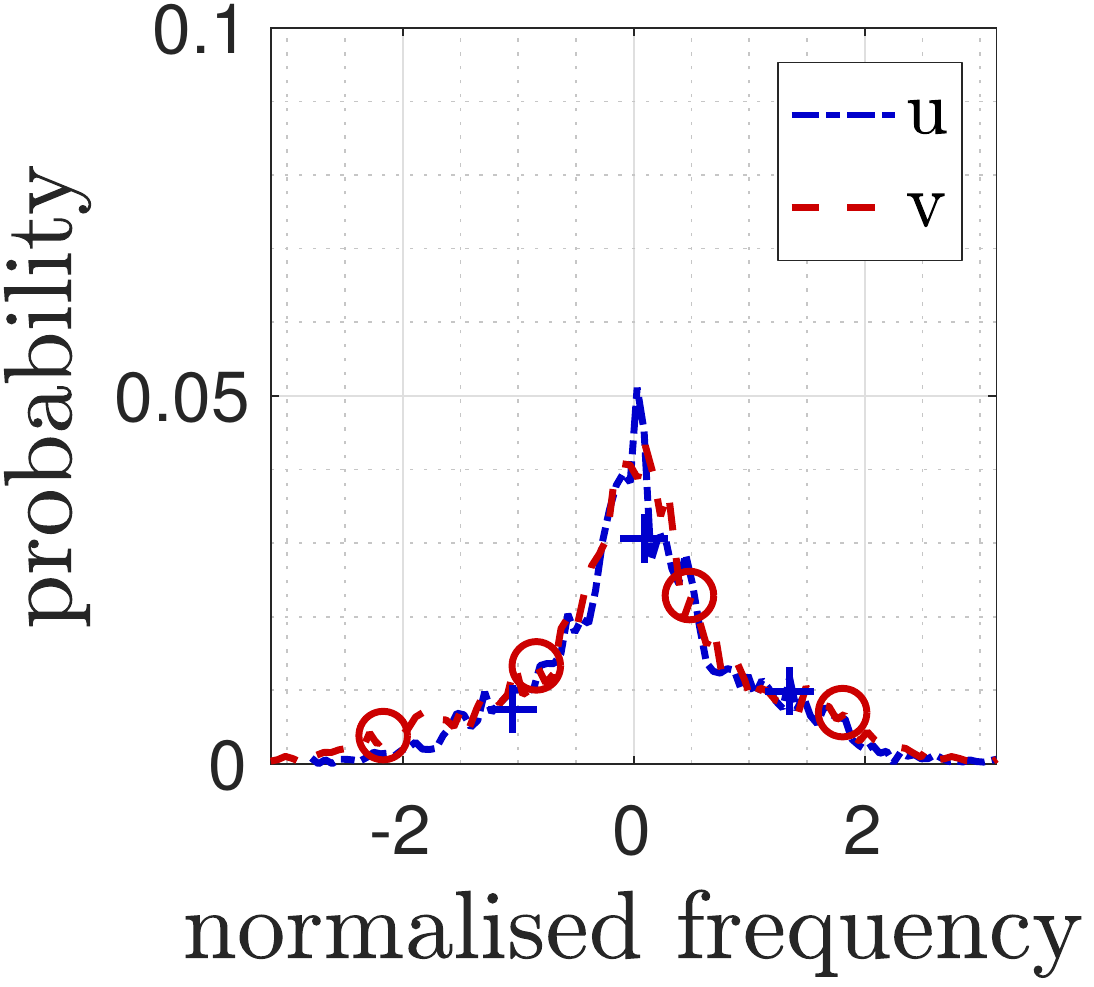}
	\end{minipage}%
	
	\caption{Evolution of the $\rm DR$ for the \ac{ppd} and \ac{pd} and \ac{admm} algorithms for the reconstruction of the 3C129 radio galaxy.
The shape of the distribution of the $u$ and $v$ normalised coordinates is presented next to the graph portraying the evolution of the $\rm DR$. 
\ac{ppd} performed a number of sub-iteration $n_{\rm itr} = 5$.}
	\label{real-data-dr}
\end{figure}

For the real data scenario, we study the reconstruction quality of \ac{ppd} in comparison with \ac{ms-clean} using observations of the 3C129 radio galaxy performed with the \ac{vla}. The reconstructed images are illustrated in $\log_{10}$ scale in Figure \ref{real-data}.
We note that \ac{ppd} achieves better quality in terms of both resolution and sensitivity.
It is able to better recover the faint emissions towards the tail of the main emission and has very little noise incorporated in the image.
In comparison, \ac{ms-clean} includes multiple spurious components in the model map and due to the post processing achieves a poor resolution, especially around the main bright source that generates the two emission plumes.
The resolution is much worse when the natural weighting is used since the size of the \bc \sw{clean} beam \ec used is larger. The \sw{clean} model is also lower resolution than in the uniform weighting case.

To better visualise the reconstruction quality, we provide in all images enlarged sections of the main source in the two boxes on the left and of the fainter point sources, from the lower part of the recovered image, in the two boxes on the right.
The faint emission showcased enlarged in the right, upper box for the \ac{ppd} reconstruction is most likely the source $\rm C$ reported by \cite{Lane2002}.
This is the faintest emission \ac{ppd} can detect without introducing noise and deconvolution artefacts.
Note that this source, as well as the emission tail of 3C129 are around $2.5$ orders of magnitude fainter than the brightest source.
\ac{ms-clean} is unable to recover these emissions well and has brighter spurious components around the main emission.
Setting the deconvolution threshold lower for \ac{ms-clean}, in order to extract more of the signal from the measurements, greatly increases the amount of spurious components detected.

As a last figure, we present the evolution of the $\rm DR$ for \ac{ppd} and \ac{pd} as a function of the number of iterations in Figure \ref{real-data-dr}.
This serves to validate the acceleration also on real data.
Here, \ac{ppd} is shown to be faster than \ac{pd}.
The distribution of the $u$--$v$ coordinates, also reported in Figure \ref{real-data-dr}, is not that extreme in this case and the speed up is small, of the order of $1.5$, which is consistent with the previous simulations.
This test serves to prove that the preconditioning works on real data.
For more unbalanced sampling profiles we expect a larger acceleration, as demonstrated through simulations.
\bc Also, since the number of sub-iterations performed by \ac{ppd} to approximate the preconditioned proximity operator is small, $n_{\rm itr} = 5$, the complexity per iteration is similar to that of \ac{pd}. \ec

\section{Conclusions}
\label{sec-conc}

We proposed an acceleration of the \ac{pd} algorithmic framework for solving the \ac{ri} imaging problem.
Building on the highly parallelizable structure of the \ac{pd} algorithm, the accelerated \ac{ppd} algorithm, benefits from all the flexibility of the \ac{pd}, allowing for an efficient distributed implementation, by using full splitting and randomised updates.
The analogy between the \sw{clean} major-minor loop and the forward-backward iterations used by the method, 
can portray \ac{ppd} as being composed of sophisticated \sw{clean}-like iterations running in parallel in multiple data, prior, and image spaces.

The proposed approach reconciles natural and uniform weighting of \sw{clean} algorithms. It optimises resolution by accounting for the correct noise statistics, leveraging natural weighting in the definition of the minimisation problem for image reconstruction. It also optimises sensitivity by enabling accelerated convergence through a preconditioning strategy incorporating sampling density information \`a la uniform weighting.

We study the acceleration through extensive simulations with realistic $u$--$v$ coverages and using real visibilities from the observation of the 3C129 radio galaxy with the \ac{vla}.
The preconditioning strategy is able to increase the convergence speed by up to an order of magnitude for highly non-uniformly sampled coverages.
We also showcase the reconstruction quality in comparison with \ac{ms-clean} for this data, exemplifying the improved resolution and sensitivity the \ac{ppd} method offers.

Our Matlab code is available online on GitHub, \url{http://basp-group.github.io/ppd-for-ri/}.
In the near future we intend to provide an efficient implementation in the \sw{purify} \sw{c++} package for a distributed computing infrastructure.

\section*{Acknowledgements}

This work was supported by the UK Engineering and Physical Sciences Research Council (EPSRC, grants EP/M011089/1 and EP/M008843/1).
We would like to thank Federica Govoni and Matteo Murgia for providing the simulated galaxy cluster image.

\bibliographystyle{mnras.bst}
\balance

\bibliographystyle{mnras.bst}
\balance

% \bibliography{abrev,convex-opt,radio-interferometry,signal-proc}

\bsp
\label{lastpage}
\end{document}